%% file: lang-paper.tex
\documentclass{article}
\usepackage{amssymb,latexsym}
\usepackage{graphicx,amsmath,amssymb}
\usepackage{amsthm}
\usepackage{microtype}
\usepackage{setspace}
\usepackage{bussproofs}
\usepackage{mathrsfs}
\usepackage{color}
\usepackage{fullpage}
\usepackage{amssymb}
\usepackage{hyperref}
\usepackage[all]{xy}
\usepackage{tikz}
\usetikzlibrary{arrows.meta}
\usepackage{verbatim}
\usepackage{ifthen}
\usepackage{xargs}
\usepackage{listings}

\usepackage{color}
\definecolor{keywordcolor}{rgb}{0.7, 0.1, 0.1}   
\definecolor{tacticcolor}{rgb}{0.0, 0.1, 0.6}    
\definecolor{commentcolor}{rgb}{0.4, 0.4, 0.4}   
\definecolor{symbolcolor}{rgb}{0.0, 0.1, 0.6}    
\definecolor{sortcolor}{rgb}{0.1, 0.5, 0.1}      
\definecolor{attributecolor}{rgb}{0.7, 0.1, 0.1} 

\begingroup
\makeatletter
\lst@InputCatcodes
\input{lstlean.tex}
\endgroup
\makeatletter
\lst@AddToHook{Init}{\interlinepenalty=10000\relax}
\lst@AddToHook{OnNewLine}{\ifnum\lst@newlines=1\par\nobreak\fi}
\makeatother

\newcommand{\axref}[1]{(\hyperref[ax:#1]{#1})}

\newcommand{\newref}[4][]{
\ifthenelse{\equal{#1}{}}{\newtheorem{h#2}[hthm]{#4}}{\newtheorem{h#2}{#4}[#1]}
\expandafter\newcommand\csname r#2\endcsname[1]{#3~\ref{#2:##1}}
\expandafter\newcommand\csname R#2\endcsname[1]{#4~\ref{#2:##1}}
\newenvironmentx{#2}[2][1=,2=]{
\ifthenelse{\equal{##2}{}}{\begin{h#2}}{\begin{h#2}[##2]}
\ifthenelse{\equal{##1}{}}{}{\label{#2:##1}}
}{\end{h#2}}
}

\newref[section]{thm}{Theorem}{Theorem}
\newref{lem}{Lemma}{Lemma}
\newref{prop}{Proposition}{Proposition}
\newref{cor}{Corollary}{Corollary}

\theoremstyle{definition}
\newref{defn}{Definition}{Definition}
\newref{example}{Example}{Example}

\theoremstyle{remark}
\newref{remark}{Remark}{Remark}

\newcommand{\fs}[1]{\mathsf{#1}}
\newcommand{\lcon}{\fs{left}}
\newcommand{\rcon}{\fs{right}}
\newcommand{\Path}{\fs{Path}}
\newcommand{\pcon}{\fs{path}}
\newcommand{\I}{\mathbf{I}}
\newcommand{\coe}{\fs{coe}}
\newcommand{\iso}{\fs{iso}}
\newcommand{\id}{\fs{id}}
\newcommand{\U}{\mathcal{U}}

\newcommand{\lam}[1]{\lambda{#1}.}

\newcommand{\Int}{\fs{Int}}
\newcommand{\pos}{\fs{pos}}
\newcommand{\negg}{\fs{neg}}
\newcommand{\con}{\fs{con}}
\newcommand{\Nat}{\fs{Nat}}
\newcommand{\suc}{\fs{suc}}

\newcommand{\Prop}{\fs{Prop}}
\newcommand{\Set}{\fs{Set}}
\newcommand{\Type}{\fs{Type}}
\newcommand{\PI}{\fs{PI}}
\newcommand{\LEM}{\fs{LEM}}

\newcommand{\arendlib}{\fs{arend\text{-}lib}}

\renewcommand{\colon}{\, :\,}

\numberwithin{figure}{section}

\begin{document}

\title{Theorem prover Arend}
\author{Fedor Part \qquad Valery Isaev \qquad Sergey Sinchuk\\[0.4em]
  \small JetBrains Research}
\date{}

\maketitle

\begin{abstract}
This paper is an exposition of the theorem prover Arend developed at JetBrains Research.
Arend language has been designed to facilitate the formalization of mathematics within the framework of univalent foundations and with the focus on constructive mathematics. It is based on a variant of homotopy type theory with cubical syntax that is more closely aligned with Martin-Löf Type Theory (MLTT) than fully computational cubical type theories, while still retaining some of their computational properties useful in practice. Among the most notable features of the Arend language is its powerful system of classes and records supporting subsumptive subtyping, partial implementations and anonymous extensions, which provides extensive flexibility for constructing hierarchies of definitions. Formalization in Arend is streamlined by the IntelliJ Arend plugin, which transforms IntelliJ IDEA into a full-fledged integrated development environment (IDE) for the Arend language. The main
Arend library $\arendlib$ includes formalizations in constructive algebra, topology and synthetic homotopy theory.
\end{abstract}

\section{Introduction}

Interactive theorem provers (ITPs) are software tools designed to assist in the creation, verification, and exploration of formal proofs by providing a rigorous framework for encoding and checking logical reasoning. They leverage formal systems, such as type theory or higher-order logic, to provide an interactive environment where users can encode theorems, construct proofs, and validate their correctness with absolute precision. By eliminating human error, ITPs have become invaluable not only in formalizing complex mathematical results but also in verifying the correctness of critical software and hardware systems, bridging the gap between pure mathematics and practical applications. Modern ITPs, such as Rocq~\cite{CoqArt}, Agda~\cite{NorellAgda}, Lean~\cite{Lean4}, and HOL-based systems like Isabelle/HOL~\cite{IsabelleHOL} and HOL4~\cite{HOL4}, are used both to verify algorithms and protocols and to formalize mathematics. Earlier milestones in mathematical formalization include the Four-Color Theorem~\cite{GonthierFourColor} and the Feit--Thompson Theorem~\cite{GonthierOddOrder}, both verified in Rocq.

Recent Lean formalizations include the main theorem on liquid vector spaces
(the Liquid Tensor Experiment)~\cite{LiquidTensorExperiment} and the
polynomial Freiman--Ruzsa theorem in characteristic~$2$~\cite{PFRFormalization}.
AI-assisted formalization has also reached large mathematical developments:
in September 2026, Anthropic released a complete Lean formalization of
Fermat's Last Theorem, produced with Claude agents and building on Mathlib,
the Imperial College FLT project, and the formalization for regular
primes~\cite{FermatLean2026}.
In the same month, OpenAI released Lean formalizations of finite-time blowup
for the three-dimensional Navier--Stokes equations with smooth forcing and
for the unforced Euler equations~\cite{OpenAINavierStokes2026}.
Its October 2026 release contains over 700 mathematical manuscripts, grouped
into 372 families; the accompanying repository reports that about $42\%$ of
its main results are formalized in Lean~\cite{OpenAIMathematics2026}.

AlphaProof and Aristotle demonstrated automated proof search on formalized
Olympiad problems~\cite{AlphaProof,Aristotle}.
DeepSeek-Prover-V2 provides open models combining informal reasoning and
Lean proof generation~\cite{DeepSeekProverV2}. Leanstral targets
proof development in existing Lean repositories~\cite{Leanstral}.
In these systems, feedback from the prover guides proof search and revision,
and completed proof terms are checked against the formal statements.

Central to many ITPs is dependent type theory, a formalism that unifies programs and proofs within a single framework. Systems like Rocq, Lean and Agda, built upon the Calculus of Inductive Constructions (CIC) or Martin-Löf Type Theory (MLTT)~\cite{MartinLof1984}, exemplify the power of dependent type theory in encoding rich mathematical structures and reasoning about them formally. Whereas Rocq and Lean (based on CIC) do not particularly favor constructivism (Lean even has the axioms of choice and excluded middle built-in), Agda (based on MLTT)
leans more towards constructive mathematics. HOL-based systems, rooted in classical higher-order logic, offer an alternative approach that is more familiar to traditional mathematicians and computer scientists. However, dependent type theories surpass HOL in expressiveness by allowing types to depend on values, enabling constructive reasoning with computational interpretations via term reductions, and unifying proofs and programs within a single framework, capabilities that HOL's classical framework lacks.

In the early 2000s, the fields of formalized mathematics and dependently typed languages — previously limited mostly to communities of computer scientists and logicians — began to draw significant attention from pure mathematicians.  A pivotal moment in this shift was the formulation of \emph{Univalent Foundations} (UF) of mathematics by
Vladimir Voevodsky between 2006 and 2009~\cite{VoevodskyOrigins}. This new approach, inspired by the homotopy semantics of MLTT, interprets types as homotopy types\footnote{The canonical geometric example of a homotopy type is a topological space up to weak homotopy equivalence. Other combinatorial examples include presheaves such as simplicial or cubical sets up to a suitable equivalence.} rather than as sets~\cite{SSetsVoev}. 
In particular, the equality type $a =_A b$ is interpreted as the type of paths 
between $a$ and $b$ in $A$, for $p, q : a =_A b$, the type $p =_{a =_A b} q$ is the type of
homotopies between $p$ and $q$ in $A$, and so on. The homotopy semantics suggested that
MLTT can serve as a
basis for a formal system where the notion of a homotopy type is taken as primitive
instead of the notion of a set. This led to formulation of Homotopy Type Theory (HoTT) as
an extension of MLTT with \emph{univalence axiom} $(A \cong B) \cong (A = B)$, which
allows to identify two equivalent
types, and \emph{higher inductive types}, a certain generalization of inductive types
allowing higher path constructors~\cite[Chapters~2 and~6]{hottbook}.

The impact of HoTT/UF on formalization of mathematics is two-fold:

\begin{enumerate}
\item Many standard constructions and theorems in homotopy theory can now be elegantly formalized using HoTT, where homotopy types serve as primitives rather than sets.
This method, known as \emph{synthetic homotopy theory}, utilizes the equality type to encapsulate complex homotopy structures~\cite[Chapter~8]{hottbook}. Set-theoretic versions of these formalizations tend to be more cumbersome. This synthetic approach not only simplifies the formalization of existing theories but also has the potential to reshape many areas of modern mathematics by adopting a higher categorical perspective over traditional set-theoretic approaches.

\item HoTT/UF provides a framework for set-theoretic formalization where sets are 
homotopy types with trivial homotopy structure and propositions are sets with at most one 
element. One can do formalizations completely confined to this set theory 
fragment of HoTT. As such a framework HoTT/UF augments MLTT with a fair degree of
\emph{extensionality}:
\begin{enumerate}
\item Quotients can be easily defined in HoTT using higher inductive types~\cite[Section~6.10]{hottbook}. Meanwhile
quotients are notoriously hard to deal with in pure MLTT due to intensionality of equality~\cite{quotientsLi}.
\item Univalence axiom implies propositional and function extensionality: any two
equivalent propositions are equal and any two pointwise equal functions are equal~\cite[Sections~3.3 and~4.9]{hottbook}.
\item Univalence axiom implies \emph{structure identity principle}: isomorphic
structures (such as groups, topological spaces and so on) are equal~\cite[Section~9.8]{hottbook}.
\end{enumerate}
\end{enumerate}

This approach can, in principle, be implemented using an ITP based on MLTT, such as Rocq or Agda\footnote{An important caveat is that the ITP should not incorporate built-in features that contradict the univalence axiom. For instance, Lean is unsuitable for UF as it incorporates a strong form of the axiom of choice built-in.}, by merely appending univalence and higher inductive types as axioms devoid of any computational content. However, this severely impairs the computational properties of MLTT, which are crucial for practical formalization efforts. Additionally, there are several other problematic issues stemming from the fact that these ITPs were not originally designed with native support for HoTT/UF. For example, the universe of propositions $\Prop$ in Rocq or Agda does not correspond to the type of propositions as defined in HoTT/UF. 

\textbf{Theorem prover Arend.} The development of Arend began in 2015 at JetBrains Research with the goal of creating a modern ITP featuring \emph{native} support for HoTT/UF. Its two principal components are the Arend language, which is based on a variant of MLTT specifically tailored to integrate seamlessly with HoTT/UF, and a suite of tooling provided by the Arend plugin for IntelliJ IDEA~\cite{ArendSite}. The key components
of the Arend ITP are listed below. The language and command-line interface
described here are those of Arend 1.12.\footnote{The checks reported here use source revision
\texttt{3adf367b9c0b} of \url{https://github.com/arend-lang/Arend}.}

\begin{enumerate}
\item \textbf{Type theory.} At the core of the type theory of Arend is the \emph{homotopy type theory with
interval type} (HoTT-I)~\cite{IsaevHoTTI,IsaevModels}, formulated by Valery Isaev in 2014. It consists of two components:
\begin{enumerate}
\item \textbf{Variant of MLTT with the interval type:} a minor, more extensional modification of MLTT that incorporates a primitive type
$\I$ for the interval. The equality type (or path type) is defined in terms of $\I$.
Function extensionality is derivable and is fully computational in this theory. 
This setup gives rise to a variant of cubical syntax: $n$-dimensional homotopies
in a type $A$ are naturally represented in HoTT-I as functions $\I^n \to A$.

\item \textbf{Univalence axiom:} a form of univalence axiom, which is a built-in axiom with some computational rules.
\end{enumerate}

The native support for HoTT/UF in the type theory of Arend is further amplified by 
the following extensions of HoTT-I:
\begin{enumerate}
\item \textbf{Inductive types with conditions.} Using the primitive $\I$, a 
path constructor
can be defined simply as a constructor $\mathsf{pcon}\, (i : \I)$ with an interval type parameter, or multiple parameters for higher-dimensional homotopies. The only
difference with an ordinary constructor is that there could be \emph{conditions} 
that dictate how $\mathsf{pcon}$ evaluates at the ends of the interval to
other constructors: 
$\mathsf{pcon}\, \lcon \Rightarrow \mathsf{con_1}$,
$\mathsf{pcon}\, \rcon \Rightarrow \mathsf{con_2}$. In particular, it implies 
$con_1 = con_2$ but does not lead to contradictions as in the case of standard inductive types in MLTT, thanks to a modified elimination principle that requires the preservation of conditions. 
This provides support for higher inductive types and, in particular, quotients with computational behavior~\cite{hitsInArend}. 

\item \textbf{Universes $\Prop$ and $\Set$ matching h-propositions and h-sets.} 
Arend incorporates an \emph{impredicative} universe $\Prop$ that captures the logic of h-propositions and a predicative hierarchy of universes $\Set\, 0 \subset \Set\, 1 \subset \dots$ for h-sets. The universes $\Prop$ and $\Set$ provide a \emph{syntactic} framework that facilitates working within the set theory domain, akin to the elementary topos of h-sets, and distinctively delineate the set-level aspect of the type theory from the higher levels, ensuring that set-theoretic formalization is confined strictly to the subtheory prescribed by these universes.

\item \textbf{Universes stratified by homotopy level.} Alongside predicative universe polymorphism, Arend provides universes of propositions, sets and higher types. A definition can be placed in the appropriate universe using a proof of its homotopy level. In Arend 1.12 homotopy levels are constants, not polymorphic parameters; cumulativity allows a type to be used at a higher homotopy level.
\end{enumerate}

The above outlines the core fragment of Arend's type theory as it relates to HoTT/UF. Its key properties include: 

\begin{enumerate}
\item \textbf{Computational behavior.} Unlike MLTT augmented merely with axioms for univalence and higher inductive types, Arend's type theory includes a number of computational reduction rules for involved terms, which prove crucial for practical usage. However, it lacks sufficient computational rules to render the theory fully computational: the MLTT property that every closed term of type $\Nat$ evaluates to a canonical number does not hold in Arend. The core theory of Arend represents one of the earliest and simplest forms of \emph{cubical type theory}, in which the interval still functions as an ordinary type. Later cubical type theories make univalence derivable and recover canonicity by adding a language of dimensions and faces together with uniform Kan composition~\cite{CCHM,HuberCanonicity}; some Cartesian variants additionally separate fibrant and non-fibrant types~\cite{CHTTIII}. This machinery is substantially more elaborate than the $\coe$/$\iso$ interface of HoTT-I; see Section~\ref{sec:cubical-comparison}. The type theory of Arend could potentially be adjusted to enhance its computational aspects, but practical experiences in formalization have not demonstrated a significant need for such enhancements.

\item \textbf{Constructivity.}  The type theory of Arend is inherently constructive. In its core theory, the law of excluded middle does not hold, and the axiom of choice is valid only in the form of \emph{unique choice}. While the axioms of classical logic can be consistently added and utilized in libraries, the main Arend library, $\mathsf{arend\text{-}lib}$, is dedicated to constructive mathematics, and no standard metas utilize any classical axioms. 
\end{enumerate}

Finally, the type theory of Arend includes the following extensions, which are of significant practical importance for formalization and yet are not present in Rocq, Lean, or Agda: 
\begin{enumerate}
\item \textbf{Records with partial implementations and anonymous extensions.}
In Arend, type classes and records support multiple inheritance with \emph{partial implementations}. Specifically, if record $D$ extends record $C$, it is possible to implement an arbitrary subset of $C$'s fields within $D$. For instance, for a field $f$ of $C$, its extension $D$ may include an expression $f \Rightarrow a$ that specifies a value $a$ for $f$. This flexibility blurs the traditional distinction between parameters and fields in classes and records, enabling more versatile hierarchical structure development. Additionally, Arend allows for the creation of records and classes on-the-fly as an \emph{anonymous extension}, denoted as $C \{f_1 \Rightarrow a_1, \ldots, f_k \Rightarrow a_k\}$. This feature facilitates dynamic customization of types without the need to define fully specified new classes or records, thus enhancing modularity and reusability. 

\item \textbf{Array types.} Arend introduces a specialized type for arrays, denoted as $\mathsf{Array}\, A$, where $A : \Type$. This type encompasses several concepts including 
the type of lists of elements of type $A$, the type of vectors of elements
of type $A$ of fixed length $n$ and the type $\mathsf{Fin}\, n \to A$ of functions from 
a finite set of cardinality $n$ to $A$. The type $\mathsf{Array}$ is a record with
fields $A : \Type$, $\mathsf{len} : \Nat$ and 
$\mathsf{at} : \mathsf{Fin}\, \mathsf{len} \to A$. The anonymous extension $\mathsf{Array}\, A$ describes arrays of various lengths with elements of type $A$, and the extension $\mathsf{Array}\, A\, n$ specifies arrays of a fixed length $n$. A crucial feature of the array type in Arend is its computational extensionality: if two arrays of equal length are elementwise computationally identical, then they are considered computationally equivalent as entire structures. 
\end{enumerate}

\item \textbf{Proof terms.} In Arend, proof terms do not need to be fully detailed. Users can take advantage of the ``inference of implicit arguments'' mechanism, which allows them to omit some arguments in expressions (more concretely, the ones that can be inferred via the ``inference of implicit arguments mechanism'' built into the Arend type checker). Additionally, users can employ metas such as $\mathsf{rewrite}$ to construct particularly complicated parts of expressions. Metas should be seen as ``expression-level tactics'', meaning they can be seamlessly interwoven with standard Arend term constructs. In other words, unlike Rocq, Arend does not have separate ``proof'' and ``term'' levels – everything (including proofs) in Arend happens on the ``term'' level.

\item \textbf{IntelliJ IDEA plugin.} The easiest way to use Arend is through the
tooling provided by
the Arend plugin for IntelliJ IDEA. IntelliJ IDEA, developed by JetBrains, is a versatile IDE with intelligent code assistance, robust refactoring tools, and seamless integration with languages like Java, Kotlin, and Python. It has a variety of built-in
tools for version control, debugging and testing.

Arend plugin supports the following functionality:

\begin{enumerate}
\item \textbf{Incremental type checking.} Under normal circumstances the Arend plugin uses the so-called ``on-the-fly'' (or smart) mode of type checking. In this mode, the type checker operates in a background thread, automatically reprocessing each code modification made by the user.
\item \textbf{Arend Messages tool window.} The Arend Messages tool window serves as the primary interactive element in Arend, providing essential insight into proof goals, expected types, and error notifications. While writing terms the user can 
insert hole expressions \verb|{?}| for arbitrary subterms, which would introduce
goals shown in Arend Messages. This allows for a convenient way of writing proof terms
by gradually breaking down the original proof goal into several subgoals.
\item \textbf{Editor features.} The Arend editor supports a variety of features that 
greatly simplify working with Arend code such as: auto-completion of identifiers;
quick fixes for Arend errors; intention actions such as adding missing clauses
in pattern matching; refactorings such as handling consequences of changing
signatures of definitions or moving definitions; navigation tools such as Go 
to Declaration or Find Usages; quick documentation popups supporting LaTex;
parameter hint tooltip which can be invoked in application expressions.

\item \textbf{Debugger for metas.}
\end{enumerate}

\end{enumerate}

\textbf{Arend CLI and LLM-assisted formalization.} Besides the IDE plugin, Arend
provides a command-line interface exposing the type checker and several library
retrieval tools. This makes it possible to use Arend together with AI coding agents
such as Claude Code; a collection of ready-made agent skills for Arend formalization
is also available (see Section~\ref{sec:claude-code}).

\textbf{The library arend-lib.}
The main Arend library $\arendlib$ covers three broad areas. Its main
developments are constructive; results using excluded middle or choice
are separated into modules that explicitly assume these principles.

\begin{enumerate}
\item \textbf{Constructive mathematics.} This is the main part of the library. The 
mathematical landscape in the constructive setting is richer than in
the classical setting since classically equivalent definitions often become 
inequivalent constructively. 

This part contains the following: schemes via locally ringed locales; principal ideal domains
and the proof that they are Smith domains of Krull dimension at most one; splitting fields of polynomials
and algebraic closure for countable, decidable fields; constructive forms of the
Fundamental Theorem of Algebra, including Richman's root-multiset construction
and algebraic closedness of the algebraic numbers; connection between zero-dimensional
and integral extensions; matrices over commutative rings, determinants, characteristic
polynomials, Cayley-Hamilton theorem; linear algebra over Smith domains; integral ring extensions;
polynomials over one or several variables; Nakayama's lemma; derivative over topological 
rings; directed limits for sequences and functions; series and power series; natural,
integer, rational, real and complex numbers and various structures on them; categories,
functors, adjoint functors, Kan extensions, (co)limits; elementary topoi and Grothendieck
topoi; topological spaces, locales, uniform and cover spaces, and their completions.

The main references are the books \cite{mines87, lombardi15}, Richman's
constructive FTA \cite{RichmanFTA}, and Isaev's paper on constructive
complete spaces \cite{isaev24}.

\item \textbf{Synthetic homotopy theory.} All of the synthetic homotopy theory
of the HoTT Book~\cite{hottbook} has been formalized in $\arendlib$, partly
with AI assistance. It includes pointed types, loop spaces and homotopy
groups; spheres, suspensions, joins and pushouts; truncations and
connectedness; the fundamental group of the circle; the Hopf fibration
and long exact sequences; the Freudenthal, van Kampen and Whitehead
theorems; Eilenberg--MacLane spaces $K_1(G)$; localizations, modalities
and the generalized Blakers--Massey theorem
(Section~\ref{sec:lib-hott}).

\item \textbf{Computer science.} This part includes abstract reduction
systems, confluence and termination, higher-order term rewriting with the
$\lambda$-calculus as an example, modularity of confluence, and first-order
algebraic theories with their models and free models.
\end{enumerate}

\textbf{Future directions.} So far, developments in synthetic homotopy theory in HoTT
have been limited. One reason is that types in HoTT are necessarily $\infty$-groupoids,
a special kind of $\infty$-categories where all morphisms are invertible.
This is because equality is symmetric.
Attempts to define more general $\infty$-categorical structures in terms of 
$\infty$-groupoids in HoTT have led to the still-open question of definability of (semi-)simplicial types in HoTT\cite{simplTypesNote}. Recently, proposals for \emph{directed} HoTT \cite{dirHoTTbook, simplHoTT},
where types are $\infty$-categories, have emerged. A variant of directed HoTT is
being developed for Arend; its implementation is currently in the
\texttt{cat} branch.\footnote{\url{https://github.com/arend-lang/Arend/tree/cat}}

\textbf{The structure of this paper.} Section~\ref{sec:TypeTheory} introduces
HoTT-I and Arend's extensions, explains their use in set-theoretic formalization
and synthetic homotopy theory, and compares their computational behavior with
other cubical type theories, including open problems concerning additional
computation rules. Section~\ref{sec:arend-lang} presents universe polymorphism,
classes and records, and arrays through code examples. Section~\ref{sec:ide}
describes the IntelliJ IDEA plugin's type checking, editing, navigation and
refactoring tools. Section~\ref{sec:claude-code} explains how the Arend CLI
supports LLM-assisted formalization with Claude Code.
Section~\ref{sec:lib} surveys $\arendlib$, focusing on constructive category
theory, topology and algebra, followed by synthetic homotopy theory and
further topics.

\section{Underlying type theory of Arend}\label{sec:TypeTheory}

The type theory of Arend is an extension of the 
\emph{homotopy type theory with interval type} (HoTT-I). 
It is a simple modification
of Martin-Löf type theory (MLTT) which allows for native support of homotopy type theory and has certain important advantages 
over MLTT even for formalisations unrelated to homotopy theory.

\subsection{Homotopy type theory with interval type}

\begin{enumerate}
\item \textbf{Interval type} $\I$: a new primitive which serves as a basis for
the definition of equality type $x =_A y$.
Type $\I$ is defined as a data type with two constructors $\lcon\, :\, \I$,
$\rcon\, :\, \I$ and the elimination principle given by the function $\coe$: 

\begin{center}
\AxiomC{$\Gamma \vdash$}
\UnaryInfC{$\Gamma \vdash \I$}
\DisplayProof
\quad
\AxiomC{$\Gamma \vdash$}
\UnaryInfC{$\Gamma \vdash \lcon : \I$}
\DisplayProof
\quad
\AxiomC{$\Gamma \vdash$}
\UnaryInfC{$\Gamma \vdash \rcon : \I$}
\DisplayProof
\end{center}

\medskip
\begin{center}
\AxiomC{$\Gamma, x : \I \vdash A$}
\AxiomC{$\Gamma \vdash a : A[x := \lcon]$}
\AxiomC{$\Gamma \vdash i : \I$}
\TrinaryInfC{$\Gamma \vdash \coe(\lam{x} A, a, i) : A[x := i]$}
\DisplayProof
\end{center}

\[ \coe(\lam{x} A, a, \lcon) \equiv a,\ \ \coe(\lam{x} A, a, i) \equiv a \text { if } x \notin \fs{FV}(A) \]

The second of these is called \textbf{regularity}: coercion in a family that
does not mention the interval variable is the identity.  It is referred to by
that name below.

Functions from $\I$ to a type $A$ can be interpreted as paths in $A$. In HoTT-I, the equality type is defined as the type of paths $\Path(\lam{x} A, a, a')$, which are maps
$f$ from $\I$ to $A$ with fixed endpoints $f(\lcon)\equiv a$ and $f(\rcon)\equiv a'$:

\medskip
\begin{center}
\AxiomC{$\Gamma, x : \I \vdash A$}
\AxiomC{$\Gamma \vdash a : A[x := \lcon]$}
\AxiomC{$\Gamma \vdash a' : A[x := \rcon]$}
\TrinaryInfC{$\Gamma \vdash \Path(\lam{x} A, a, a')$}
\DisplayProof
\end{center}

\medskip
\begin{center}
\AxiomC{$\Gamma, x : \I \vdash a : A$}
\UnaryInfC{$\Gamma \vdash \pcon(\lam x a) : \Path(\lam x A, a[x := \lcon], a[x := \rcon])$}
\DisplayProof
\end{center}

\medskip
\begin{center}
\AxiomC{$\Gamma \vdash p : \Path(\lam x A, a, a')$}
\AxiomC{$\Gamma \vdash i : \I$}
\BinaryInfC{$\Gamma \vdash p\ @_{a,a'}\ i : A[x := i]$}
\DisplayProof
\end{center}

\begin{align*}
& \pcon(\lam x t)\ @_{a,a'}\ i \equiv t[x := i],
\ \ \ \pcon(\lam x p\ @\ x) \equiv p \text{ if } x \notin \fs{FV}(p) \\
& p\ @_{a,a'}\ \lcon \equiv a,
\ \ \ p\ @_{a,a'}\ \rcon \equiv a'
\end{align*}

The standard $\mathsf J$-eliminator for equality types in MLTT can be derived from
these rules. 

\item \textbf{Univalence axiom:} the function $\iso$ witnessing the principle ``isomorphic types are equal'' which implies the \textbf{structure identity principle} (see \ref{sec:SIP} below).
Given types $A$, $B$, mutually inverse functions $f:A\to B$, $g:B\to A$ and proofs $p$, $q$ of
$g\circ f = \mathsf{id}_A$ and
$f\circ g = \mathsf{id}_B$, the function $\iso$ constructs a family of types $\iso(A, B, f, g, p, q, i)$ parameterized
by interval $i\colon \I$, such that $\iso(A, B, f, g, p, q, \lcon) \equiv A$ and
$\iso(A, B, f, g, p, q, \rcon) \equiv B$: 
\medskip
\begin{center}
\Axiom$\fCenter \Gamma \vdash A$
\noLine
\UnaryInf$\fCenter \Gamma \vdash B$
\def\extraVskip{1pt}
\Axiom$\fCenter \Gamma, x : A \vdash b : B$
\noLine
\UnaryInf$\fCenter \Gamma, y : B \vdash a : A$
\Axiom$\fCenter \Gamma, x : A \vdash p : a[y := b] = x$
\noLine
\UnaryInf$\fCenter \Gamma, y : B \vdash q : b[x := a] = y$
\def\extraVskip{2pt}
\AxiomC{$\Gamma \vdash i : \I$}
\QuaternaryInfC{$\Gamma \vdash \iso(A, B, \lam x b, \lam y a, \lam x p, \lam y q, i)$}
\DisplayProof
\end{center}

\begin{align*}
& \iso(A, B, \lam x b, \lam y a, \lam x p, \lam y q, \lcon) \equiv A, 
\ \ \iso(A, B, \lam x b, \lam y a, \lam x p, \lam y q, \rcon) \equiv B \\
& \coe(\lam i \iso(A, B, \lam x b, \lam y a, \lam x p, \lam y q, i), a_0, \rcon) \equiv b[x := a_0] \text { if } i \notin \fs{FV}(A\ B\ b\ a\ p\ q)
\end{align*}

\end{enumerate}

\subsection{Extensions of HoTT-I in Arend}\label{sec:ExtTT}

The core type theory of Arend includes a number of extensions of HoTT-I:

\begin{enumerate}

\item \textbf{Inductive types with conditions.} Definitions of inductive types in the core type theory of Arend extend the standard definitions with
\emph{conditions} on constructors. Conditions allow to glue different constructors when evaluated at
certain values of parameters. For example, for the type $\Int$ of integers with two constructors $\pos\, (n\colon \Nat)$
and $\negg\, (n\colon \Nat)$ a condition might be $\pos\, 0\equiv \negg\, 0$. In general a condition on a constructor
$\con\, (a_1\, :\, A_1)\dots (a_n\, :\, A_n)$ is almost the same as a definition of a partial function by pattern matching which specifies
how $\con$ evaluates. The most interesting examples of conditions involve pattern matching on the interval (such pattern matching is not allowed 
for functions): it can be used to glue arbitrary constructors $\con_1$ and $\con_2$ by means of the constructor $\con_=\, (i\colon \I)$ with
conditions $\con_= \ \lcon \equiv \con_1$ and $\con_= \ \rcon \equiv \con_2$. This allows to define quotient types and higher inductive types
(see \ref{sec:quot} and \ref{sec:HITs} below).

The mechanism of inductive types with conditions in Arend has been analysed in the paper
\cite{hitsInArend}.  

\item \textbf{Propositions and sets.} Type theory of Arend contains the type $\Prop$ -- an impredicative universe of all propositions, and a predicative
hierarchy of universes of sets $\Set\ 0 \subseteq \Set\ 1 \subseteq \dots$. Universes $\Prop$ and $\Set\ n$ constitute a fragment of the type theory
which represents a variant of set theory and provide a framework for conventional 
set-theoretic formalization.

The following properties characterize the logic and the universe $\Prop$:

\begin{enumerate}
\item The operations $\forall$, $\wedge$, $\to$, $\neg$ as well as the constants 
$\mathsf{True}$, $\mathsf{False}$ are defined as usual in propositions-as-types logic,
corresponding to $\prod$, $\times$, $\to$, $\cdot \to \textbf{0}$ and the types
$\textbf{1}$, $\textbf{0}$, respectively. Operations $\exists$, $\vee$ can be defined
as inductive types or via $\sum$, $+$ as 
${\exists(x : A) P(x) :=} \mathsf{nonempty}(\sum_{(x : A)}P(x))$ and 
$P \vee Q := \mathsf{nonempty}(P + Q)$, where $\mathsf{nonempty}(A)$ is an inductive
type (see $(\ref{itm:main-prop})$ below). 

\item $\Prop$ is \textbf{impredicative} and the logic of $\Prop$ is by default \textbf{constructive}: the law of excluded middle ${\LEM := \forall (P : \Prop)\, P \vee \neg P}$ and general axiom of choice do not hold. Propositions are sets: $\Prop \subseteq \Set\ 0$ and $\Prop\colon \Set\ 0$.

\item Propositions $P\, :\, \Prop$ satisfy \textbf{propositional proof irrelevance} $\PI(P) :=\\ {\prod_{(p, q : P)}\, p = q}$. This is one of the main properties that distinguish propositions from other types: a proposition conveys no more information
than a truth value. 

\item \textbf{Propositional extensionality} holds for $\Prop$: if $P \leftrightarrow Q$ for $P, Q\colon \Prop$ then $P = Q$. This is a consequence of proof irrelevance and
the univalence axiom.

\item\label{itm:main-prop} The universe $\Prop$ defines the syntactic notion of proposition: 
propositions are types in $\Prop$ and the main property of types in $\Prop$ is proof irrelevance. But there
is also a more semantic notion of an \textbf{h-proposition}, which is any type $A$
satisfying proof irrelevance, namely such that $\mathsf{isProp}(A):= \PI(A)$ is provable.

In the type theory of Arend any definition of an inductive type $P : \Type$, which is an 
h-proposition, can be supplemented with a proof 
$p : \mathsf{isProp}(P)$ in which case $P$ gets placed in $\Prop$ and becomes a
proposition syntactically. This applies also to
record types and there are analogous
extensions for function definitions which ensure, for example, that $\mathsf{isProp}(A)$
is in $\Prop$.
This mechanism allows to define the predicate $\mathsf{nonempty}(A) : \Prop$\footnote{
$\mathsf{nonempty}(A)$ is also called \emph{propositional truncation} and denoted $\|A\|_{-1}$} as inductive type with constructors $\mathsf{in}\, (a : A)$ and 
$\mathsf{trunc}\, (a\, b : \mathsf{nonempty}(A))\, (i : \I)$ with conditions 
$\mathsf{trunc}\, a\, b\, \lcon\equiv a$,  
$\mathsf{trunc}\, a\, b\, \rcon\equiv b$ and an obvious proof 
$\lam x \lam y\mathsf{path}(\mathsf{trunc}\, x\, y) : \mathsf{isProp}(\mathsf{nonempty}(A))$.

Another key mechanism: for an inductive type $P$ in $\Prop$ and an h-proposition $A$ 
functions $P\to A$ can be defined by simplified elimination where constructors containing
parameters of type $\I$ are omitted. This implies $\PI(P)$ for any $P : \Prop$:
with this elimination one can construct a function $\mathsf{nonempty}(P) \to P$ which
is inverse to the obvious function $P \to \mathsf{nonempty}(P)$ and prove that they are
mutually inverse.

As an alternative to the mechanism for placement of inductive types $P : \Type$
with $p : \mathsf{isProp}(P)$ in $\Prop$ Arend supports a mechanism for marking inductive
definitions as \textbf{truncated  definitions} in $\Prop$ and limiting definitions of functions
$P \to A$ by elimination only to h-propositions $A$. For example, propositions $\exists(x : A) P(x) : \Prop$ and $P \vee Q : \Prop$ can be defined directly as truncated inductive types rather than via $\mathsf{nonempty}$. We call \textbf{h-prop elimination} this
kind of elimination from truncated inductive types in $\Prop$. If $A$ is not already inferred to lie in $\Prop$, the eliminator includes a proof $p\colon \mathsf{isProp}(A)$.

In the presence of univalence elimination from truncated definitions in $\Prop$ cannot be extended to all types. Full, unrestricted elimination from propositions to arbitrary 
types would imply the existence of a global choice operator 
$\mathsf{gc}\colon \Pi_{A : \Set\, n}\mathsf{nonempty}(A) \to A$, which is inconsistent
with the univalence axiom (\cite{rijkeBook}, Corollary 17.5.3). Moreover propositional proof irrelevance for $P\colon \Prop$ cannot be strengthened to the computational one 
$p\equiv q$, $p,q\in P$. Although it is probably consistent, it causes some issues
in a type theory with h-prop elimination.

The difference from Rocq and Lean can be illustrated by elimination from
$\mathsf{nonempty}(A)$. Given $f\colon A\to B$ and a proof that any two
elements of $B\colon\Type$ are equal, Arend defines
\[
  \bar f\colon\mathsf{nonempty}(A)\to B,
  \qquad \bar f(\mathsf{in}(a))\equiv f(a).
\]
The proof that $B$ is an h-proposition suffices; $B$ need not already be
classified in $\Prop$. In Rocq and Lean, the ordinary eliminator for the
corresponding proposition expressing nonemptiness requires a target in
$\Prop$. Supplying a proof that all elements of a target $B\colon\Type$
are equal does not change this restriction.

Rocq and Lean do allow elimination into arbitrary types from certain other
inductive propositions: for example, a proof of falsehood gives an element
of any type, and a proof of equality allows transport between types.
These exceptions depend on the form of the source inductive definition;
they do not apply to nonemptiness of an arbitrary type $A$. In Arend, the
restriction to h-propositional targets applies to definitions declared
truncated. An ordinary inductive definition placed in $\Prop$ using a proof
of its homotopy level retains its ordinary eliminator.

See Section~\ref{sec:arend-lang} for examples in Arend language.

\item As a consequence of h-prop elimination holds a weak form of the axiom of
choice: the \textbf{principle of unique choice} $\mathsf{UC}$. The principle 
$\mathsf{UC}$ essentially
says that, for sets $X,Y$ and a proposition-valued relation $P$, if
${\forall (x : X)}\, {\exists! (y : Y)}\ P(x, y)$, where $\exists!$ means
``exists and unique'', then $\exists (f : X\rightarrow Y)\, \forall (x : X)\, P(x, f(x))$.
Here existential quantifiers are propositionally truncated.
h-prop elimination is a stronger form of $\mathsf{UC}$, which, in particular, defines
functions which compute at constructors. This implies, for example, that every bijection has an inverse with computational 
properties. 

The \textbf{axiom of choice for sets} $\mathsf{AC}$ obtained by dropping
uniqueness in this formulation is not provable in Arend, but can be
consistently assumed. The restriction to sets matters when comparing this
statement with choice principles for higher types~\cite[Section 3.8]{hottbook}.

\item h-prop elimination and the univalence axiom imply that $\Prop$ is isomorphic to
totalities of h-propositions:
$\Prop \cong \sum_{(A : \Set\, n)}\, \mathsf{isProp}(A)$ for all $n\in \mathbb{N}$.
It follows from provability of 
$\mathsf{isProp}(A)\to {\mathsf{nonempty}(A) = A}$
using h-prop elimination and the univalence axiom. Moreover, with $\LEM$ the universe $\Prop$ is isomorphic to $\mathsf{Bool}$. 

\end{enumerate}

To sum up, the logic of Arend provides a basis for either constructive or classical univalent formalization of mathematics where:

\begin{enumerate}
\item Univalence axiom and the structure identity principle hold.
\item The principle of unique choice $\mathsf{UC}$ holds in the computational form of h-prop elimination for $\Prop$. 
\item The axiom of choice $\mathsf{AC}$ and the law of excluded middle $\LEM$ are not
accepted in $\arendlib$, but can be consistently assumed in some formalizations.
\end{enumerate}

Universes $\Set\ n$ are characterized by the following properties:

\begin{itemize}
\item For $X\colon \Set\ n$ and $x , y\colon X$ the type $x = y$ lies in $\Prop$. 
In particular, sets satisfy uniqueness of identity proofs 
$\mathsf{UIP}(X) := \prod_{(x, y\colon X)}\PI(x = y)$.
\item $\Set\ n \subset \Set\ (n+1)$, $\Set\ n \colon \Type\ (n+1,1)$:
the universe of sets is a $1$-type, not itself a set. Here the second
parameter denotes the homotopy level, as explained below.
There is also a hierarchy of universes $\Type\ n$ of all types satisfying 
$\Set\ n \subset \Type\ n$.
\item As in the case of propositions, there is a semantic notion of an 
\textbf{h-set}, which is any type $A$ for which $\mathsf{UIP}(A)$ is provable.
And similarly to propositions, definitions with a proof of $\mathsf{isSet}(A) := \mathsf{UIP}(A)$ can be placed in $\Set$, and the elimination for truncated inductive
types in $\Set\ n$ is confined to h-sets (\textbf{h-set elimination}). This implies 
$\Set\ n \cong \sum_{(A : \Type\, n)}\, \mathsf{isSet}(A)$.
\end{itemize}

\item \textbf{Homotopy levels.} Sets in Arend are types $A$ such that 
their equality type $x =_A y$ is a proposition. This construction can be iterated 
by considering
$1$-types such that their equality type is a set, $2$-types such that their equality type
is a $1$-type and so on. This leads to the notion of a \textbf{homotopy level} of a type:
we say that propositions are of homotopy level $-1$ and a type is of homotopy level 
$k\in \mathbb{N}$ if its equality type is of homotopy level $k - 1$. We say that a type is
of homotopy level $\infty$ if no such $k$ exists.

In Arend a universe is parameterized by a homotopy level $k$ in addition to the usual
predicative level $n$; we write $\Type\ (n, k)$ for the universe of $k$-types of
predicative level $n$. As in the case of $\Prop$ and $\Set\ n$ there is
a universe placement mechanism for definitions, $k$-type elimination and an equivalence between 
$\Type\ (n, k)$ and the type of all semantic $k$-types:
$\Type\ (n, k)\cong \sum_{(A : \Type\ n)}\mathsf{isOfHLevel}(A, k)$.

The two parameters are, however, not on the same footing. Definitions are polymorphic in
the predicative level: such a level can be a variable, and level expressions are built
from level variables, $0$, $\suc$ and $\max$. The homotopy level, by contrast, is always a
constant --- either $-1$, or a natural number, or $\infty$. So although there is a universe
$\Type\ (n, k)$ for each $k$, there is no polymorphism over $k$: a definition cannot be
abstracted over a homotopy level the way it is over a predicative one. Earlier versions of
Arend did admit an implicit homotopy level parameter alongside the predicative one; this
has since been dropped.

\item \textbf{Subtyping and dependent record types}. For some pairs $(A, B)$ of types in Arend 
holds $A < B$ (subsumptive subtyping), which means that the typing relation $a\, :\, A$ implies 
the typing relation $a\, :\, B$.
Subtyping is involved in two mechanisms: in cumulativity of the 
hierarchy of universes and, more significantly, in the system of dependent record types.

Cumulativity means that a type can be used at larger predicative and homotopy
levels: $\Type\, (p,h)<\Type\, (p',h')$ when $p\leq p'$ and $h\leq h'$.
In particular,
\[\Prop < \Set\,0 < \Set\,1 < \cdots.\]

\textbf{Dependent record types} provide a basis for working with hierarchies of mathematical structures by
means of bundling data and creating new structures as extensions or specifications of other structures.
Record types $C \{x_1 : A_1, x_2 : A_2, \ldots, x_n : A_n\}$ 
(where $A_i$ may depend on $x_1,\ldots,x_{i-1}$) are similar to sigma-types: their elements
are also dependent tuples $(a_1 : A_1, a_2 : A_2[a_1/x_1], \ldots, a_n : A_n[a_1/x_1,
\ldots,a_{n-1}/x_{n-1}])$. Variables $x_1,\ldots,x_n$ are called the \emph{fields} of $C$.
The difference from sigma-types is that a new record type $D$ can be 
formed as an extension of $C$ in which case $D$ becomes a subtype of $C$. Record type $D$
can extend $C$ with new fields $x_{n+1},\ldots, x_{n+k}$ or with \textbf{manifest fields} of the
form $x_i \Rightarrow a_i$, $i\in \{1,\ldots, n\}$, which assign a specific value for fields of $C$.
If $D$ contains $x_i \Rightarrow a_i$ then its objects would be tuples where $i$-th component 
is computationally equal to $a_i$.

Some of the properties of record types:

\begin{enumerate}

\item Record type extensions model the concept of inheritance in programming
languages. In Arend multiple inheritance is allowed: a record type can be an extension and 
therefore a subtype of several record types. Multiple inheritance together with manifest fields 
provide a flexible framework for grouping data associated to mathematical objects using sharing, extensions and subtyping.

\item Record extensions consisting of just manifest fields can be anonymous. 
This implies that records' fields also behave like parameters: if 
$C \{x_1 : A_1, x_2 : A_2, \ldots, x_n : A_n\}$ is a record, the application 
$C\, a_1\, \ldots\, a_k$ is the anonymous extension 
$C \{ x_1 \Rightarrow a_1, \ldots, x_k \Rightarrow a_k\}$.

\item If $D$ is an extension of $C$ and $x : A$ is a field of $C$ then the type of the
field $x$ can be overridden in $D$ and changed to a subtype $B$: $x : B$, $B < A$.

For example, a record $\mathsf{MonoidHom}$ can have fields
$\mathsf{Dom},\mathsf{Codom}:\mathsf{Monoid}$. An extension
$\mathsf{GroupHom}$ can override these fields with type $\mathsf{Group}$,
since $\mathsf{Group}$ extends $\mathsf{Monoid}$. This is distinct from
the coercion that sends a group or monoid record to its underlying set.

\item Although propositions in Arend are not computationally proof irrelevant in general, 
a form of computational proof irrelevance holds for propositions inside record types. Namely,
some fields whose type is a proposition can be declared \textbf{properties} in which case 
their value will be irrelevant in comparing elements of records.

For example, a record $\mathsf{Cat}$ for categories would have data fields for objects,
morphisms, identity
morphism and composition and property fields for associativity of composition and neutrality of
identity morphism. One consequence of this would be that if $C : \mathsf{Cat}$ then $C$ is computationally equal to $(C^{op})^{op}$.

\end{enumerate}

\item \textbf{Arrays.} Arend has built-in type $\mathsf{Array}$ of arrays which subsumes
the standard inductive type $\mathsf{List}\, (A : \Type)$ of lists of varying length,
the standard indexed inductive type $\mathsf{Vec}\, (A : \Type)\, (n : \Nat)$ of 
lists of fixed length $n$ and the type of functions 
$\mathsf{Fin}\, \mathsf{len} \to A$, where 
$\mathsf{Fin}\, (\suc\, n) : \Set$ is
the type corresponding to the finite set $\{0, \ldots, n\}$.
The type $\mathsf{Array}$ is a special kind of record with the fields: $A : \Type$ for elements, $\mathsf{len} : \Nat$ for length and an indexing function 
$\mathsf{at} : \mathsf{Fin}\, \mathsf{len} \to A$.
Types $\mathsf{List}$ and $\mathsf{Vec}$ can be 
defined via $\mathsf{Array}$ as anonymous extensions: 
$\mathsf{List}\, (A : \Type) := \mathsf{Array}\, A$, 
$\mathsf{Vec}\, (A : \Type)\, (n : \Nat) := \mathsf{Array}\, A\, n$. The type 
$\mathsf{Array}\, A\, n$ is obviously isomorphic to the type of functions
$\mathsf{Fin}\, n \to A$, but it satisfies additional computational equalities
which is the central motivation for introducing $\mathsf{Array}\, A\, n$:

\begin{center}
\Axiom$\fCenter \Gamma \vdash A$
\noLine
\UnaryInf$\fCenter \Gamma \vdash n : \Nat$
\def\extraVskip{1pt}
\Axiom$\fCenter \Gamma \vdash f : \mathsf{Fin}\, (n + 2) \to A$
\noLine
\UnaryInf$\fCenter \Gamma \vdash g : \mathsf{Fin}\, (n + 2) \to A$
\Axiom$\fCenter \Gamma \vdash f\, (n+1) \equiv g\, (n+1)$
\noLine
\UnaryInf$\fCenter \Gamma, x : \mathsf{Fin}\, (n + 1) \vdash f\, x \equiv g\, x$
\def\extraVskip{2pt}
\TrinaryInfC{$\Gamma \vdash f \equiv g$}
\DisplayProof
\end{center}

There is also dependent version 
$\mathsf{DArray}$ with $A : \mathsf{Fin}\, \mathsf{len} \to \Type$ and 
$\mathsf{at} : \prod_{(i : \mathsf{Fin}\, \mathsf{len})} A\, i$. 

\end{enumerate}

We now explain what these modifications bring to MLTT and how it affects formalisation in general, not necessarily in the context of homotopy theory. We then outline some nice properties of the type theory of Arend specific for formalisations in homotopy theory.

\subsection{Properties of the theory of Arend relevant for set-theoretic formalization}

\subsubsection{Function extensionality}\label{sec:funext}

Function extensionality principle says that if two functions are pointwise equal then they are equal:

\medskip
\begin{center}
\Axiom$\fCenter \Gamma \vdash A$
\noLine
\UnaryInf$\fCenter \Gamma \vdash B:A\to\Type$
\def\extraVskip{1pt}
\Axiom$\fCenter \Gamma \vdash f : \prod\limits_{x:A}B\, x$
\noLine
\UnaryInf$\fCenter \Gamma \vdash g : \prod\limits_{x:A}B\, x$
\Axiom$\fCenter \Gamma, x : A \vdash p : f\, x = g\, x$
\def\extraVskip{2pt}
\TrinaryInfC{$\Gamma \vdash \mathsf{FunExt}(f, g,\lam x p) : f = g$}
\DisplayProof
\end{center}

This principle cannot be proven in MLTT and has to be added as an axiom. 
It is known to be provable from univalence axiom in HoTT, but the proof is not completely straightforward (\cite{hottbook}, Section 4.9).
Definition of equality type in HoTT-I allows to define $\mathsf{FunExt}$ very easily:
$$\mathsf{FunExt}(f, g, \lam x p):\equiv \mathsf{path}(\lam i\lam x p @ i)$$ 

Moreover, this definition of $\mathsf{FunExt}$ is well behaved computationally.  
As a map between pointwise equalities to equalities
between functions $\mathsf{FunExt}$ admits an obvious inverse 
$\lam p \lam x \mathsf{path}(\lam i (p @ i)\, x)$ such that composition on both sides is computationally the identity.

Derivation of $\mathsf{FunExt}$ in cubical type theories is also elementary (for example,
in Cubical Agda \cite{cubicalAgda}). In Lean $\mathsf{FunExt}$ is derived using built-in
quotient primitives, including the quotient soundness axiom.
  
\subsubsection{Quotient types}\label{sec:quot} 

Given a set $A$ and a relation $R : A\to A\to \Prop$\footnote{Here $\Prop$ stands for
the totality of propositions, which can be the universe of propositions like in Arend, Rocq, Lean or just sort $\Type$ in pure MLTT} the \textbf{quotient type} $A / R$ is
the set obtained from $A$ by identifying all $a, b : A$ such that $R(a, b)$ holds. More formally $A / R$ can be characterised as a set
satisfying the following properties:
\begin{enumerate}
\item There is a function $c : A\to A / R$.
\item (Universal property) For any set $B$ any function $f : A \to B$ respecting $R$ (that is
such that $f(a) = f(b)$ whenever $R(a, b)$) factorizes uniquely through $c$: there exists a unique 
$h : A / R \to B$ such that $f = h\circ c$. The map $h$ corresponds to an eliminator for $A / R$.
\end{enumerate}

In the category of sets $A / R$ is the coequalizer of two projections
$\sum_{(a, b\colon A)} R(a, b)\rightrightarrows A$. The concept of quotient type is well known
to be hard to deal with in MLTT due to intensionality of equality \cite{quotientsLi,
hof95ext,alt99ext}. 
Although quotient 
types can be defined in MLTT in some cases (say, for rationals) in general they are not definable in MLTT (say, for Cauchy reals)\cite{quotientsLi}.
We briefly outline some of the proposed ways to introduce arbitrary quotient types:

\begin{enumerate}
\item \textbf{Setoids.} Instead of working with types one can work with types coupled with an equivalence relation \cite{hof95ext,alt99ext,setoidsInTT}. If we think of types
as sets, this corresponds to replacing sets with setoids. Quotient type would correspond to quotient setoid, which is defined simply
by taking the equivalence relation generated by the original relation and the
additional identifications. Rocq, for example, has a well developed setoid
infrastructure that helps to adopt this approach. However, this approach is rather impractical
because of all the complications of working with a custom term rewriting system for equivalence relations (as opposed to built-in one for the equality type). 

\item\label{AxiomsDefQT} \textbf{Axioms for existence of quotient types.} Quotient types can be introduced purely axiomatically by simply assuming 
the statement phrasing that there exists a type $A / R$ satisfying the universal property.
Such existence axioms alone supply no reduction rules for the quotient
eliminator. Computations using the postulated operations can therefore become
stuck, even when the axioms are logically constructive. Computational
behaviour depends on the additional rules: Lean, for example, has a built-in
reduction for lifting a function and applying it to a quotient constructor.

\item\label{PropDefQT} \textbf{Quotients of h-sets in the logic of h-propositions in HoTT.}
Quotients can be defined for h-sets using 
subuniverse $\mathsf{hProp}$ of h-propositions and univalence (or extensionality for 
functions and h-propositions): if $R$ is $\mathsf{hProp}$-valued equivalence relation on
an h-set $A : \mathsf{hSet}$,
then $A / R$ can be defined as the type of all $P : A \to \mathsf{hProp}$ that 
define an equivalence class of $R$ in $A$. This definition raises the universe
level which can be dealt with by means of \emph{resizing rules} (\cite{hottbook}, Section 6.10).

\item\label{HITdefQT} \textbf{Higher inductive types.} An appealing perspective on dealing with constructs like quotient types systematically was brought
by HoTT in the form of a mechanism to introduce equalities
in definitions of inductive types. Definitions of higher inductive types (HITs) in HoTT apart from ordinary constructors contain constructors for
equalities called path constructors, and the modified elimination principle ensures the universal property (see Section~\ref{sec:HITs} below and Chapter 6 in \cite{hottbook}). 

First, this allows to define the operation $\Vert A\Vert_0$ of \emph{set-truncation} for types 
$A : \Type$ which produces h-set by trivializing the structure of the identity type on $A$ 
and enforces $\mathsf{UIP}$ on $\Vert A\Vert_0$. The h-set $\Vert A\Vert_0$ can be 
defined as the higher inductive type with ordinary constructor 
$\mathsf{in}_{\Set} \, (x : A) : \Vert A\Vert_0$ and path constructor 
$\mathsf{trunc}\, (x, y : \Vert A\Vert_0)\, (p, q : x = y) : p = q$.

Second, for an h-set $A$ and $\mathsf{hProp}$-valued relation $R$ the quotient h-set $A / R$
can be defined as $A / R := \Vert Q(A, R)\Vert_0$ where $Q(A, R)$
is the higher inductive type with ordinary constructor 
$\mathsf{in}_Q \, (x : A) : Q(A, R)$ and path constructor 
$\mathsf{equiv}\, (x, y : A)\, (p : R\, x\, y)\, :\, \mathsf{in}_Q\, x = \mathsf{in}_Q\, y$
\footnote{The untruncated type $Q(A, R)$ is not an h-set in general}.
To define $g\colon Q(A,R)\to B$, one supplies a function $f\colon A\to B$
and paths $h_{x,y,p}\colon f(x)=f(y)$ for $p\colon R\,x\,y$. The rules are
\[
  g(\mathsf{in}_Q\,x)\equiv f(x),
  \qquad
  \mathsf{ap}_g(\mathsf{equiv}\,x\,y\,p)=h_{x,y,p},
\]
where $\mathsf{ap}_g$ is the action of $g$ on paths. In the presentation of
HITs in the HoTT Book (\cite{hottbook}, Section~6.2), the first equation is
judgemental and the second is propositional. Thus the second equation can
be used in proofs, but does not itself reduce the expression on its left.
\end{enumerate}

The type theory of Arend allows for particularly convenient definitions following the last
two approaches.

Definition (\ref{PropDefQT}) can be stated in Arend with the universe $\Prop$ in place of 
$\mathsf{hProp}$ and $\Set$ in place of $\mathsf{hSet}$.
Since $\Prop$ in Arend is impredicative, resizing rules are unnecessary in this case.
Implementing this construction in another prover requires checking its
universe and elimination rules. In particular, elimination into an arbitrary
type proved to be an h-proposition is not generally supplied by the
empty and singleton elimination exceptions of Rocq and Lean.

Arend also improves on the approach (\ref{HITdefQT}). It turns HITs into ordinary inductive types with conditions. Instead of the path constructor
$\mathsf{equiv}\, (x, y : A)\, (p : R\, x\, y) : \mathsf{in}\, x = \mathsf{in}\, y$ for
$A / R : \Set$ we can define an ordinary constructor 
$\mathsf{equiv}\, (x, y : A)\, (p : R\, x\, y)\, (i : \I) : A / R$ with conditions
$\mathsf{equiv}\, x\, y\, p \, \lcon \equiv \mathsf{in}\, x$ and
$\mathsf{equiv}\, x\, y \, p \, \rcon \equiv \mathsf{in}\, y$. Moreover instead 
of using set-truncation HIT Arend allows to simply declare $A / R$ to be a
truncated inductive type in $\Set$ with elimination restricted to h-sets.
These constructor and eliminator equations do not give general computation
rules for coercion through the quotient type; the limitation is discussed
in Section~\ref{sec:cubical-comparison}. The same mechanism also allows
quotients to be presented directly as inductive types with conditions,
as in the case of integers (see the beginning
of Section~\ref{sec:ExtTT}) or the type of univariate polynomials in $\arendlib$ 
(see Section~\ref{sec:lib-alg}).

\subsubsection{Isomorphic types are equal}\label{sec:SIP}

One important property of MLTT is that it cannot distinguish between isomorphic types
\footnote{We prefer to call ``isomorphism'' here what is usually called 
``equivalence'' in HoTT since we prefer to turn to pre-HoTT terminology while discussing 
applications to set-theoretic formalization}. Namely,
if $A$ and $B$ are isomorphic types and $P\, X$ is a property of types then $P\, A$ and the 
negation of $P\, B$ cannot be simultaneously proven in MLTT. This can be shown by constructing a homotopical model of MLTT where existence
of isomorphism between $A$ and $B$ implies existence of elements in $A = B$ \cite{SSetsVoev}.

In fact, a stronger statement holds for MLTT: all properties of types are stable under 
isomorphism. Namely, for any predicate $P\, X$ on types which is defined also for $X$ outside
of universes\footnote{In particular, $P\, X$ can be arbitrary closed predicate}
there is a function 
$[\![ P ]\!](A, B : \Type) : (A \cong B) \to (P\, A \cong P\, B)$. This is called
\emph{external univalence} - a metatheoretic property of theories capturing preservation of 
isomorphisms. It is metatheoretic the construction of $[\![ P ]\!]$ is dependent on
the structure of $P$ and is not uniform.
External univalence can be shown using homotopical inverse diagram construction 
\cite{KapLumHomIDiag} or univalent parametricity translation \cite{UnivParametr}. 
It has 
been proven in the general setting of second-order generalized algebraic theories in
\cite{ExternalUniv} which
implies that it holds for any type theory with identity types and any standard choice
of type formers like $\sum$-, $\prod$-types, universes, inductive types and so on.

It is thus natural to avoid the routine of proving stability for each particular $P\, X$
and pass to an extension or a modification of MLTT where univalence holds \emph{internally} 
or more precisely where $A \cong B \to A = B$ holds. 

The simplest minimal such extension of 
MLTT would be just MLTT with $\mathsf{ua} : A \cong B \to A = B$ added as an axiom. This is
essentially the way univalence is introduced to HoTT-I: an isomorphism between types can be converted to equality between types by built-in function $\iso$. A small advantage of HoTT-I 
over MLTT with the axiom is the presence of built-in
computational rules for $\iso$ that make the use of univalence a bit more convenient.

\textbf{Example: types of unary and binary natural numbers.} A simple classical
example of isomorphic types, where a mechanism for transferring proofs of properties
from one to the other would be desirable, is given by different representations of natural
numbers: unary and binary. This example is rather relevant for the formal verification:
binary numbers are computationally efficient and are used in formal descriptions of
algorithms whereas unary numbers are easier to reason about.

Unary natural numbers are defined as
the standard inductive type $\Nat$ with constructors $0 : \Nat$ and 
$\suc : \Nat \to \Nat$. This definition is convenient for proving properties of
natural numbers via standard induction but is computationally inefficient.

The type $\Nat_{\mathsf{Bin}}$ of natural numbers represented as binary strings without
trailing zeros can be defined as an inductive type with constructors 
$0_\mathsf{Bin} : \Nat_{\mathsf{Bin}}$ and 
$\mathsf{pos} : \mathsf{Bin}_{+} \to \Nat_{\mathsf{Bin}}$ where $\mathsf{Bin}_{+}$
is an inductive type with constructors $1_{\mathsf{Bin}} : \mathsf{Bin}_{+}$ and 
$\mathsf{digit} : \mathsf{Fin}\, 2 \to \mathsf{Bin}_{+} \to \mathsf{Bin}_{+}$. 
For example, the binary number $1010$ is represented as the term
$\mathsf{pos}\, (\mathsf{digit}\, 0\, (\mathsf{digit}\, 1 (\mathsf{digit}\, 0\, 
1_{\mathsf{Bin}}))) : \Nat_{\mathsf{Bin}}$. In contrast to the unary definition 
$\Nat$ above, 
this definition of natural numbers is computationally efficient, but proving even
basic properties of $+_{\mathsf{Bin}}$ and $*_{\mathsf{Bin}}$ defined in
computationally efficient way becomes rather involved.

Problems such as enabling $\Nat$ to be used for proving properties of 
$\Nat_{\mathsf{Bin}}$ have been central to a long line of research on mechanisms for transferring proofs of theorems between isomorphic types \cite{changingIndTypes, freeRefinements, transferThmCoq}. At the heart of many approaches lies the concept 
of parametricity \cite{Reynolds83} which is closely tied to developments on free 
theorems about
observational equivalences \cite{freeThm}, data refinements for free \cite{freeRefinements} and proofs for free \cite{freePrf}. 

Recent work \cite{UnivParametr} demonstrates how
certain limitations of parametricity-based approaches can be addressed by 
combining parametricity and univalence. Parametricity enables a \emph{white-box} translation of definitions and proofs of theorems between equivalent types by leveraging the internal structure of the terms being translated and their types. In contrast, univalence provides a universally applicable \emph{black-box} translation through transport along equalities, such as $\mathsf{ua}(f) : A = B$ for $f : A \cong B$. Each approach, however, has its limitations when used in isolation. 

For instance, transporting a function $\Nat_\mathsf{Bin}\to\Nat_\mathsf{Bin}$
to $\Nat\to\Nat$ does not by itself preserve its complexity: converting
between representations can be costly, and transport must itself compute.
Induction on $\Nat$ remains available, but the transported function need not
have useful judgemental equations for that induction. Parametric transport
can instead use correspondences between chosen implementations, such as $+$
and $+_{\mathsf{Bin}}$. In dependent types this must also account for
type-level computation: a family $A(n)$ defined using the eliminator for
$\Nat$ satisfies computational
rules which would be lost for its analogue for $\Nat_{\mathsf{Bin}}$ and that would
create problems when translating functions $\prod_{n : \Nat}A(n)$ defined using
eliminator for $\Nat$. 

As suggested in \cite{UnivParametr}, these problems can be addressed by devising
combined approach of univalent parametricity, which intertwines white-box and black-box translations in a complementary and effective manner.

\bigskip

\textbf{Isomorphic structures.} MLTT likewise cannot distinguish between various isomorphic structures on top of types such as groups, rings,
topological spaces and so on. In other words MLTT cannot distinguish between isomorphic objects in various categories admitting a natural forgetful
functor to the category of sets. And typically the function $\iso$ can be used to derive equality of isomorphic objects and freely transfer properties
of the one to properties of the other. This is known as the \textbf{structure identity principle} in Univalent Foundations (\cite{hottbook}, Section 9.8).

\subsubsection{Logic and set theory}\label{sec:SetTh}

For a fixed predicative level $n$, impredicative $\Prop$ and univalence make
the category of sets in $\Set\,n$ an elementary topos. With $\LEM$ and
choice for sets it satisfies Lawvere's Elementary Theory of the Category of
Sets~\cite[Section 10.1]{hottbook}. Interpretations of material set theories
require a separate construction of sets and membership: the cumulative
hierarchy $V$ of the HoTT Book gives a model of ZFC with choice
\cite[Section 10.5]{hottbook}. Its construction uses a universe and a higher
inductive type, with $V$ in a larger universe. Interpretations of CZF use
different constructions and additional hypotheses~\cite{thesisCZF}; they
do not follow merely from the elementary-topos statement.

\subsubsection{Hierarchies of structures}\label{sec:hierachies}

Formalization of mathematics crucially 
relies on mechanisms for grouping data and properties of mathematical structures. In 
languages based on variants of MLTT this is often done by means of dependent record
types supplemented with typeclasses and instance inference mechanism. 
Dependent record types and typeclasses in such languages are closely 
related to module systems and typeclasses in functional programming languages
\cite{ModulesXavier,ModulesHarp,typeclassesHaskell}.
Here we focus on record types since the mechanism of typeclasses does not require any extensions
of the core type theory, we discuss it in more detail in Section~\ref{sec:classesLang}.

Record types can be used, for example, to define the type of monoids as follows: 
\begin{flalign*}
\mathsf{Monoid}\, \{ & E\, :\, \Set,\, \mathsf{ide}\, : \, E,\, *\, :\, E\to E \to E, \\
& \mathsf{ide\text{-}left}\, (x : E)\, :\, \mathsf{ide} * x = x,\, 
\mathsf{ide\text{-}right}\, (x : E)\, :\, x * \mathsf{ide} = x, 
\\
& *\text{-}\mathsf{assoc}\, (x\, y\, z : E) : (x * y) * z = x * (y * z)\}
\end{flalign*}
Elements $X : \mathsf{Monoid}$ are essentially tuples containing as data the carrier set $X.E$, 
monoid identity $X.\mathsf{ide}$ and binary operation $X.*$ together with proofs that this data defines
a monoid structure on $X.E$. 

The challenges that arise when designing a dependently typed language with record types 
include the support of multiple inheritance, subtyping, partial implementations and sharing. 
For example, a field is both a commutative local ring and
a GCD domain so it is natural to define the type $\mathsf{Field}$ of fields as an extension of
$\mathsf{LocalCRing}$ and $\mathsf{GCDDomain}$ which requires multiple inheritance. Apart 
from making $\mathsf{Field}$
inherit the data and properties of $\mathsf{LocalCRing}$ and $\mathsf{GCDDomain}$ it involves making $\mathsf{Field}$ a
subtype of the two. Finally, in the constructive setting definition of a domain contains apartness predicate $x \# 0$ as part of data\footnote{With $\LEM$ $x \# 0$ must be the same as $x\neq 0$} which in $\mathsf{Field}$ can be defined as invertibility\footnote{A constructive (Heyting) field is by definition a commutative local ring
such that non-invertible elements equal to $0$} $x \# 0\Rightarrow \mathsf{isInvertible}(x)$ and this requires partial implementations. The tightness of the apartness relation then
exactly says that non-invertible elements equal to $0$ and is left as unimplemented 
field. \space We discuss
this example in more detail in Section~\ref{sec:classesLang}.
  
In Arend these challenges are addressed both at the level of the core type theory and at a
higher level. As described in Section~\ref{sec:TypeTheory} the type theory of Arend has extensions for 
subsumptive subtyping $A < B$ and manifest fields $x \Rightarrow a$ for record types where the
latter provides a mechanism for partial implementations and sharing. From theoretical point of
view such extensions of MLTT were studied, for example, in \cite{RecordsInLF}. 
There are also implementations: manifest fields are supported, for example, by the 
proof assistant Matita \cite{Matita}. 
These extensions are
extremely useful for working with hierarchies of so-called \emph{bundled} or
\emph{semi-bundled} definitions where all or almost all
operational data of a mathematical structure and its defining predicates are placed in
fields of a 
record (and the rest of data is in parameters). For example, the definition of 
$\mathsf{Monoid}$ above is fully bundled, its less bundled variants
would be records with parameters $\mathsf{Monoid}'\, (E : \Set)$, 
$\mathsf{Monoid}''\, (E : \Set)\, (ide : E)$, 
$\mathsf{Is\_Monoid}\, (E : \Set)\, (ide : E)\, (op : E\to E\to E)$. Definitions 
like $\mathsf{Is\_Monoid}$ are typically classified as \emph{unbundled} since all its
data components are parameters. Unbundled definitions
allow to reduce partial implementations to usual applications, though at the price of
losing key benefits of record types such as being a type of tuples with named projections. 
Still with the mechanism of typeclasses and instance inference unbundled approach
can be made viable as suggested, for example, in \cite{SpittersUnbundled}. 

Without manifest fields possibilities for defining record extensions become much more 
limited even in unbundled case. 
For example, manifest fields allow to define record $\mathsf{Ring}$ as an extension of 
$\mathsf{Semiring}$. Some of defining properties of $\mathsf{Semiring}$, namely 
$x * 0 = 0$ and $0 * x = 0$,
become derivable in $\mathsf{Ring}$ and therefore these fields should be implemented (see also Section~\ref{sec:classesLang}). Without manifest fields, in bundled or unbundled case,
typically one has to define more of the small, atomic records and combine them to define
various composite records, the option of constructing composite records as specializations of
other composite records becomes unavailable. 

Arend language has numerous high level constructs on top of record types which we mainly
discuss in Section~\ref{sec:classesLang}: typeclasses, coercive subtyping, anonymous extensions and so on. Parameters of records is also a high level construct in Arend:
the mechanism of manifest fields allows to translate parameters to fields in the core type theory.
  
The development of major theorem provers from MLTT family such as Lean, Rocq and Agda has been avoiding subsumptive subtyping and manifest fields
in the core theory thus making record types the same as sigma types. Subsumptive subtyping there is replaced with coercive subtyping $A <_c B$ in which case typechecker inserts an
application $c\, a\, :\, B$ of some function $c\, :\, A\to B$ whenever a term 
$a\, :\, A$ is used in a context where a term of type $B$ is expected. There are 
also ways to deal with manifest fields without extending the core type theory as studied, 
for example, in \cite{ManifestLuo}.

\subsubsection{Arrays}\label{sec:arrays}

The primary application of the type $\mathsf{Array}$ is that it significantly
simplifies recursion and induction for nested inductive types involving inductive types 
$\mathsf{List}\, A$ and $\mathsf{Vec}\, A\, n$. To illustrate this consider inductive type 
of terms $\mathsf{Term}\, (F : \Set)\, (a : F \to \Nat)$ with the constructor
$\mathsf{fun}\, (f : F)\,
 (v : \mathsf{Vec}\, (\mathsf{Term}\, F\, a)\, (a\, f))$ -- constructor for the function application term $f\, v_1\, \ldots\, v_{a(f)}$, where $f$ is a functional symbol
of arity $a(f)$ and $v_1$, $\ldots$, $v_{a(f)}$ are terms. 

In this case the nesting occurs when the type $\mathsf{Term}$, which is being defined,
is used as a parameter of type $\mathsf{Vec}$ in its constructor $\mathsf{fun}$.
By $\mathsf{Vec}\, (A : \Type)\, (n : \Nat)$ here we mean the standard type of
lists of length $n$ represented as indexed 
inductive type generated by the constructors
 $\mathsf{nil} : \mathsf{Vec}\, A\, 0$
and $\mathsf{cons}\, (a : A)\, (v : \mathsf{Vec}\, A\, n) :
 \mathsf{Vec}\, A\, (\suc\, n)$.
 
A useful induction principle must supply an induction hypothesis for every
entry of $v$. Ordinary inductive vectors have a definable indexing operation
$\mathsf{get}$, but the induction principle for this nested definition must
account for the vector structure as well. With arrays, the recursive
arguments are directly presented as a function on $\mathsf{Fin}\,(a(f))$,
and the induction hypotheses take the form
$\prod_{i:\mathsf{Fin}\,(a(f))}T(v\,i)$ for a motive $T$.
Arrays also support constructor pattern matching and judgemental equality
by recursive comparison of their entries, as in the rule in
Section~\ref{sec:ExtTT}. These are the computational conveniences of the
built-in array presentation.

\subsection{Type theory of Arend under types as homotopy types viewpoint}\label{sec:HITs}

\begin{enumerate}
\item \textbf{Higher inductive types.} Higher inductive types (HITs) provide a mechanism for inductive definitions of types with
non-trivial homotopical structure on them. HITs are akin to CW complexes which are defined by specifying
cells of various dimensions and the way they glued. The original approach to defining HITs in HoTT was as follows:
a HIT $D$ contains ordinary constructors, which are generating points of $D$, and path constructors, which are generating paths of $D$
and have types of the form $l =_D r$, $s =_{(l =_D r)} t$ and so on 
(\cite{hottbook}, Chapter 6).

For example, the circle $\mathbb{S}^1$ can be defined as HIT with the following constructors:

\begin{enumerate}
\item Point constructor $\mathsf{base} : \mathbb{S}^1$ for the base point.
\item Path constructor $\mathsf{loop} : \mathsf{base} =_{\mathbb{S}^1} \mathsf{base}$ for the loop.
\end{enumerate}

A definition of the torus $\mathbb{T}^2$ would involve cells of dimension $2$, for example:

\begin{enumerate}
\item Point constructor $\mathsf{point} : \mathbb{T}^2$.
\item Path constructors 
$\mathsf{loop}_1,\, \mathsf{loop}_2 : \mathsf{point} =_{\mathbb{T}^2} \mathsf{point}$ for two loops.
\item Path constructor 
$\mathsf{face} : \mathsf{loop}_1\cdot \mathsf{loop}_2 =_{(
\mathsf{point} =_{\mathbb{T}^2} \mathsf{point})} \mathsf{loop}_2\cdot \mathsf{loop}_1$ defining
a cell of dimension $2$, where $\mathsf{loop}_1\cdot \mathsf{loop}_2$ is concatenation of paths.
\end{enumerate}

The recursion principle for HITs is a straightforward generalisation of the one for ordinary inductive types. For example,
in order to define a function from $\mathbb{S}^1$ to a type $A$ one needs to specify a point $a : A$
and a loop $l : a =_A a$ in $A$ and the recursor gives a function 
$\mathsf{rec}_{\mathbb{S}^1}^{a, l} : \mathbb{S}^1 \to A$ such that
$\mathsf{rec}_{\mathbb{S}^1}^{a, l}(\mathsf{base})\equiv a$
and $\mathsf{pmap}\ \mathsf{rec}_{\mathbb{S}^1}^{a, l}\ \mathsf{loop} = l$, where 
$\mathsf{pmap}$ is an obvious function mapping a path between arguments to a path between images of a function
. Dependent case, that is the induction
principle, is also not hard to state, but it is a bit more subtle (\cite{hottbook}, 
Section 6.2).

With this approach to HITs it is far from obvious how to formulate a rigorous schema of general inductive definitions of this kind. 

The interval type in Arend allows to use ordinary \textbf{constructors with conditions} instead of path constructors. For example, 
$\mathbb{T}^2$ can be defined in Arend as an \textbf{inductive type with conditions} with the following constructors:

\begin{enumerate}
\item $\mathsf{point} : \mathbb{T}^2$.
\item $\mathsf{line}_1\, (i : \I) : \mathbb{T}^2,\, \mathsf{line}_2\, (i : \I) : \mathbb{T}^2$ with conditions $\mathsf{line}_{1(2)}\, \lcon \equiv \mathsf{point}$ and
$\mathsf{line}_{1(2)}\, \rcon \equiv \mathsf{point}$.
\item $\mathsf{face}\, (i : \I)\, (j : \I) : \mathbb{T}^2$ with conditions:
\begin{itemize}
\item $\mathsf{face}\ \lcon\ j \equiv \mathsf{line}_2\ j$.
\item $\mathsf{face}\ \rcon\ j \equiv \mathsf{line}_2\ j$.
\item $\mathsf{face}\ i\ \lcon \equiv \mathsf{line}_1\ i$.
\item $\mathsf{face}\ i\ \rcon \equiv \mathsf{line}_1\ i$.
\end{itemize}
\end{enumerate}

The recursion and induction principles are just the standard ones with the requirement that conditions must be preserved by the function being
defined. The general schema of inductive definitions of this kind can be easily formulated: it is the standard schema plus specification of valid forms of conditions, which is 
rather simple. Definitions of HITs
in Arend are pretty similar to definitions of HITs in cubical type theories, however there
are some differences (\cite{hitsInArend}, Section 2.4.4).

Elimination principles for HITs, as formulated in \cite{hottbook}, can be derived in Arend,
so all HoTT book style proofs can be carried out in Arend. But often inductive types with
conditions allow for much simpler proofs. To illustrate this consider a proof of 
$\mathbb{T}^2 \cong \mathbb{S}^1 \times \mathbb{S}^1$ in Arend. The definition of mutually
inverse functions $f : \mathbb{T}^2 \to \mathbb{S}^1 \times \mathbb{S}^1$ and 
$g : \mathbb{S}^1 \times \mathbb{S}^1 \to \mathbb{T}^2$ is completely straightforward:

\begin{enumerate}
\item Using elimination for $\mathbb{T}^2$ the function $f$ is defined by its values 
on constructors: 
$f(\mathsf{point}) :\equiv (\mathsf{base}, \mathsf{base})$,
$f(\mathsf{line}_1\, i) :\equiv (\mathsf{loop}\, i, \mathsf{base})$,
$f(\mathsf{line}_2\, j) :\equiv (\mathsf{base}, \mathsf{loop}\, j)$,
$f(\mathsf{face}\, i\, j) :\equiv (\mathsf{loop}\, i, \mathsf{loop}\, j)$.
\item Using double elimination for $\mathbb{S}^1$ the inverse $g$ is defined by its values
on constructors: $g(\mathsf{base}, \mathsf{base}) :\equiv \mathsf{point}$,
$g(\mathsf{loop}\, i, \mathsf{base}) :\equiv \mathsf{line}_1\, i$,
$g(\mathsf{base}, \mathsf{loop}\, j) :\equiv \mathsf{line}_2\, j$,
$g(\mathsf{loop}\, i, \mathsf{loop}\, j) :\equiv \mathsf{face}\, i \, j$.
\end{enumerate}

The identities $g(f\, x) = x$ and $f(g\, y) = y$ are proven by reflexivity
for each of the constructors and thus can be proven trivially using elimination. 
In contrast, the HoTT book style proof of $\mathbb{T}^2 \cong \mathbb{S}^1 \times \mathbb{S}^1$
is rather involved (\cite{hottbook}, Exercise 6.3).

\end{enumerate}

\subsection{Comparison with other cubical type theories}\label{sec:cubical-comparison}

Arend's interval and the interval of the cubical type theories serve the same
purpose, namely presenting an $n$-dimensional cube as a term depending on $n$
coordinates, but they are different kinds of object, and that difference decides
what computes.  This subsection states the difference, measures how far
computation actually goes by means of a concrete example, describes what the
cubical theories do instead, and says what would have to be added here.

\paragraph{Interval variables.}
In HoTT-I the interval is a type and $i \colon \I$ is an ordinary assumption of
the same standing as $n \colon \Nat$.  A substitution may replace $i$ by an
arbitrary term of type $\I$, including one containing $\coe$, path application,
or a function defined by any means whatever.  In the cubical type theories a
dimension is a separate syntactic class, rather than a term of an ordinary
fibrant type. Dimension expressions are
generated by a fixed grammar --- names and the two endpoints, together with
$\wedge$, $\vee$ and $\neg$ in the De Morgan case --- and only expressions of
that grammar may be substituted for a dimension variable. This distinction
constrains the available computation rules, although regularity and the
particular constructor clauses are further design choices. HoTT-I's
interval is an ordinary fibrant type, while the cubical dimension interval
is not.  We explain the \hyperref[par:interval-fibrancy]{fibrancy obstruction}
after introducing faces and composition below.

\paragraph{Cube operations are available.}
It is not the case that HoTT-I lacks the operations on cubes.  The cartesian
ones cost nothing at all, precisely because interval variables are ordinary
variables: a face is the substitution $i:=\lcon$, a degeneracy is weakening, a
diagonal is $i:=j$, and an exchange of coordinates is a permutation of the
context.  None of these needs a rule, and each is automatically stable under
substitution, being an operation of the ambient theory rather than of a separate
syntactic class.

The operations that a cartesian cube category does not have are also available,
as ordinary definitions rather than as primitive dimension syntax.  Reversal of
paths is definable, so with $\mathsf{segment}:=\pcon(\lam i\, i)$ of type
$\Path(\lam x\, \I,\lcon,\rcon)$ the term
$\mathsf{inv}(\mathsf{segment})\ @\ k$ exchanges the endpoints of $\I$.
The connection $\mathsf{squeeze}$ is also definable in HoTT-I: iterating
$\coe$ in path types gives horn fillers from which it can be
constructed~\cite[Section~2.1]{IsaevHoTTI}.  Its judgemental boundary equations
have the form of those for a minimum:
\[
 \mathsf{squeeze} \colon \I \to \I \to \I,
 \qquad
 \begin{aligned}
 \mathsf{squeeze}(\lcon,j)&\equiv\lcon, &
 \mathsf{squeeze}(\rcon,j)&\equiv j,\\
 \mathsf{squeeze}(i,\lcon)&\equiv\lcon, &
 \mathsf{squeeze}(i,\rcon)&\equiv i.
 \end{aligned}
\]
Read as a square, with $i$ running horizontally and $j$ vertically, it has two
constant sides and two that are the interval itself:
\[
\xymatrix@=3pc{
 \lcon \ar[r]^{\lam i\, i} & \rcon \\
 \lcon \ar[r]_{\lam i\, \lcon} \ar[u]^{\lam j\, \lcon}
 & \lcon \ar[u]_{\lam j\, j}
}
\]
The dual connection $\mathsf{squeezeR}\colon\I\to\I\to\I$ can be
constructed explicitly from $\mathsf{squeeze}$ using one coercion in a nested
path type.  Put
\[
 \begin{aligned}
 q_k&:=\pcon(\lam j\,\mathsf{squeeze}(k,j)),\\
 Q_k&:=\Path\bigl(\lam i\,
          \Path(\lam j\,\I,\mathsf{squeeze}(k,i),k),\
          q_k,\mathsf{idp}_k\bigr).
 \end{aligned}
\]
An element of $Q_k$ is a square whose left side is $q_k$, whose right and
top sides are constant at $k$, and whose bottom side is
$\lam i\,\mathsf{squeeze}(k,i)$.  The picture shows these boundary values,
with $i$ horizontal and $j$ vertical; write $0:=\lcon$ and $1:=\rcon$
in the diagrams.
\begin{center}
\begin{tikzpicture}[
  x=1cm,y=1cm,font=\small,
  connection edge/.style={draw=black!65,line width=.7pt,
    -{Latex[length=1.7mm]}},
  varying edge/.style={connection edge,draw=blue!65!black},
  every node/.style={inner sep=3pt}
]
  \fill[blue!4] (0,0) rectangle (3,1.6);
  \node (q00) at (0,0) {$0$};
  \node (q10) at (3,0) {$k$};
  \node (q01) at (0,1.6) {$k$};
  \node (q11) at (3,1.6) {$k$};
  \draw[varying edge] (q00)--(q10)
    node[midway,below] {$\mathsf{squeeze}(k,i)$};
  \draw[varying edge] (q00)--(q01)
    node[midway,left] {$q_k\ @\ j$};
  \draw[connection edge] (q01)--(q11) node[midway,above] {$k$};
  \draw[connection edge] (q10)--(q11) node[midway,right] {$k$};
  \node[text=black!65] at (1.5,.8) {a square in $Q_k$};
\end{tikzpicture}
\end{center}
At $k=\lcon$ all four sides are
constant at $\lcon$, so $\pcon(\lam i\,\mathsf{idp}_{\lcon})$
inhabits $Q_{\lcon}$.  Define
\[
 H:=\coe\bigl(\lam k\,Q_k,\
         \pcon(\lam i\,\mathsf{idp}_{\lcon}),\ \rcon\bigr),
 \qquad
 \mathsf{squeezeR}(i,j):=(H\ @\ i)\ @\ j.
\]
The coercion transports this constant square to one with the required boundary:
\begin{center}
\begin{tikzpicture}[
  x=1cm,y=1cm,font=\small,
  connection edge/.style={draw=black!65,line width=.7pt,
    -{Latex[length=1.7mm]}},
  varying edge/.style={connection edge,draw=blue!65!black},
  every node/.style={inner sep=3pt}
]
  \fill[black!3] (0,0) rectangle (2.5,1.6);
  \node (s00) at (0,0) {$0$};
  \node (s10) at (2.5,0) {$0$};
  \node (s01) at (0,1.6) {$0$};
  \node (s11) at (2.5,1.6) {$0$};
  \draw[connection edge] (s00)--(s10) node[midway,below] {$0$};
  \draw[connection edge] (s00)--(s01) node[midway,left] {$0$};
  \draw[connection edge] (s01)--(s11) node[midway,above] {$0$};
  \draw[connection edge] (s10)--(s11) node[midway,right] {$0$};
  \node at (1.25,.8) {$0$};
  \node at (1.25,2.3) {$Q_0$: constant square};

  \draw[-{Latex[length=2mm]},line width=.8pt] (3.35,.8)--(5.85,.8)
    node[midway,above] {$\coe(\lam k\,Q_k,-,1)$};

  \begin{scope}[xshift=6.7cm]
    \fill[blue!4] (0,0) rectangle (2.5,1.6);
    \node (t00) at (0,0) {$0$};
    \node (t10) at (2.5,0) {$1$};
    \node (t01) at (0,1.6) {$1$};
    \node (t11) at (2.5,1.6) {$1$};
    \draw[varying edge] (t00)--(t10) node[midway,below] {$i$};
    \draw[varying edge] (t00)--(t01) node[midway,left] {$j$};
    \draw[connection edge] (t01)--(t11) node[midway,above] {$1$};
    \draw[connection edge] (t10)--(t11) node[midway,right] {$1$};
    \node at (1.25,.8) {$\mathsf{squeezeR}(i,j)$};
    \node at (1.25,2.3) {$Q_1$: the square $H$};
  \end{scope}
\end{tikzpicture}
\end{center}
Since $q_{\rcon}\equiv\mathsf{segment}$ and
$\mathsf{squeeze}(\rcon,i)\equiv i$, the type of $H$ is
\[
 \Path\bigl(\lam i\,\Path(\lam j\,\I,i,\rcon),\
             \mathsf{segment},\mathsf{idp}_{\rcon}\bigr).
\]
The endpoint equations for the outer path give the first row below, and
those for the inner path give the second:
\[
 \begin{aligned}
 \mathsf{squeezeR}(\lcon,j)&\equiv j, &
 \mathsf{squeezeR}(\rcon,j)&\equiv\rcon,\\
 \mathsf{squeezeR}(i,\lcon)&\equiv i, &
 \mathsf{squeezeR}(i,\rcon)&\equiv\rcon.
 \end{aligned}
\]
Operations of this kind are what the cubical literature calls \emph{connections},
the De Morgan ones being $\wedge$ and $\vee$.  What they are for is contracting a
path to reflexivity.  Given $p$ from $a$ to $b$, the path
$\pcon(\lam j\, p\ @\ \mathsf{squeeze}(i,j))$ is the initial segment of $p$ up to
the point $i$: it is constant at $a$ when $i$ is $\lcon$, and it is $p$ itself
when $i$ is $\rcon$.  That family is exactly the data the $\mathsf J$-eliminator
needs, and with it $\mathsf J$ is defined in the familiar way,
\[
 \mathsf J(B,b,p)
 :=
 \coe\bigl(\lam i\, B(p\ @\ i,\ \pcon(\lam j\, p\ @\ \mathsf{squeeze}(i,j))),\
 b,\ \rcon\bigr),
\]
its reflexivity computation holding judgementally by regularity, the second
$\coe$ equation of \S\ref{sec:TypeTheory}.  On this last
point the comparison runs the other way round.  In CCHM and in Cubical Agda the
corresponding equation holds only up to a path~\cite{CCHM,cubicalAgda}.
Transport is computed there by recursion on the head of the type, with one
clause per type former, so it reduces whenever the type is concrete; but the
family occurring in $\mathsf J$ has a variable as its head, and no clause
applies.  This costs no canonicity --- a closed transport always meets a type
former eventually, and canonicity for CCHM is a theorem~\cite{HuberCanonicity}
--- and what it does cost is the definitional equation for an arbitrary family.
HoTT-I obtains that equation from regularity, whose side condition is of a
different kind: it asks whether the bound variable occurs in the family, and not
what the family looks like.  The $\mathsf J$ instance is exactly the case that
separates the two conditions.  Substituting $p:=\mathsf{idp}$ leaves the family
$\lam i\, B(a,\mathsf{idp})$, since $\mathsf{idp}\ @\ i\equiv a$ and
$\pcon(\lam j\, \mathsf{idp}\ @\ \mathsf{squeeze}(i,j))\equiv\mathsf{idp}$.  Its
head is the variable $B$, so no clause keyed to a type former can fire; but $i$
does not occur in it, so regularity does.  Iterating
$\coe$ in path families in the same style yields fillers for every finite full
cubical horn~\cite{IsaevHoTTI}.  So the distinction between HoTT-I and a
structural cubical theory is not the availability of connections, reversal,
$\mathsf J$, or box filling.

\paragraph{What the interval does not carry.}
What is absent is an equational theory on the interval.  In a De Morgan theory
the dimension terms form a free De Morgan algebra and are compared by a decision
procedure for that algebra, so that $i\wedge i$ and $i$ are the same term before
any rule is consulted.  Here $\mathsf{squeeze}$ is an ordinary function: one can
prove $\mathsf{squeeze}(i,i)=i$ using the two connections, but the two sides
are not judgementally equal, and a term of $\I$ has in
general no normal form as a dimension expression at all.

One consequence is immediate.  There is no \emph{strict} lattice of faces.
Internally one may of course form the proposition
$\Path(\lam y\, \I, \lcon, i)$ and combine such propositions with products and
truncated sums, but no such family can play the role of the face $(i=0)$: a face
must be inhabited at $i:=\lcon$ and empty at $i:=\rcon$, whereas
$\pcon(\lam j\, \mathsf{squeeze}(i,j))$ inhabits the displayed family for
\emph{every} $i$, the interval being contractible.  The other consequences are
the subject of the rest of this subsection.

\paragraph{How far does computation go?}
The question is best asked of an example.  Take the circle $\mathbb S^1$ of
\S\ref{sec:HITs}, presented in the Arend style by a point $\mathsf{base}$ and a
constructor $\mathsf{loop}$ with an interval argument, subject to the conditions
$\mathsf{loop}\,\lcon\equiv\mathsf{base}\equiv\mathsf{loop}\,\rcon$.  The two are
different things and both are needed below: $\mathsf{loop}$ is a function
$\I\to\mathbb S^1$, while the path it generates is
\[
 \ell:=\pcon(\lam i\, \mathsf{loop}\, i)
 \ \colon\ \mathsf{base}=\mathsf{base}.
\]
Write $p\cdot q$ for concatenation of paths and $p^{-1}$ for inversion, both
defined from $\mathsf J$ in the usual way: for $p \colon \Path(\lam y\, A,a,b)$
and $q \colon \Path(\lam y\, A,b,c)$ one has
$p\cdot q \colon \Path(\lam y\, A,a,c)$ and
$p^{-1} \colon \Path(\lam y\, A,b,a)$, with
\[
 p\cdot\mathsf{idp}\equiv p,
 \qquad
 \mathsf{idp}^{-1}\equiv\mathsf{idp}
\]
judgementally, by the reflexivity computation of $\mathsf J$.  Let $\mathsf{next}$ and $\mathsf{prev}$ be the
successor and predecessor of $\Int$.
The universal cover is the function $\mathsf{cover} \colon \mathbb S^1\to\Set_0$
given on constructors by
\[
 \mathsf{cover}(\mathsf{base})\equiv\Int,
 \qquad
 \mathsf{cover}(\mathsf{loop}\, i)
 \equiv\iso(\mathsf{next},\mathsf{prev},\ldots,i),
\]
and the winding number of a loop is obtained by transporting $0$ along it,
\[
 \mathsf{winding}(p)
 :=\coe(\lam i\, \mathsf{cover}(p\ @\ i),\ 0,\ \rcon)
 \ \colon\ \Int .
\]
This is the standard test of computational univalence, and it is a fair one:
each of the equations below is provable in Arend, so the question is only
whether the kernel reduces the term or whether one has to prove the equation by
hand.  Normalizing in Arend 1.12 gives the following.

\begin{center}
\begin{tabular}{@{}ll@{}}
\hline
term & normal form \\
\hline
$\mathsf{winding}(\mathsf{idp})$ & $0$ \\
$\mathsf{winding}(\ell)$ & $1$ \\
$\mathsf{winding}(\ell\cdot\ell)$
 & no reduction: a residual $\coe$ through $\ell\cdot\ell$ \\
$\mathsf{winding}(\ell^{-1})$
 & no reduction: a residual $\coe$ through $\ell^{-1}$ \\
$\mathsf{winding}\bigl(\pcon(\lam i\, \mathsf{loop}(\mathsf{squeeze}(i,i)))\bigr)$
 & no reduction: a residual $\coe$ through
   $\iso(\ldots,\mathsf{squeeze}(i,i))$ \\
\hline
\end{tabular}
\end{center}

\noindent
So the univalence round trip computes once, and stops there.  Note that two
separate transports do compose: transporting $0$ along $\ell$ and then
transporting the result along $\ell$ again reduces to $2$.  It is only the
transport along the \emph{composite path} that does not.

\paragraph{What goes wrong.}
All three failures have one cause.  The computation rule for $\iso$ in
\S\ref{sec:TypeTheory} applies only when three conditions hold together:

\begin{itemize}
\item the family is $\lam i\, \iso(A,B,\ldots,i)$, its last argument being the
  bound variable itself;
\item the isomorphism data do not mention $i$;
\item the target is $\rcon$.
\end{itemize}

\noindent
The third condition already excludes a generic target.  The failures in the
table, however, all violate the first, and for two different reasons.

For the composite $\ell\cdot\ell$, the path is built by $\mathsf J$, so
$(\ell\cdot\ell)\ @\ i$ is not an application of $\mathsf{loop}$ to anything
and $\mathsf{cover}((\ell\cdot\ell)\ @\ i)$ never exposes an $\iso$ line.
The inverse $\ell^{-1}$ fails for the same reason.  In
the fifth row the family is $\lam i\, \iso(\ldots,\mathsf{squeeze}(i,i))$, which
does have $\iso$ as its head and the correct endpoints, since
$\mathsf{squeeze}(\lcon,\lcon)\equiv\lcon$ and
$\mathsf{squeeze}(\rcon,\rcon)\equiv\rcon$; but its last argument is a term of
$\I$ that does not reduce to the bound variable, so the rule still does not
apply, and there is no equational theory on $\I$ to normalize
$\mathsf{squeeze}(i,i)$ to $i$ first. In a De Morgan cubical theory the
corresponding reparametrization uses a dimension expression, and
$i\wedge i$ is already $i$. Cartesian dimension syntax has no such
connection operation.

It should be said plainly what this does and does not mean.  Nothing is
unprovable: the winding number of $\ell\cdot\ell$ is $2$, and one proves it in a
line from the standard transport laws.  What fails is computation by reduction,
and with it canonicity, since
$\mathsf{winding}(\ell\cdot\ell)$ is a closed term of $\Int$ in normal form
that is not a numeral.

\paragraph{Composition is coercion in a path type.}
Two of the three failures have a common source, and it is not composition as
such.  Concatenation and inversion are defined from $\mathsf J$, and $\mathsf J$,
by the definition displayed above, is a coercion; so they are coercions.  It is
worth carrying out the two unfoldings.

For concatenation, take
\[
 p \colon \Path(\lam y\, A,a,b),
 \qquad
 q \colon \Path(\lam y\, A,b,c),
 \qquad
 B := \lam x\, \lam z\, \Path(\lam y\, A, a, x),
\]
noting that the second argument $z$ does not occur in the body of $B$.  The path
argument of $\mathsf J$ is $q$, so the family appearing in its definition is
\[
 B\bigl(q\ @\ i,\ \pcon(\lam j\, q\ @\ \mathsf{squeeze}(i,j))\bigr)
 \ \equiv\ \Path(\lam y\, A,\ a,\ q\ @\ i),
\]
the witness being discarded, and therefore
\[
 \mathsf J(B,p,q)
 \ \equiv\ \coe\bigl(\lam i\, \Path(\lam y\, A,\ a,\ q\ @\ i),\ p,\ \rcon\bigr).
\]
The types are as they should be: the base case is $B(b,\mathsf{idp})$, which is
$\Path(\lam y\, A,a,b)$, and that is where $p$ goes, while the result is
$B(c,q)$, which is $\Path(\lam y\, A,a,c)$.

For inversion, with $p \colon \Path(\lam y\, A,a,b)$, take $B$ to be
$\lam x\, \lam z\, \Path(\lam y\, A, x, a)$ and the path argument to be $p$
itself.  The same two steps give
\[
 B\bigl(p\ @\ i,\ \pcon(\lam j\, p\ @\ \mathsf{squeeze}(i,j))\bigr)
 \ \equiv\ \Path(\lam y\, A,\ p\ @\ i,\ a),
\]
\[
 \mathsf J(B,\mathsf{idp},p)
 \ \equiv\ \coe\bigl(\lam i\, \Path(\lam y\, A,\ p\ @\ i,\ a),\
 \mathsf{idp},\ \rcon\bigr),
\]
with base case $B(a,\mathsf{idp})$, that is $\Path(\lam y\, A,a,a)$, and result
$B(b,p)$, that is $\Path(\lam y\, A,b,a)$.  In the first coercion the right
endpoint of the path family moves and the left is fixed; in the second the left
moves and the right is fixed.

The two computation rules recorded above are recovered at once.  Put
$q:=\mathsf{idp}$ in the first coercion: then $\mathsf{idp}\ @\ i\equiv b$, so
the family is $\lam i\, \Path(\lam y\, A,a,b)$, which does not depend on $i$, and
regularity gives $p$.  Put $p:=\mathsf{idp}$ in the second and the same argument
gives $\mathsf{idp}$.

With these coercion-based implementations, the concatenation and inversion
examples expose the same gap: there is no computation rule for $\coe$ in a
family of path types.
The fifth row concerns reparametrized univalence: it calls for either an
equation on the interval or a coercion rule that handles the reparametrized
family directly.

\paragraph{Faces, partial elements and composition.}
The cubical theories build everything on one further notion, which the dimension
grammar makes available and which is worth describing before $\mathsf{Glue}$.  A
\emph{face formula} $\varphi$ is a formula of that grammar, built from equations
$(i=0)$ and $(i=1)$ by conjunction and disjunction, describing part of the
boundary of a cube.  A \emph{partial element} is a term defined only where
$\varphi$ holds.  The basic operation consumes one: given a family $A$ along a
direction $i$, a partial element $u$ on $\varphi$, and a starting point $a_0$
agreeing with $u$ at $i=0$, composition produces the value at $i=1$,
\[
\frac{\Gamma,i:\mathbb I,\varphi\vdash u:A\qquad
      \Gamma\vdash a_0:A[i:=0]\qquad
      \Gamma,\varphi\vdash u[i:=0]\equiv a_0}
     {\Gamma\vdash \mathsf{comp}^{i}A[\varphi\mapsto u]a_0:A[i:=1]},
\qquad
\Gamma,\varphi\vdash
\mathsf{comp}^{i}A[\varphi\mapsto u]a_0\equiv u[i:=1],
\]
and the result restricts to $u$ where $\varphi$ holds.  The bracket is the
partial element itself: $[\varphi\mapsto u]$ is the element whose value is $u$
where $\varphi$ holds, and it may carry several clauses, one for each face on
which a value is prescribed.  For a small example,
the concatenation of $p \colon \Path(\lam y\, A,a,b)$ and
$q \colon \Path(\lam y\, A,b,c)$ is a square whose bottom is $p$ and whose two
vertical sides are the constant path at $a$ and the path $q$, and its top is
their concatenation:
\[
\xymatrix@=3pc{
 a \ar[r]^{p\cdot q} & c \\
 a \ar[r]_{p} \ar[u]^{a} & b \ar[u]_{q}
}
\]
\[
 p\cdot q
 \ =\
 \lam i\, \mathsf{comp}^{j} A
   \bigl[\,(i=0)\mapsto a,\ \ (i=1)\mapsto q\ j\,\bigr](p\ i),
\]
path application being written by juxtaposition as in that notation.  The
partial element is given only on $(i=0)\vee(i=1)$, the two disjuncts do not
overlap, and the premise of the rule is discharged by the judgemental equations
$p\ 0\equiv a$ and $p\ 1\equiv b\equiv q\ 0$.  The face $j=1$ is not prescribed;
it is the result.

Two degenerate cases have names.  Taking $\varphi$ false leaves no partial
element and recovers transport, that is $\coe$.  Taking the family $A$ not to
depend on the filling direction $j$ gives \emph{homogeneous composition}
$\mathsf{hcomp}$, which fills a box inside one fixed type and so needs no
transport along $A$; the example just given is of that form.  Conversely $\coe$
and $\mathsf{hcomp}$ together reconstruct dependent composition, by transporting
the cap along $A$ and then filling homogeneously in the fibre over $j=1$.  CCHM
takes $\mathsf{comp}$ as primitive, Cubical Agda exposes $\mathsf{transp}$ and
$\mathsf{hcomp}$, and cartesian cubical type theory uses $\coe$ and
$\mathsf{hcom}$~\cite{CCHM,cubicalAgda,CartesianSyntax}.

This apparatus is what HoTT-I does not have, and the reason is the one given
above: $\varphi$ ranges over a lattice of strict faces, and no internal family
of propositions can be one.  Arend's constructor conditions prescribe faces for
a particular inductive constructor, as the definition of the torus in
\S\ref{sec:HITs} illustrates, but they are syntactic conditions attached to a
constructor and not a general calculus of partial elements.  What HoTT-I has in
place of the general operation is full horn filling, which is the special case
in which the partial element is given on all faces of a cube but one; the
example above is of that special case, which is why path concatenation is
available here without any of this machinery.

\paragraph{Fibrancy of the dimension interval.}\label{par:interval-fibrancy}
Composition also explains why cubical dimension variables are kept separate
from ordinary type-theoretic variables.  A fibrant type must support the
prescribed filling operations, uniformly under substitution.  The dimension
interval itself does not support these operations.  In the standard cubical
models, this interval is the representable presheaf whose elements are
exactly the expressions allowed by the dimension grammar.  The obstruction
therefore depends on that grammar.

In cartesian cubical syntax, a dimension expression is either $0$, $1$, or
a dimension variable; there is no operation $x\mapsto 1-x$~\cite{CartesianSyntax}.
If the dimension interval were fibrant, every compatible open box in it
would have a filler given by such an expression.  Consider the three sides
of a square prescribed by
\[
 F(x,0)=0, \qquad F(0,y)=y, \qquad F(1,y)=0,
\]
which agree at the two corners they share.  A filler supplies the missing side
$q(x):=F(x,1)$, and the two vertical sides force $q(0)=1$ and $q(1)=0$:
\[
\xymatrix@=3pc{
 1 \ar@{-->}[r]^{q} & 0 \\
 0 \ar[r]_{0} \ar[u]^{y} & 0 \ar[u]_{0}
}
\]
Here $q(x)$ must be a dimension expression with $x$ as its only variable,
so the only possibilities are $0$, $1$, and $x$.  Their endpoint pairs
$(q(0),q(1))$ are respectively $(0,0)$, $(1,1)$, and $(0,1)$.
The required pair $(1,0)$ is absent, so this open square has no filler
in the dimension interval.

De Morgan syntax does contain a reversal $\neg x$, so that particular box can be
filled, and in fact every individual box can.  What CCHM asks for is stronger:
the fillers must be chosen \emph{uniformly}, that is, so as to commute with
every substitution of dimensions, diagonals and connections included.  It is
uniformity that fails.  The witness is a box with outer dimensions $i,j$ and
filling direction $k$, whose cap at $k=0$ is $0$ and whose tube is given only on
the two disjoint corner faces
\[
 (i=0)\wedge(j=1)\ \mapsto\ 0,
 \qquad
 (i=1)\wedge(j=0)\ \mapsto\ k,
\]
each of which agrees with the cap at $k=0$.  In the cube below, the tube
consists of the two colored vertical edges; the other side faces are
unspecified.  The whole shaded bottom face has value $0$.
\begin{center}
\begin{tikzpicture}[
  x=1cm,y=.82cm,font=\small,
  interval frame/.style={draw=black!45,line width=.5pt},
  interval missing/.style={draw=black!45,line width=.5pt,densely dashed},
  interval zero/.style={draw=blue!65!black,line width=1.6pt},
  interval varying/.style={draw=orange!85!black,line width=1.6pt},
  interval axis/.style={-{Latex[length=1.5mm]},line width=.5pt},
  every node/.style={inner sep=2pt}
]
  \coordinate (a) at (0,0);
  \coordinate (b) at (3,0);
  \coordinate (c) at (4.3,1.3);
  \coordinate (d) at (1.3,1.3);
  \coordinate (at) at (0,3);
  \coordinate (bt) at (3,3);
  \coordinate (ct) at (4.3,4.3);
  \coordinate (dt) at (1.3,4.3);
  \path[interval frame,fill=black!9] (a)--(b)--(c)--(d)--cycle;
  \path[interval missing,fill=black!2] (at)--(bt)--(ct)--(dt)--cycle;
  \draw[interval missing] (a)--(at);
  \draw[interval missing] (c)--(ct);
  \draw[interval zero] (d)--(dt)
    node[pos=.48,left,text=blue!65!black] {$0$};
  \draw[interval varying] (b)--(bt)
    node[pos=.60,right,text=orange!85!black] {$k$};
  \fill[blue!65!black] (d) circle (1.6pt) (dt) circle (1.6pt);
  \fill[orange!85!black] (b) circle (1.6pt) (bt) circle (1.6pt);
  \node at (2.15,.65) {$0$};
  \node at (2.15,3.65) {$p(i,j)$};
  \node[anchor=west] at (4.45,1.3) {$k=0$};
  \node[anchor=west] at (4.45,4.3) {$k=1$};
  \node[anchor=south east,text=blue!65!black] at (1.2,4.4)
    {$p(0,1)=0$};
  \node[anchor=west,fill=white,text=orange!85!black] at (3.15,3)
    {$p(1,0)=1$};
  \draw[interval axis] (-1.25,.25)--(-.35,.25) node[right] {$i$};
  \draw[interval axis] (-1.25,.25)--(-.66,.84) node[above right] {$j$};
  \draw[interval axis] (-1.25,.25)--(-1.25,1.35) node[above] {$k$};
\end{tikzpicture}
\end{center}
Writing $p(i,j)$ for the resulting top face at $k=1$, the tube forces
$p(0,1)=0$ and $p(1,0)=1$.  Substitute
$j:=i$.  The face formula becomes false, so the box turns into the empty box
with cap $0$, whose filler is one chosen element not depending on $i$;
uniformity therefore forces $p(i,i)$ to be that constant.  No element of the
free De Morgan algebra on $i,j$ meets these three requirements at
once~\cite{SattlerInterval}.  Cubical Agda reflects all of this by placing
$\mathbb I$ in the non-fibrant universe $\mathsf{IUniv}$, without
$\mathsf{transp}$ or $\mathsf{hcomp}$~\cite{cubicalAgda}.

In HoTT-I, by contrast, fillers in $\I$ may contain coercions and path
applications; they are not confined to a grammar of dimension expressions.
Its interval is an ordinary contractible fibrant type.  A model may take for
it any contractible fibrant object whose endpoint inclusion is a
cofibration~\cite{IsaevModels}.  This provides the full horn fillers used
earlier, without supplying the strict face lattice needed by cubical
composition.

\paragraph{Glue types and univalence.}
In CCHM, a univalence line is constructed from a more general type former,
$\mathsf{Glue}$~\cite{CCHM}.  Its purpose is to extend a type given on a face
to the whole context, using an equivalence to a type already defined there.

For a concrete picture, consider a square of parameters $(i,j)$ and the face
$\varphi=(i=0)\vee(j=0)$ consisting of its left and bottom edges.  A type $A$
is given over the whole square, while a type $T$ and an equivalence
$e\colon T\simeq A$ are given only on these two edges.  The construction
extends $T$ to a type $G$ over the square, together with an equivalence
$\mathsf{unglue}\colon G\to A$ extending $e$:
\begin{center}
\begin{tikzpicture}[
  x=1cm,y=.85cm,font=\small,
  glue sheet/.style={draw=black!50,fill=black!3,line width=.45pt},
  glue face/.style={draw=blue!65!black,line width=1.4pt},
  glue map/.style={-{Latex[length=1.7mm]},line width=.5pt},
  every node/.style={inner sep=2pt}
]
  \node[font=\small\bfseries] at (1.75,4.05) {Given data};
  \path[glue sheet] (0,0)--(2.6,0)--(3.5,.95)--(.9,.95)--cycle;
  \node at (1.75,.48) {$A$};
  \draw[glue face] (.9,.95)--(0,0)--(2.6,0);
  \draw[glue map] (0,2.25)--(0,.10) node[midway,left] {$e$};
  \draw[black!40,densely dotted] (.9,3.25)--(3.5,3.25)--(2.6,2.3);
  \draw[glue face] (.9,3.25)--(0,2.3)--(2.6,2.3);
  \node[anchor=south,text=blue!65!black] at (1.3,2.35) {$T\text{ on }\varphi$};
  \node[anchor=north] at (1.75,-.25) {$A\text{ defined everywhere}$};
  \node[anchor=east,font=\scriptsize,text=black!65] at (.4,2.87) {$i=0$};
  \node[anchor=north,font=\scriptsize,text=black!65] at (1.3,2.27) {$j=0$};

  \draw[glue map] (4.15,2.05)--(5.7,2.05)
    node[midway,above] {$\mathsf{Glue}$};

  \begin{scope}[xshift=6.5cm]
    \node[font=\small\bfseries] at (1.75,4.05) {Result};
    \path[glue sheet] (0,0)--(2.6,0)--(3.5,.95)--(.9,.95)--cycle;
    \node at (1.75,.48) {$A$};
    \draw[glue face] (.9,.95)--(0,0)--(2.6,0);
    \draw[glue map] (0,2.25)--(0,.10) node[midway,left] {$e$};
    \draw[glue map] (1.75,2.15)--(1.75,1.10)
      node[midway,right] {$\mathsf{unglue}$};
    \path[glue sheet] (0,2.3)--(2.6,2.3)--(3.5,3.25)--(.9,3.25)--cycle;
    \draw[glue face] (.9,3.25)--(0,2.3)--(2.6,2.3);
    \node at (2.1,3.02) {$G$};
    \node[anchor=south,text=blue!65!black] at (1.3,2.33) {$G\equiv T\text{ on }\varphi$};
    \node[anchor=north] at (1.75,-.25) {$\mathsf{unglue}\equiv e\text{ on }\varphi$};
  \end{scope}
\end{tikzpicture}
\end{center}
The squares depict the parameter domain, and the thick edges mark $\varphi$.
In the left panel, the dotted edges indicate the part of the domain on which
no type $T$ has been specified.  In the right panel, $G$ is defined throughout
and agrees judgementally with $T$ on the marked edges.  The map to $A$ is an
equivalence throughout, restricting to the given $e$ on those edges.

In general, suppose that $A\colon\U$ is defined in $\Gamma$, while
$T\colon\U$ and
$e\colon T\simeq A$ are defined in the restricted context $\Gamma,\varphi$.
Then
\[
  G:=\mathsf{Glue}[\varphi\mapsto(T,e)]A
\]
is a type in $\Gamma$.  It comes with a projection
$\mathsf{unglue}\colon G\to A$, which is an equivalence.  For $u\colon G$,
the restriction equations are
\[
  \Gamma,\varphi\vdash G\equiv T,
  \qquad
  \Gamma,\varphi\vdash\mathsf{unglue}(u)\equiv e(u).
\]
Here and below, $e$ also denotes the underlying function of the equivalence.
Thus $G$ has the same homotopy type as $A$, while its restriction to the chosen
face is judgementally the specified type $T$.

The introduction form makes the elements of $G$ explicit.  Given $a\colon A$
in $\Gamma$ and $t\colon T$ on $\varphi$, with $e(t)\equiv a$ on that face,
one forms
\[
  \mathsf{glue}[\varphi\mapsto t]a\colon G.
\]
The compatibility condition is a judgemental equality; no path witnessing it
is an additional argument.  The computation rules are
\[
  \mathsf{unglue}(\mathsf{glue}[\varphi\mapsto t]a)\equiv a,
  \qquad
  \Gamma,\varphi\vdash\mathsf{glue}[\varphi\mapsto t]a\equiv t.
\]
Conversely, an element $u\colon G$ supplies both its projection to $A$ and,
on $\varphi$, an element of $T$; reconstructing it from these data gives
$\mathsf{glue}[\varphi\mapsto u](\mathsf{unglue}(u))\equiv u$.

For an equivalence $e\colon A\simeq B$, take $B$ as the base and glue $A$
along $e$ at the left endpoint and $B$ along the identity equivalence at the
right endpoint:
\[
  G_i:=\mathsf{Glue}[(i=0)\mapsto(A,e),\ (i=1)\mapsto(B,\id)]\,B,
  \qquad
  \mathsf{ua}(e):=\lam i\,G_i.
\]
The restriction rule gives $G_0\equiv A$ and $G_1\equiv B$, so
$\mathsf{ua}(e)$ is a path from $A$ to $B$ in the universe.  The projections
to the base fit into the diagram
\[
\xymatrix@R=2.5pc@C=3pc{
  A \ar@{--}[r]^{\mathsf{ua}(e)} \ar[d]_{e}
    & B \ar[d]^{\id} \\
  B \ar@{-}[r]_-{\text{constant}} & B.
}
\]
The horizontal lines are families of types, and the vertical maps are
equivalences.  At a variable $i$, the family retains the $\mathsf{Glue}$
structure, including the projection to $B$ and the prescribed endpoint data.

To obtain computational univalence, one also needs composition in this family.
The composition rule for $\mathsf{Glue}$ first projects the given partial
element and starting point to the base type and composes there.  It then uses
the equivalence on the target face to obtain compatible data in the partial
type, corrects the base component, and applies $\mathsf{glue}$.  The
equivalence witnesses and the composition operations of the constituent types
provide these steps, with the required agreement on the prescribed faces.
In particular, transport along $\mathsf{ua}(e)$ is computed by this general
rule.  In CCHM its agreement with the underlying map of $e$ is propositional:
\[
  \coe(\lam i\,G_i,\ a,\ 1)=_B e(a).
\]
The computation can retain transports in a constant type, which are connected
to the identity by paths but do not disappear by a general regularity rule.
This propositional agreement suffices to prove univalence~\cite{CCHM}.

The advantage over the $\iso$ rule discussed above is that composition is
defined for every family whose outer type former is $\mathsf{Glue}$.  If $d(k)$
is a cubical dimension expression, substituting $i:=d(k)$ in $G_i$ changes the
face formulas.  The result either reduces on a true face or retains
$\mathsf{Glue}$ as its outer type former, so the same composition algorithm
applies whenever the restriction rule does not already determine the type.
This avoids the
specific obstruction in HoTT-I, where a coercion rule recognizes only the
canonical $\iso$ line.

Cartesian cubical type theory implements a related construction with
$V$-types, which use a single distinguished face.  They satisfy
$V_0(A,B,e)\equiv A$ and $V_1(A,B,e)\equiv B$, and have introduction,
projection, coercion and homogeneous composition rules~\cite{CHTTIII}.
Here $\mathsf{coe}^{0\to r}_{i.C_i}(c):C_r$ denotes the primitive cubical
coercion of $c:C_0$ along the family $i\mapsto C_i$; $i$ is bound in
the family, and $r$ is the target dimension.
For fixed $A$, $B$, an equivalence $e:A\simeq B$ and $a:A$, the coercion
rule from the left endpoint has the form
\[
 \mathsf{coe}^{0\to r}_{i.V_i(A,B,e)}(a)
 \longmapsto
 \mathsf{vin}_r\bigl(a,\ \mathsf{coe}^{0\to r}_{i.B}(e(a))\bigr).
\]
At a variable $r$, the constructor retains the source element and its image
transported in $B$.  Its endpoint rules are
$\mathsf{vin}_0(a,b)\equiv a$ and $\mathsf{vin}_1(a,b)\equiv b$.
Consequently, at $r=1$ the displayed rule returns the remaining coercion in
$B$, and yields $e(a)$ whenever that coercion reduces to the identity.
Rules for other source endpoints and for homogeneous composition complete
the computational structure.

\paragraph{Isaev's structural computation rules.}
Isaev's manuscript~\cite{IsaevCanonicity} describes how finite horn
fillers can compute by the outer type former: pairs are filled
componentwise, functions transport their arguments backwards, and paths
expose another boundary direction.  We give the structural schemata here
and leave their completion for univalence as an open problem.  These are
\emph{not computation rules of HoTT-I}, are not implemented in Arend, and
are not proposed additions to it: practical formalization has not justified
adopting them.  They illustrate a possible more computational variant;
we do not claim canonicity for this extension.

Write $0:=\lcon$, $1:=\rcon$.  Consider ordinary interval variables
$(i,x_2,\ldots,x_n)$ and a family $A(i,\vec x)$, where
$\vec x=(x_2,\ldots,x_n)$.  The filling direction is $i$: the given
face at $i=0$ is the \emph{cap}, the missing face at $i=1$ is the
\emph{lid}, and the other $2(n-1)$ faces are the \emph{side walls}.
A horn consists of a cap $a(\vec x):A(0,\vec x)$ and terms
\[
 b_k^\varepsilon(i,\vec x_{\hat k})
 :A(i,\vec x)[x_k:=\varepsilon]
 \qquad(2\leq k\leq n,\quad\varepsilon\in\{0,1\}).
\]
The subscript $\hat k$ means that $x_k$ is omitted.  These terms must
agree judgementally at every intersection: for example,
$b_k^\varepsilon(0,\vec x_{\hat k})\equiv a(\vec x)[x_k:=\varepsilon]$,
and the two substitutions describing any intersection of side walls
must give the same term.  All these are ordinary typing and equality
judgements after endpoint substitution.  No face formulas or restricted
contexts are added.

In the square below, $x_2=j$; in the cube, $(x_2,x_3)=(j,k)$.
The blue cap and grey side walls are given; the dashed orange lid is computed.

\begin{center}
\begin{tikzpicture}[
  x=1cm,y=1cm,font=\small,
  horn cap/.style={draw=blue!70!black,line width=1.1pt},
  horn wall/.style={draw=black!60,line width=.75pt},
  horn lid/.style={draw=orange!80!black,line width=1pt,dashed},
  horn axis/.style={draw=black!55,-{Latex[length=1.5mm]},line width=.45pt},
  every node/.style={inner sep=2pt}
]
  \node[font=\small\bfseries] at (1.4,2.9) {Square horn};
  \fill[black!2] (0,0) rectangle (2.8,1.9);
  \draw[horn cap] (0,0)--(2.8,0)
    node[midway,below,text=blue!70!black] {cap $a$};
  \draw[horn wall] (0,0)--(0,1.9)
    node[midway,left] {$b_2^0(i)$};
  \draw[horn wall] (2.8,0)--(2.8,1.9)
    node[midway,right] {$b_2^1(i)$};
  \draw[horn lid] (0,1.9)--(2.8,1.9)
    node[midway,above,text=orange!80!black] {lid};
  \draw[horn axis] (-1.25,.3)--(-1.25,1.5)
    node[midway,left] {$i$};
  \draw[horn axis] (.85,-.65)--(1.95,-.65)
    node[midway,below] {$j$};

  \begin{scope}[xshift=7.1cm]
    \node[font=\small\bfseries] at (1.65,2.9) {Cube horn};
    \path[fill=black!3] (.85,.6)--(3.25,.6)--(3.25,2.4)--(.85,2.4)--cycle;
    \path[fill=black!3] (0,0)--(.85,.6)--(.85,2.4)--(0,1.8)--cycle;
    \path[horn cap,fill=blue!12] (0,0)--(2.4,0)--(3.25,.6)--(.85,.6)--cycle;
    \path[fill=black,fill opacity=.025] (0,0)--(2.4,0)--(2.4,1.8)--(0,1.8)--cycle;
    \path[fill=black,fill opacity=.045] (2.4,0)--(3.25,.6)--(3.25,2.4)--(2.4,1.8)--cycle;
    \draw[horn wall] (0,0)--(0,1.8);
    \draw[horn wall] (2.4,0)--(2.4,1.8);
    \draw[horn wall] (3.25,.6)--(3.25,2.4);
    \draw[horn wall,densely dotted] (.85,.6)--(.85,2.4);
    \draw[horn lid] (0,1.8)--(2.4,1.8)--(3.25,2.4)--(.85,2.4)--cycle;
    \node[text=blue!70!black] at (1.62,.27) {cap $a$};
    \node[text=orange!80!black] at (1.62,2.1) {lid};
    \node[text=black!70] at (1.55,1.08) {side data};
    \draw[horn axis] (-.55,.3)--(-.55,1.45)
      node[midway,left] {$i$};
    \draw[horn axis] (.65,-.65)--(1.75,-.65)
      node[midway,below] {$j$};
    \draw[horn axis] (2.65,-.35)--(3.4,.18)
      node[midway,below right] {$k$};
  \end{scope}
\end{tikzpicture}
\end{center}

Abbreviate the finite list of cap and side arguments by
$\mathcal H=(a;(b_k^\varepsilon))$.  Isaev's operation returns the lid
at arbitrary terms $\vec j=(j_2,\ldots,j_n)$ of type $\I$:
\[
 \mathsf{Fill}^n_{(i,\vec x).A}(\mathcal H)(\vec j)
 :A(1,\vec j).
\]
The notation $\mathcal H$ is only an abbreviation for this list of
arguments.  On each boundary of the lid the operation reduces to the
corresponding side wall at $i=1$:
\[
 \mathsf{Fill}^n_{(i,\vec x).A}(\mathcal H)
   (j_2,\ldots,\varepsilon,\ldots,j_n)
 \longmapsto b_k^\varepsilon(1,\vec j_{\hat k}).
\]
For $n=1$ there are no side walls or output coordinates, and
$\mathsf{Fill}^1_{i.A}(a):=\coe(\lam i\,A,a,1)$.
Although the higher operations are definable using coercion in nested
path types in base HoTT-I, here they are primitive operations with the
displayed reductions, not transparent abbreviations for those definitions.

The whole filler, including its value at an arbitrary $r:\I$, is obtained
by shortening the filling interval. As in Isaev's computational presentation,
take $\mathsf{sq}:=\mathsf{squeeze}$ as primitive with its four endpoint
equations; its earlier HoTT-I definition is not an unfolding rule here. Define
\begin{equation}\label{eq:tau-filler}
 \mathsf{F}^n_A(\mathcal H)(r,\vec j):=
 \mathsf{Fill}^n_{(s,\vec x).A(\mathsf{sq}(s,r),\vec x)}
   (\mathcal H[i:=\mathsf{sq}(s,r)])(\vec j).
\end{equation}
The substitution changes the side arguments; the cap is unchanged.
We retain \emph{regularity}: transport in a family independent of the
transport variable reduces to its input.  Its horn version returns
$a(\vec j)$ when both $A$ and every side wall are independent of $i$.
A constant ambient type alone is insufficient when the side walls move.
Together with the four endpoint equations of $\mathsf{sq}$, these rules give
\[
 \mathsf{F}^n_A(\mathcal H)(0,\vec j)\equiv a(\vec j),\qquad
 \mathsf{F}^n_A(\mathcal H)(1,\vec j)
   \equiv\mathsf{Fill}^n_{(i,\vec x).A}(\mathcal H)(\vec j),
\]
and the prescribed side values at every $r$.

\paragraph{Pairs, paths and natural numbers.}
In the following structural clauses, the cap and side arguments are
pairs, path abstractions, or lambda abstractions, respectively.
For a horn of pairs in $\Sigma(y:A(i,\vec x))B(i,\vec x,y)$,
let $\mathcal H.1$ and $\mathcal H.2$ denote the componentwise
projections of all its arguments.  First fill the first components over
the entire box, obtaining
$a^*(i,\vec x):=\mathsf{F}^n_A(\mathcal H.1)(i,\vec x)$.
Then fill the second components in the family determined by $a^*$:
\begin{equation}\label{eq:tau-sigma}
 \mathsf{Fill}^n_{(i,\vec x).\Sigma(y:A)B}(\mathcal H)(\vec j)
 \longmapsto
 \bigl(a^*(1,\vec j),\,
   \mathsf{Fill}^n_{(i,\vec x).B(i,\vec x,a^*(i,\vec x))}
     (\mathcal H.2)(\vec j)\bigr).
\end{equation}
The boundary equations of $a^*$ ensure that each second component has
exactly the type required by the recursive filling.

For paths, applying every cap and side argument to a fresh coordinate
$k$ gives a horn in the underlying family.  The two endpoint sections
$l$ and $r$ become its two additional side walls:
\begin{equation}\label{eq:tau-path}
 \mathsf{Fill}^n_{(i,\vec x).\Path(\lam k\,D(i,\vec x,k),l,r)}
   (\mathcal H)(\vec j)
 \longmapsto
 \pcon\bigl(\lam k\,
   \mathsf{Fill}^{n+1}_{(i,\vec x,k).D}
      (\mathcal H\ @\ k;\,l,r)(\vec j,k)\bigr).
\end{equation}
Thus the one-dimensional instance makes the square implicit in path
transport explicit:
\[
 \coe\bigl(\lam i\,\Path(\lam k\,D(i,k),l_i,r_i),p,1\bigr)
 \longmapsto
 \pcon\bigl(\lam k\,
   \mathsf{Fill}^2_{(i,k).D}(p\ @\ k;\,l_i,r_i)(k)\bigr).
\]
Iteration handles any finite nest of paths, retaining all its boundaries.
For natural numbers the constructor clauses are
\begin{equation}\label{eq:tau-natural}
 \mathsf{Fill}^n_{(i,\vec x).\Nat}(0)(\vec j)\longmapsto0,
 \qquad
 \mathsf{Fill}^n_{(i,\vec x).\Nat}(\suc\mathcal H)(\vec j)
 \longmapsto\suc\bigl(
   \mathsf{Fill}^n_{(i,\vec x).\Nat}(\mathcal H)(\vec j)\bigr).
\end{equation}
Here $0$ means that the cap and every side are zero, and
$\suc\mathcal H$ means that every argument is a successor.

\paragraph{Functions.}
For a horn of functions in $\Pi(y:A(i,\vec x))B(i,\vec x,y)$,
fix an output point $(1,\vec j)$ and an argument $y:A(1,\vec j)$.
The argument must be transported from this point to every point of the
box before the cap and side functions can be applied.
For a family $k\mapsto D_k$ and endpoints $s,t:\I$, write
$\coe_{2,k.D_k}(s,a,t):D_t$ for transport of $a:D_s$ from $s$ to $t$;
the notation $k.D_k$ binds the family coordinate $k$.
Contractibility of $\I$ supplies chosen paths
$\gamma_{s,t}:\Path(\lam u\,\I,s,t)$, so this operation can be defined by
\[
 \coe_{2,k.D_k}(s,a,t):=
 \coe(\lam u\,D_{\gamma_{s,t}\ @\ u},a,1):D_t
 \qquad(a:D_s).
\]
Isaev writes $\coe^n_A(\mathbf s,y;\mathbf t)$ for the iteration of
this operation along the coordinates in order:
\[
\begin{aligned}
 \coe^0_A(a)&:=a,\\
 \coe^{m+1}_A((\mathbf s,s),a;(\mathbf t,t))
 &:=\coe_{2,k.A(\mathbf t,k)}
   \bigl(s,\coe^m_{A(-,s)}(\mathbf s,a;\mathbf t),t\bigr).
\end{aligned}
\]
Put $v(i,\vec x):=\coe^n_A((1,\vec j),y;(i,\vec x))$.
Write $\mathcal H[v]$ for the horn obtained by applying each given
function to the corresponding restriction of $v$.  The function clause is
\begin{equation}\label{eq:tau-pi}
 \mathsf{Fill}^n_{(i,\vec x).\Pi(y:A)B}(\mathcal H)(\vec j)
 \longmapsto
 \lam{y:A(1,\vec j)}\,
 \mathsf{Fill}^n_{(i,\vec x).B(i,\vec x,v(i,\vec x))}
   (\mathcal H[v])(\vec j).
\end{equation}
It transports the argument backwards, applies the given functions,
and fills their values forwards, including when the domain depends on
the transverse coordinates $\vec x$.
For the result to have the required type $B(1,\vec j,y)$ judgementally,
this presentation requires
\[
 \coe_{2,k.D_k}(s,a,s)\equiv a,
 \qquad\text{hence}\qquad v(1,\vec j)\equiv y.
\]
The path-based definition above alone does not establish this diagonal
computation. Thus the function clause is conditional on an implementation
of $\coe_2$ with this equation and compatible reductions.

\paragraph{Open problem: computation for $\iso$ and the universe.}
The structural clauses do not supply a general rule for
\[
 \mathsf{Fill}^n_{(i,\vec x).
   \iso(A,B,f,g,p,q,k)}(\mathcal H)(\vec j),
\]
where all the displayed data, including $k:\I$, may depend on
$(i,\vec x)$.  The existing $\iso$ rule covers only transport along the
canonical line with fixed isomorphism data.  A general rule must also
handle varying data, arbitrary interval arguments, and every supplied
higher boundary.

The universe requires attention as well.  In dimension one regularity gives
$\mathsf{Fill}^1_{i.\U}(A)\longmapsto A$, but a higher universe horn may
have moving type-valued sides.  One possibility is to retain its lid as a
new type expression, instead of reducing it to an existing type former.
For example, let $\mathcal H_s$ be compatible square horns in $\U$,
varying with an ordinary interval variable $s:\I$.  For $j:\I$ we obtain
\[
\begin{aligned}
 T(s,j)&:=\mathsf{Fill}^2_{(i,x).\U}(\mathcal H_s)(j):\U,\\
 \coe(\lam s\,T(s,j),a,1)&:T(1,j)\qquad(a:T(0,j)).
\end{aligned}
\]
The first line constructs a type; the second transports an element of
that type.  Retaining the first expression does not yet give a reduction
for the second.  The clauses for pairs, functions and paths do not apply,
since the type family now has $\mathsf{Fill}$ as its outermost operation.
A computation rule for filling in these new types is therefore needed,
already for ordinary transport.

There is no fixed bound on the nesting that this rule must handle.
A later universe horn may use $T(s,j)$ among its cap and side data.
Its lid is then another type expression containing the earlier universe
filler, and this construction can be repeated.  Filling in the resulting
types must compute at every finite nesting depth, with reductions that
agree with the boundary equations and regularity.  Isaev's manuscript
explicitly leaves this nested case unfinished.

\emph{Open problem.}  Complete the structural schemata with general
computational rules for $\iso$ and universe fillers, in ordinary HoTT-I
contexts and without a separate language of dimensions or face conditions.
Include the diagonal computation needed by the function clause and check
its compatibility with the other reductions. The rules should respect all prescribed boundaries, retain regularity,
and be stable under substitution of arbitrary terms of type $\I$.
Establish subject reduction, compatibility of overlapping rules,
and canonicity for the resulting system; normalization and decidability
of judgemental equality are further metatheoretic questions.
The structural schemata alone do not establish these properties.
Higher inductive types
would also require their own compatible computation rules.

\section{Arend language}\label{sec:arend-lang}

In this section we present the most essential language constructs of Arend.

\subsection{Levels and universe polymorphism}

Since there is no way to assume consistently the existence of the set $\Set$ of all sets any construction that aims at producing 
a set out of the totality of all sets has to be applied to a universe of small sets $\Set\ 0$ producing a set in a universe of
large sets $\Set\ 1$. Then such constructions can typically be iterated and produce a set in the universe $\Set\ i+1$ of level $i+1$ out
of the universe $\Set\ i$ of level $i$. And typically this routine is straightforward and can be automated.

Arend language allows 
to use $\Set$ and $\Type$ in practice as if they were types of
types and delegate the routine of dealing with levels to the typechecker. Typechecker replaces 
$\Set$ and $\Type$ with hierarchies of
universes, converts definitions accordingly and checks if definitions are not contradictory because of a circularity in
universe hierarchy. This is achieved by the following mechanisms:

\begin{enumerate}
\item \textbf{Universe polymorphism.} Definitions are made polymorphic on universe levels: there are implicit level parameters
associated with each definition.
\item \textbf{Level inference.} In each occurrence of $\Set$ or $\Type$ typechecker infers level according to some natural procedure.  
\end{enumerate}

Universe management in Arend is rather similar to the one in Lean and recent versions of Rocq.
It is quite different in Agda: Agda has considerably less automation, one has to be 
much more explicit about levels. Also, unlike level expressions in Agda where they are 
terms of type \verb|Level|, level expressions in Arend are not typed.

\subsubsection{Homotopy levels}\label{sec:hom_levels_lang} 

The universe \verb|\Prop| and the hierarchy \verb|\Set n| are two members of a family of
universes indexed additionally by the homotopy level \verb|k| discussed in
Section~\ref{sec:ExtTT}. Since the homotopy level is always a constant, each member of the
family has its own notation: \verb|\Prop| is the universe of $(-1)$-types, \verb|\Set n|
that of $0$-types, \verb|\k-Type n| --- with \verb|k| a literal numeral --- that of
$k$-types for $k \geq 1$, and \verb|\Type n| that of all types. Thus \verb|\Set n| and
\verb|\0-Type n| are the same universe. The library's Eilenberg--MacLane
type \verb|K1|, mentioned in Section~\ref{sec:lib-hott}, is a \verb|\1-Type|. 

\subsubsection{\texorpdfstring{Definitions in $\Prop$, $\Set\, n$ and $k$-type universes}{Definitions in Prop, Set n and k-type universes}}\label{sec:use_level_lang}

By default a definition is inferred to be of some homotopy level according to
certain rules and is placed in the corresponding universe.
However one often needs to use a definition which is not inferred by the typechecker 
to be in \verb|\Prop|, \verb|\Set| or \verb|\k-Type| to define a proposition, a set
or a $k$-type in the corresponding universe. There are two ways to do this in Arend: 

\begin{enumerate}
\item \textbf{Use level}. Assume \verb|D| is a definition of a record, class or inductive type 
is provably a $k$-type which means that the predicate \verb|isOfHLevel D k| is provable
(in $\arendlib$ this predicate is spelled \verb|D ofHLevel_-1+ k|, with a variant
\verb|ofHLevel_-2+| for the other indexing convention, and the two special cases
\verb|isProp| and \verb|isSet|). In
that case
\verb|D| can be turned into a definition in \verb|\k-Type n| by means of extending it with
a proof of \verb|isOfHLevel D k| which is done in Arend via \verb|\use \level| construct. 
In particular, extending \verb|D| with a proof of \verb|isProp D| or \verb|isSet D| would
place \verb|D| in \verb|\Prop| or \verb|\Set| respectively. For example, the predicate
\verb|nonempty A| --- which in $\arendlib$ is the propositional truncation \verb|TruncP A|
with constructors \verb|inP| and \verb|truncP| --- can be defined in \verb|\Prop| as
follows:

\begin{lstlisting}
\data nonempty (A : \Type)
  | in A
  | trunc (a a' : nonempty A) : a = a' -- just another syntax for the constructor
                                       -- trunc (a a' : nonempty A) (i : I) with
                                       -- conditions trunc a a' left => a
                                       --            trunc a a' right => a'
  \where {
    \use \level levelProp {A : \Type} (a a' : nonempty A) : a = a' => path (trunc a a')
  }
\end{lstlisting}

As mentioned in Section~\ref{sec:TypeTheory} elimination to h-propositions from inductive
types in \verb|\Prop| does not involve constructors with parameters of the type \verb|I|.
This means that such constructors can be omitted in definitions of
functions by pattern matching in this case. Eliminator corresponding to the recursion 
principle for \verb|nonempty| can be defined using pattern matching as follows:

\begin{lstlisting}
\lemma rec {A B : \Type} (p : isProp B) (t : nonempty A) (f : A -> B) : B \level p \elim t
      | in a => f a
\end{lstlisting}

This simplified pattern matching allows to derive proof irrelevance for types in \verb|\Prop|
using \verb|nonempty|:

\begin{lstlisting}
\lemma prop-pi {A : \Prop} {a a' : A} : a = a'
  => path \case trunc (in a) (in a') __ \with { -- \case f __ is unfolded to
    | in x => x                                 -- \lam i => \case f i
  }
\end{lstlisting}

\item \textbf{Truncated inductive types}. Any definition of an inductive type can be placed
in a universe of any given homotopy level by marking the definition as \emph{truncated}. This
marking is done by means of the keyword \verb|\truncated| in front of a definition. 
It places the definition in a universe \verb|\k-Type n|, but at the cost
of restricting elimination to types \verb|A| satisfying \verb|isOfHLevel A k|. In other
words, pattern matching cannot be used to define functions from truncated inductive types
in \verb|\k-Type n| to a type \verb|A| unless there is a proof of \verb|isOfHLevel A k|.
For example, the quotient set can be defined as follows:

\begin{lstlisting}
\truncated \data Quotient {A : \Type} (R : A -> A -> \Type) : \Set
  | in~ A
  | ~-equiv (x y : A) (R x y) : in~ x = in~ y -- again, this is unfolded to 
  										      -- ~-equiv (x y : A) (R x y) (i : I) with conditions
\end{lstlisting}

The recursion principle for \verb|Quotient| would be:

\begin{lstlisting}
\sfunc rec {A B : \Type} (R : A -> A -> \Type) (p : isSet B)
           (f : A -> B) (f-R : \Pi (x y : A) (R x y) -> f x = f y) (x : Quotient R) : B \level p
   \elim x
      | in~ a => f a
      | ~-equiv a a' r i => f-R a a' r @ i 
\end{lstlisting}

\end{enumerate} 

\subsection{Typeclasses and records}\label{sec:classesLang}

Here we highlight some aspects of the system of typeclasses and records in Arend. This system
is built upon the system of records in the core type theory discussed in
Section~\ref{sec:TypeTheory}.

\begin{enumerate}

\item \textbf{Partial implementations.} 
As described in Section~\ref{sec:TypeTheory} the mechanism of manifest fields allows to
partially implement parent records or classes\footnote{The difference between classes and 
records is not relevant at this point. We discuss specifics of classes in more detail below}
inside a child record or class. For example, as mentioned in
Section~\ref{sec:hierachies} the type of rings can be defined in Arend as an extension of the type of
semirings using partial implementation:

\begin{lstlisting}
\class PseudoSemiring \extends AbMonoid, Semigroup {
  | ldistr {x y z : E} : x * (y + z) = x * y + x * z
  | rdistr {x y z : E} : (x + y) * z = x * z + y * z
  | zro_*-left {x : E} : zro * x = zro
  | zro_*-right {x : E} : x * zro = zro
}

\class PseudoRing \extends PseudoSemiring, AbGroup {
  | zro_*-left {x} => {?} -- proof of 0 * x = 0 using invertibility wrt '+'
  | zro_*-right {x} => {?} -- proof of x * 0 = 0 using invertibility wrt '+'
}

-- Adding a multiplicative unit on top of the two definitions above gives
\class Semiring \extends PseudoSemiring, Monoid { -- ... }
\class Ring \extends PseudoRing, Semiring
\end{lstlisting}

Here \verb|PseudoSemiring| extends \verb|AbMonoid| and \verb|Semigroup| which are additive commutative
monoidal and multiplicative semigroup structures on the same
carrier \verb|E|\footnote{In arend-lib there are two hierarchies for groups and monoids: one for
$+$ and one for $*$. }. Since \verb|AbGroup| extends 
\verb|AbMonoid| with the group structure, the type \verb|PseudoRing| is defined as a pseudo-semiring
where \verb|+| operation is strengthened to a group operation. 

Another example: tight apartness relation \verb|x # y|. It is constructively better behaved variant
of inequality \verb|Not (x = y)| since it satisfies the tightness condition
\verb|Not (x # y) -> x = y| (whereas \verb|Not (x = y)| constructively does not). Such 
relations are useful, for example, for defining domains constructively as rings satisfying 
\verb|x # 0 -> y # 0 -> (x * y) # 0| (the field \verb|#0-*| of the class \verb|Domain|).
It is natural to define the base class \verb|Set#| of sets with 
apartness relation and various types of structures with apartness as its extensions. However,
for example, in case of groups specifying the full binary relation \verb|x # y| is redundant since
it can be expressed in terms of \verb|x # 0|. This can be done using partial implementations:

\begin{lstlisting}
-- groups with '+' operation (not necessarily commutative) and apartness relation '#'.
-- The class below lives in the \where block of 'AddGroup' and is referred to
-- from the outside as 'AddGroup.With#'
\class With# \extends AddGroup, Set# 
    | \fix 8 #0 : E -> \Prop
    | #0-zro : Not (zro `#0)
    | #0-negative {x : E} : x `#0 -> negative x `#0
    | #0-+ {x y : E} : (x + y) `#0 -> x `#0 || y `#0
    | #0-tight {x : E} : Not (x `#0) -> x = zro

    | # x y => (x - y) `#0
    -- we omit implementations of properties of x # y
\end{lstlisting}

Finally, another example from Section~\ref{sec:hierachies} - the type of Heyting fields which is by definition a
commutative local ring where all noninvertible elements are equal to zero. This example illustrates 
the usefulness
of partial implementations for defining structures which extend several branches of hierarchies
in such a way that some data and properties of one branch is implemented in terms of 
data and properties of other branches.

\begin{lstlisting}
\class Field \extends LocalCRing, GCDDomain
  | #0 x => Inv x
  | #0-+ => LocalRing.sumInv=>eitherInv -- property of local rings: if x + y is invertible 
                                        -- then either x or y is invertible.
  -- we omit implementations of isGCDDomain and other properties of apartness '#'
   
  -- the tightness '#0-tight' remains unimplemented and represents one of defining
  -- properties of the Heyting field: all noninvertible elements are equal to zero
\end{lstlisting}

\item \textbf{Anonymous extensions.} Extensions of records and classes can be defined on the
fly: if \verb|C| is a record containing a field 
\verb|f : A| then the term \verb|C { f => a }| refers to a record extension obtained from
\verb|C| by specifying \verb|f| to a term \verb|a : A|.

For example, one can use the class of all semirings to define the type of all semiring
structures on natural numbers:

\begin{lstlisting}
\func SemiringsOnNat : \Set => Semiring { E => Nat } -- or simply '=> Semiring Nat'
\end{lstlisting}

Here \verb|SemiringsOnNat| is a subtype of \verb|Semiring|.

\item \textbf{Properties.} As mentioned in Section~\ref{sec:TypeTheory} a form of 
computational proof
irrelevance holds in Arend for records and classes. Arend makes an explicit distinction
between property fields and data fields. The result of comparison of two 
objects depends only on comparison of data fields, proofs of properties are ignored. 
By default a field
is a property if its type is a proposition which in particular is the case if the type
is equality between elements of a set.

For example, the class \verb|PseudoSemiring| has four property fields:
\verb|ldistr|, \verb|rdistr|, \verb|zro_*-left| and \verb|zro_*-right|. Proofs of
these are ignored when two pseudo-semirings are compared.

\item \textbf{Static and dynamic blocks.} All fields of a record or class \verb|D|
as well as definitions of inductive types and functions declared in the inner block
of \verb|D| have an implicit parameter \verb|\this : D| which represents the instance
\footnote{Expressions $a.f$ for taking value of a field or invoking a definition
$f$ of an instance $a$ are converted to applications $f\, \{a\}$}.
Fields and definitions in this block are thus \emph{dynamic} analogously to dynamic
class methods in Java or C++. Although dynamic definitions can be replaced with external static 
definitions with an extra parameter of type \verb|D|, dynamic definitions of
basic operations and properties of \verb|D| are more concise and neat. For example,
part of dynamic block of \verb|Monoid| might look as follows:

\begin{lstlisting}
-- '\infixl 7 *' and '*-assoc' are inherited from 'Semigroup'
\class Monoid \extends Pointed, Semigroup {
  | ide-left {x : E} : ide * x = x
  | ide-right {x : E} : x * ide = x

  \func pow (a : E) (n : Nat) : E \elim n
    | 0 => ide
    | suc n => pow a n * a

  \lemma pow_+ {a : E} {n m : Nat} : pow a (n + m) = pow a n * pow a m \elim m
    | 0 => inv ide-right
    | suc m => pmap (* a) pow_+ *> *-assoc
}
\end{lstlisting}

Just like any other definition in Arend classes and records have \verb|\where| block for
associated \emph{static} definitions. These static blocks are quite useful, for example,
for defining some standard structures on top of a mathematical structure like having an
apartness relation or decidable equality. For example, for rings one use the static block
as follows:

\begin{lstlisting}
\class Ring \extends Semiring, AbGroup {
 -- ...
} \where {
-- | A ring with a tight apartness relation.
  \class With# \extends Ring, AddGroup.With# {
 		-- ... 
  }
  
  -- | A ring with decidable equality.
  \class Dec \extends AddGroup.Dec, With# {
		-- ...  
  }
}
\end{lstlisting}

\item \textbf{Type classes.} The concept of a type class originates in Haskell programming
language. The mechanism of type classes enables ad-hoc polymorphism, a form of polymorphism
similar to operator overloading in object-oriented programming: type classes specify
bundles of operations on types which can be implemented differently for different types.

In Arend type classes are definitions which start with the keyword \verb|\class|
and are similar to records. The key difference is the \emph{instance inference} mechanism
for type classes: some of objects of a type class can be declared as instances which
would enable the typechecker to infer a suitable instances whenever necessary.
For example, a semiring structure on natural numbers can be defined as an instance of
class \verb|Semiring|:

\begin{lstlisting}
\instance NatSemiring : Semiring Nat
  | zro => 0
  | + => +
  | * => *
  | ide => 1
  -- we omit proofs of semiring properties for Nat
\end{lstlisting}

At a place where multiplication of natural numbers is used typechecker can infer 
the instance \verb|NatSemiring| and make applicable
at this place everything that is applicable to semirings, for example,
a simplification tactic erasing subexpressions multiplied by $0$. 

\begin{lstlisting}
\func example (n : Nat) : n * 0 + 1 = 1
  => simplify -- the tactic 'simplify' infers NatSemiring and uses proofs
              -- of properties of +, * to derive the equality
\end{lstlisting}

With respect to the behavior of instances and instance inference Arend
is very similar to other theorem provers like Rocq, Lean or Agda.

\end{enumerate}

\subsection{Arrays}

The type $\mathsf{Array}$ is defined in $\mathsf{Prelude.ard}$ as a record:
\begin{lstlisting}
\record DArray {len : Nat} (A : Fin len -> \Type) 
              (\coerce at : \Pi (j : Fin len) -> A j)
\func Array (A : \Type) => DArray { | A _ => A }
\end{lstlisting}

There are also functions \verb|nil| and \verb|a :: l| which can be used in pattern matching
as if \verb|DArray| were an inductive type. Coercion \verb|at| allows to use arrays as
if they were functions. A function \verb|G|, as mentioned in Section~\ref{sec:arrays},
can be defined by recursion on terms as follows:

\begin{lstlisting}
\data Term (F : \Set) (a : F -> Nat) | fun (f : F) (v : Array (Term F a) (a f))
\func G {F : \Set} {a : F -> Nat} (t : Term F a) \elim t
  | fun f v => fun f (\lam i => G (v i)) 
\end{lstlisting}

\section{IDE for Arend: IntelliJ IDEA plugin}\label{sec:ide}

\subsection{Remarks on implementation}

\textbf{Scoping and dependency analysis.} Arend files can depend on other Arend files, and Arend also allows circular dependencies within a single Arend library (aka Intellij IDEA module of type “Arend”). Furthermore, the scope of any Arend definition or theorem is not restricted to the text that follows it, but rather extends throughout the entire enclosing scope. For instance, the scope of a top-level definition spans the entire Arend file to which it belongs. This behavior is similar to the scoping mechanisms found in many general-purpose programming languages and contrasts with typical proof assistants like Rocq or Mizar. On the positive side, this approach simplifies import management and code migration for the user. On the negative side, it creates a need for Arend’s type checker to track dependencies between individual definitions across different files and to handle potential circular dependencies between them, as these can result in inconsistent theory. In Arend, circular dependencies are only allowed between recursively defined functions, and even then, only if the termination checker can detect that parameters of recursive function calls become “smaller” from one call to another call. The termination checker algorithm largely follows the one outlined in A. Abel’s master’s thesis FOETUS.

\textbf{Incremental type checking.} The dependency analysis mechanism enables another key feature in Arend: incremental type checking. When a user modifies one or more definitions, the type checker selectively discards cached information related to those definitions, as well as any other definitions that depend on them, either directly or through a chain of dependencies. This ensures that unaffected parts of the code are not reprocessed, significantly speeding up the type-checking process and helping to conserve laptop battery life.

\textbf{Smart mode and Dumb mode.} Under normal circumstances the Arend plugin uses the so-called "on-the-fly" (or \textbf{Smart}) mode of type checking. In this mode, the type checker operates in a background thread, automatically reprocessing each code modification made by the user. Running the type checker in the background prevents the IDE from freezing during the process. Since the type checking can be time-consuming and may involve extensive computations, a special mechanism halts the process if it exceeds the user-defined time limit (defaulted to 5 seconds). Additionally, there is an option to disable on-the-fly inspection entirely, switching the IDE into \textbf{"Dumb mode"}. In this mode, the IDE still resolves identifiers on the fly but does not automatically perform type checking which can still be manually run via IntelliJ IDEA’s standard mechanism of run configurations.

\subsection{Editor features of Arend plugin}

\textbf{Parameter Hints.} Parameter Hints feature displays a tooltip showing the function or datatype's signature used in expression, including expected parameters (both explicit and implicit) along with their types. The tooltip also highlights the parameter corresponding to the argument where the caret is positioned. Moving the caret to a different argument updates the highlighted parameter accordingly. When the caret is moved away from the call expression, the tooltip disappears. 
For example, when invoked on the expression \verb|IsContAt {TopSub S} f x| on the
argument \verb|f| the tooltip shows the signature of \verb|IsContAt| (is continuous at a
point):
\begin{center}
\verb|{X : TopSpace}, {Y : TopSpace}, f : X -> Y, x : X|
\end{center}
with \verb|f : X -> Y| highlighted.

This feature requires the target definition to be fully type-checked, as definitions in Arend may include dynamically inferred implicit parameters that are not explicitly listed in their signatures. However, a fallback mechanism is available: it tries to match the arguments of the selected expression to the parameters of the definition, even when the definition has not been type-checked (e.g., when the typechecker is in Dumb mode).

\textbf{Auto-completion.} Auto-completion is a feature that helps identify available identifiers in the current scope and to complete partially entered identifiers. Also, it imports definitions from scopes that have not yet been imported. The current scope is determined by the set of active imports, along with surrounding local and global declarations and variables.

\textbf{Intention Actions and Quick Fixes.} Quick Fixes and Intention Actions are standard features in IntelliJ IDEA that provide suggestions for modifying code. The main distinction between Quick Fixes and Intention Actions lies in their purpose. Quick Fixes are specifically aimed at resolving errors detected by the Arend type checker, while Intention Actions offer general improvements or suggestions not directly linked to specific errors.

\begin{enumerate}
\item \textbf{Replace a meta with result.} Replaces a meta with the proof term it generates.
\item \textbf{Replace an expression with weak head normal form.} Replaces a selected expression with its partially normalized version. Recall that an expression is said to be in WHNF if it is reduced to the point where it cannot be further simplified at the outermost level, but may still contain unevaluated or partially evaluated expressions within it. WHNF is a representation of terms used internally by the Arend type checker. It is used mainly for efficiency, as fully normalizing (reducing to a normal form) can be computationally expensive.
\item \textbf{Extract code as a function.} Moves a selected part of an expression into a separate function within the same namespace, replacing the original part with a function call. This action also generates the necessary function arguments.
\item \textbf{Generate function from goal.} This action can be invoked on an unfilled goal, replacing it with a call to an auxiliary function in the namespace of the current definition or theorem. The function's type will match the goal's type, and its arguments will include the minimal subset of context variables required for the goal.
\item \textbf{Replace with constructor.} This action fills an unfilled goal with the appropriate constructor. For example, a goal with a \verb|\Pi| type will be filled with a lambda expression, a sigma type with a tuple, and an inductive type with one of its constructors (with a prompt if multiple constructors exist), etc.
\item \textbf{Implement fields quick fix.} When applied to an empty \verb|\instance| statement with a specified type class, this quick fix generates a skeleton for implementing the class. It lists all fields inherited from ancestor classes that are not yet implemented or lack a default implementation.
\item \textbf{Implement clauses quick fix and Split pattern variable intention.} If an Arend definition or pattern-matching construct is missing some or all of its clauses, the type checker suggests adding them using this quick fix. It works for both \verb|\case| expressions and pattern-matching constructs using the \verb|\elim| or \verb|\with| keywords.

The Split Pattern action can be applied to pattern variables used in pattern-matching clauses. This action performs the following 3 steps:
\begin{enumerate}
\item Type Analysis: It determines the type of the pattern variable and identifies its associated constructors.
\item Clause Duplication: It creates as many copies of the current pattern-matching clause as there are constructors.
\item Constructor Substitution: It replaces every instance of the original pattern variable with a call to the respective constructor in each of the newly created clauses.
\end{enumerate}
\item \textbf{Auto import quick fix.} If the user clicks on an unresolved identifier, this quick fix searches for definitions with this name among the imported Arend libraries. If multiple matches are found, the user can choose one. Once selected, the quick fix updates the identifier to a valid reference and adds the necessary \verb|\import| directives.
\end{enumerate}

Additionally, several other quick fixes perform simpler tasks, such as removing unnecessary elements or making basic corrections. These include adding a missing argument or class instance to a function call if the type checker can infer it, or fixing import directives in case of errors in the Arend file preamble.

\subsection{Navigating Arend projects}

This section introduces essential tools for efficiently navigating Arend projects in IntelliJ IDEA.

\textbf{Go to declaration.} The user can jump to the location where an identifier is defined.

\textbf{Find usages.} This command locates all occurrences of the identifier within the specified scope, organizes them by the type of usage and the containing file, and also displays a brief snippet of the surrounding code.

\textbf{Go to Class/File/Symbol.} The user can navigate to a definition or a file with
a given name.

\textbf{Class Hierarchy.} The Arend plugin supports IntelliJ IDEA’s Class Hierarchy feature, enabling the user to visualize and navigate the inheritance structure of an Arend class or record, showing its parent and child classes. There is a dedicated tool window displaying the hierarchy tree. The user can toggle between views of superclasses and subclasses.
The hierarchy panel is interactive, allowing the user to navigate directly to any class by clicking on it. Clicking the Visualize as Orthogonal Diagram button opens a separate window with an orthogonal diagram of the class hierarchy. The contents of the diagram depend on whether the subclasses or superclasses view is active.

\textbf{Proof search.} Nearly every proof assistant struggles with the problem of efficient location of relevant theorems or lemmas in their formal libraries. Consider, for instance, the problem of recalling the name of a lemma that encodes the property that addition is commutative. Is it labeled as \verb|add_comm|, or \verb|+-comm| in the library?

The Arend plugin helps address this issue by allowing the user to search for functions based on the definitions they reference. This functionality, called Proof Search, is comparable to the Search vernacular command in Rocq, Type-Directed Search in Idris, or Loogle in Lean.

The user needs to type a query pattern in the input field. Proof Search then lists all definitions whose parameter or result types contain a subexpression matching the specified pattern. In the simplest case this pattern consists of a single reference to an Arend definition and the Proof Search lists all definitions mentioning this identifier in their parameters or result type. 

For example, the user can enter a pattern resembling a partial application of a function or a type. Such patterns are allowed to be written in infix form.  The placeholder \verb|_| can be used as a wildcard pattern to match any arbitrary term. For instance, the query 
\verb|List Nat| will return only those definitions that explicitly include the subexpression \verb|List Nat| within the types of their parameters (or within their result type). On the other hand, the following definitions can be found using the search pattern \verb|_ + _ = _ + _|:

\begin{lstlisting}
\func foo (a b : Nat) : a + b = b + a
\func bar (a b : Int) : a + b = b + a
\end{lstlisting}

\subsection{Refactoring Arend projects}

This section introduces refactoring tools for Arend in IntelliJ IDEA.

\textbf{Code formatting.} Arend file can be automatically formatted according to Arend's text
formatting conventions.

\textbf{Optimize Imports.} Similar to other languages supported by IntelliJ IDEA, the Arend plugin provides a tool to analyze import headers in each file for potentially superfluous import directives. These unnecessary directives are highlighted in gray and can be removed using the corresponding intention action.

\textbf{Rename refactoring.} The Rename feature in IntelliJ IDEA allows the user to change the name of variables, classes, functions, data types, or other identifiers across the Arend project.

\textbf{Change Signature refactoring.} The Arend plugin supports the Change Signature refactoring feature of IntelliJ IDEA, providing a tool for modifying function or theorem signatures while preserving the correctness of call expressions and types throughout the entire Arend codebase.

This feature applies signature changes across the codebase, ensuring that type dependencies — such as variables referenced in the types of other parameters — are accurately adjusted. For example, renaming a parameter used in a type declaration automatically updates all type definitions that reference it.

\textbf{Move refactoring.} The Arend plugin implements IntelliJ IDEA’s Move Refactoring feature, enabling the user to reorganize code by moving classes, functions, data types, or fields to a new location within the same file or to a different Arend file. This operation automatically updates all references and corrects file imports.

\section{Using Claude Code with Arend CLI}\label{sec:claude-code}

The IntelliJ IDEA plugin described in the previous section is not the only way to interact with the Arend type checker. Arend also ships with a command-line interface (CLI), which opens up the possibility of using the prover outside the IDE. One use case that has recently become practical is LLM-assisted formalization: coupling the type checker with an AI coding agent such as Claude Code developed by Anthropic. An agent of this kind works in an edit--check loop: it modifies source files, invokes external tools, reads their textual output and iterates. A command-line type checker with textual feedback is therefore a natural fit for such an agent, and the absence of a graphical environment is no obstacle.

\textbf{Arend CLI.} In its default mode of operation the CLI resolves and type checks a library, a module, or a single definition in a module, and prints the list of errors and remaining goals:
\begin{lstlisting}
arend [<Library>] [<Module>[:<def>]]
\end{lstlisting}

In Arend 1.12, \verb|--serialize| persists successfully typechecked modules
as binary cache files (\verb|.arc|); ordinary one-shot invocations do not
write these files by default. Subsequent invocations can reuse valid module
caches. The flag \verb|-r| (\verb|--recompile|) forces recompilation from
source, ignoring the caches. Besides type checking, the CLI provides
retrieval tools that query the loaded library:
\begin{itemize}
\item \verb|-ss <pattern>| (symbol search) finds definitions by name pattern and prints their locations and one-line signatures;
\item \verb|-ps <pattern>| (proof search) finds lemmas by the shape of their signature, similarly to the Proof Search feature of the IDE plugin, e.g. \verb|-ps "_ + _ = _ + _"|;
\item \verb|-fu <name>| (find usages) lists the usage sites of a definition;
\item \verb|-ch <class>| (class hierarchy) prints the superclasses and subclasses of a class together with its \verb|\new|- and \verb|\instance|-sites;
\item \verb|-sc <name>| (scope info) lists everything visible in scope at the position of a given definition.
\end{itemize}
Symbol search uses an on-disk index cached by modification time. All five
retrieval commands accept \verb|--json| for machine-readable output. They
provide an agent with names, signatures, locations and usages without
requiring it to scan the entire library.

The flag \verb|-p Module:definition| prints an elaborated definition after
type checking, while \verb|-c| additionally checks the resulting core terms.
For repeated interactions, \verb|-d| starts a daemon that retains a
typechecked library in memory and persists its module caches. Subsequent
commands normally use this daemon; \verb|--no-daemon| forces an independent
invocation. Thus an agent can combine retrieval with ordinary source edits
and typechecking feedback in an edit--check loop.

\textbf{Skills.} Claude Code can be supplied with reusable domain knowledge in the form of \emph{skills}: self-contained instruction files whose frontmatter describes the situations in which the agent should load them. A collection of such skills for Arend is available in the repository \verb|arend-skills|\footnote{\url{https://github.com/sxhya/arend-skills}}, which can be installed by symlinking or copying the skill directories into the agent's skills directory. The collection consists of the following skills:
\begin{itemize}
\item \textbf{arend-formalize} --- a procedural workflow for turning an informal mathematical statement into a typechecking Arend definition or lemma: how to orient oneself in $\arendlib$ using the retrieval flags above, how to choose the classes and lemmas to reuse, and the order of operations up to the point where the only remaining errors are the unproven goals \verb|{?}|;
\item \textbf{arend-prove} --- a workflow for filling in the \verb|{?}| placeholders once the statement type checks, together with a catalog of metas (\verb|rewrite|, \verb|cases|, \verb|mcases|, \verb|simplify|, \verb|equation|, \verb|cong|, \verb|ext|, \verb|linarith|, \verb|unfold|, etc.), their semantics, decision tables and common failure modes;
\item \textbf{arend-quirks} --- a collection of notes on the ways in which the syntax and semantics of Arend differ from what a user of Rocq, Agda or Lean would naively expect;
\item \textbf{arend-error-type-mismatch} and \textbf{arend-error-extraneous-input} --- reference material for diagnosing two families of misleading error messages, in which the reported type mismatch or the reported unexpected token is merely a downstream effect of an upstream name-resolution or parsing problem.
\end{itemize}

The skills \textbf{arend-formalize} and \textbf{arend-prove} form the intended
pipeline: the first brings a file to the state where only goal errors remain,
and the second attempts to close them, introducing auxiliary lemmas when
needed. Remaining holes are unfinished work; completion requires a
typecheck with no errors or goals. The error-diagnosis skills are consulted
when a check fails. The skills also record newly learned language and meta
behaviour for later use.

\section{The library arend-lib}\label{sec:lib}

In this section we highlight some parts of $\arendlib$; listings are abridged
where marked by omissions.

\subsection{Constructive mathematics}

\subsubsection{Category theory}\label{sec:lib-cat}

\textbf{Categories and precategories.} The approach to definition of categories in 
$\arendlib$ follows the UF approach outlined in \cite{hottbook}, Chapter~9. According
to it, the term \textbf{category} is used not for any category in the usual sense, but 
for a \emph{univalent} category where isomorphic objects are equal. Arbitrary category
in the usual sense is called a \textbf{precategory}.

\begin{lstlisting}
\class Precat (Ob : \Type) {
  | Hom : Ob -> Ob -> \Set
  | id {X : Ob} : Hom X X
  | \fixl 8 o \alias \infixl 8 ∘ {X Y Z : Ob} : Hom Y Z -> Hom X Y -> Hom X Z
  | id-left {X Y : Ob} {f : Hom X Y} : id ∘ f = f
  | id-right {X Y : Ob} {f : Hom X Y} : f ∘ id = f
  | o-assoc {X Y Z W : Ob} {h : Hom Z W} {g : Hom Y Z} {f : Hom X Y} : (h ∘ g) ∘ f = h ∘ (g ∘ f)

  -- Iso {C : Precat} {dom cod : C} (f : Hom dom cod) is the type of isomorphisms;
  -- it is obtained by successively extending the record
  -- Map {C : Precat} {dom cod : C} (\coerce f : Hom dom cod) with the properties
  -- of a monomorphism, of a split monomorphism and, finally, of an isomorphism
  \func idtoiso {a b : Ob} (p : a = b) : Iso {_} {a} {b} \elim p
    | idp => idIso -- the identity isomorphism
    
}  

\class Cat \extends Precat {
  | univalence {a b : Ob} : IsEquiv (idtoiso {_} {a} {b}) -- 'idtoiso' is an equivalence
}  
\end{lstlisting}

Functors and natural transformations are defined as follows:

\begin{lstlisting}
\class Functor (C : Precat) (D : Precat) (\classifying F : C -> D) {
  -- 'F' is the classifying field of 'Functor'
  -- This implies that 'Functor' is implicitly coerced to 'C -> D'
  -- Moreover it affects the instance inference mechanism 
  | Func {X Y : C} : Hom X Y -> Hom (F X) (F Y)
  | Func-id {X : C} : Func (id {_} {X}) = id
  | Func-o {X Y Z : C} {g : Hom Y Z} {f : Hom X Y} : Func (g ∘ f) = Func g ∘ Func f
}  

\record NatTrans {C : Precat} {D : Precat} (F G : Functor C D) {
  | \coerce trans (X : C) : Hom (F X) (G X) 
     -- 'NatTrans' is coerced to a function from objects to morphisms
     -- Records do not have classifying fields and do not take part in
     -- instance inference mechanism
  | natural {X Y : C} (f : Hom X Y) : trans Y ∘ F.Func f = G.Func f ∘ trans X
}  
\end{lstlisting}

Earlier versions of $\arendlib$ declared \verb|NatTrans| with four explicit, linearly
ordered predicative level parameters, so as to allow the two categories \verb|C| and
\verb|D| to live in different universes:
\begin{lstlisting}
\record NatTrans \plevels obj >= hom >= iobj >= ihom {C : Precat \levels (iobj,ihom) _}
                 {D : Precat \levels (obj,hom) _} (F G : Functor C D)
\end{lstlisting}

The syntax for levels has since been simplified: the keywords \verb|\plevels| and
\verb|\levels| have been removed from the language, level parameters of a definition are
now declared by a suffix \verb|.{l1, ..., ln}| on its name, and level arguments are
supplied by the same suffix at the use site, as in \verb|Precat.{iobj,ihom}| or
\verb|\instance PosetCat.{u} : Cat Poset.{u}|. In the case of \verb|NatTrans| the explicit
declaration turned out to be unnecessary altogether and the current definition relies on
level inference.

\textbf{Standard theory.} The library $\arendlib$ contains formalization of the basic
theory from standard textbooks on categories: adjoint functors, Kan extensions, 
(co)limits, comma and slice categories, Yoneda lemma, cartesian closed and preadditive
categories, subobjects and factorization systems, displayed categories, reflective
subcategories and the simplex category. There is also a solver \verb|Category.Solver| for
equations between composites of morphisms.

\textbf{Topoi.} The library $\arendlib$ contains elements of the theory of elementary
and Grothendieck topoi. A \emph{site} is defined in $\arendlib$ as a precategory
together with a predicate on \emph{presieves} which is stable under pullback:

\begin{lstlisting}
\func Presieve {C : Precat} (x : C) => Set (ObOver x)
-- 'ObOver x' is the type of pairs (y : C, f : Hom y x)
-- 'Set X' is the type of subsets 'X -> \Prop' of 'X'

\class Site \extends Precat {
  | isCover (x : Ob) : Presieve x -> \Prop
  | cover-stable {x y : Ob} (h : Hom x y) : ∀ {S : isCover y} ∃ (T : isCover x)
      ∀ {px : T} ∃ (py : S) (f : Hom px.1 py.1) (py.2 ∘ f = h ∘ px.2)
}
\end{lstlisting}

The saturation of \verb|isCover| under the usual closure rules is given by a truncated
inductive type \verb|Covering|, and sheaves on a site \verb|C| with values in a category
\verb|D| form the class \verb|VSheaf D C| of \verb|D|-valued presheaves
satisfying the descent condition.

\subsubsection{Topology}\label{sec:lib-top}

\textbf{Locales.} The concept of \emph{a locale} captures the system of open
subsets $O(X)$ of a topological space $X$ without reference to its underlying
set of points. Locales align better with constructive mathematics, as many natural 
systems of ``open subsets'' lack enough points without the axiom of choice. For instance, classically, an affine scheme is defined as the topological space $\mathsf{Spec}\, R$, whose points are the prime ideals of a ring $R$, equipped with a canonical sheaf of rings. However, in a constructive setting, prime ideals may not exist. In contrast, the locale of ``open subsets'' corresponding to radical ideals of $R$ always preserves the 
meaningful structure. 
Therefore, it is more natural in constructive mathematics to define an affine scheme as this locale of radical ideals, together with its sheaf of rings. There are also reasons to use locales instead of topological spaces in the classical setting, we will discuss this
in the context of algebraic geometry in Section~\ref{sec:lib-alg}. 

Formally, $O(X)$ is defined as a \emph{frame}, which is a poset with all small joins
$\vee$ and all finite meets $\wedge$ satisfying the infinite distributive law: 
$x \wedge (\vee_i\, y_i) \leq \vee_i\, x \wedge y_i$. In particular, a frame is
a complete, distributive lattice, and hence also a Heyting algebra. A \emph{frame homomorphism}
is a homomorphism of posets preserving finite meets and small joins. This defines
the category $\mathsf{Frm}$ of frames. The category $\mathsf{Locale}$ of locales
is the dual of $\mathsf{Frm}$: $\mathsf{Locale} := \mathsf{Frm}^{\text{op}}$. 
In particular, a locale is given by the same data as a frame.

In $\arendlib$ the type of locales is defined as the class \verb|Locale|, which
extends three classes: \verb|CompleteLattice|, the class of Heyting
algebras\footnote{Spelled \texttt{HeytingAlebra} in the library.} and \verb|Site|.
The last of these makes every locale a site, whose covers are the families of opens
whose join dominates the given open (Section~\ref{sec:lib-cat}).

The class \verb|Locale| thus forms a diamond in the lattice hierarchy.
On one side, \verb|CompleteLattice| descends from \verb|BoundedLattice|, adding the
operations \verb|Join| and \verb|Meet| of type \verb|(E -> \Prop) -> E|, that is,
joins and meets of arbitrary \emph{subsets} of \verb|E|. On the other side sits the
class of Heyting algebras. It descends from the class of bounded distributive
lattices, \verb|BoundedDistributiveLattice|, and so from \verb|BoundedLattice| as
well, and it adds the implication
\verb|--> : E -> E -> E|. Two of the inherited fields are implemented in
\verb|Locale| in terms of the infinite distributivity law \verb|Join-ldistr>=|:
the implication, and the finite distributivity
\begin{lstlisting}
  | ldistr>= {x y z : E} : x ∧ (y ∨ z) <= (x ∧ y) ∨ (x ∧ z)
\end{lstlisting}

Indexed joins are recovered from \verb|Join| as the derived
operations \verb|IJoin| and \verb|SJoin|:
\begin{lstlisting}
\func IJoin {J : \Type} (f : J -> E) : E => Join (TSetIm f)
\func SJoin {A : \Type} (f : A -> E) (P : A -> \Type) : E => Join (SetIm f P)
\end{lstlisting}

The categories of frames and locales are defined as follows:

\begin{lstlisting}
\record FrameHom \extends PosetHom {
  \override Dom : Locale
  \override Cod : Locale
  -- Properties expressing that 'func', the field of PosetHom, is a homomorphism of frames.
  -- Every equality below comes with a '>=' variant containing the nontrivial inequality;
  -- the two are interderivable via '\default' implementations, so an instance may
  -- implement whichever of the two is more convenient.
  | func-top : func top = top
  | func-top>= : top <= func top
  | func-meet {x y : Dom} : func (x ∧ y) = func x ∧ func y
  | func-meet>= {x y : Dom} : func x ∧ func y <= func (x ∧ y)
  | func-Join {C : Dom -> \Prop} : func (Join C) = SJoin func C
  | func-Join>= {C : Dom -> \Prop} : func (Join C) <= SJoin func C

  \default func-Join>= => =_<= func-Join
  \default func-Join => <=-antisymmetric func-Join>= $ SJoin-univ \lam Ca => func-<= (Join-cond Ca)
  -- ...
}

-- 'FrameHom.id' and 'FrameHom.∘' are the identity and the composition of frame
-- homomorphisms; since the underlying function of a composite is a literal
-- composition, the category laws hold by reflexivity
\func FrameCat.{u} : Cat Locale.{u} \cowith
  | Hom => FrameHom
  | id => FrameHom.id
  | o => FrameHom.∘
  | id-left => idp -- proof by reflexivity
  | id-right => idp
  | o-assoc => idp
  | univalence => sip \lam {X} {S1} {S2} h1 h2 => exts Locale {
    -- 'sip' is a meta implementing the structure identity principle: it reduces
    -- univalence to the statement that two frame structures on the same carrier
    -- admitting frame homomorphisms in both directions along the identity are equal
    | <= x y => ext (h1.func-<=, h2.func-<=)
    | meet x y => h1.func-meet
    | top => h1.func-top
    | Join C => -- ...
  }

-- A homomorphism of locales is a frame homomorphism in the opposite direction
\func LocaleHom (L M : Locale) => FrameHom M L

-- Declaring 'LocaleCat' an \instance (as opposed to \func in case of 'FrameCat')
-- enables instance inference mechanism to infer 'LocaleCat' instance for locales
\instance LocaleCat.{u} : Cat Locale.{u}
  => FrameCat.op
\end{lstlisting}

The library constructs terminal objects, binary products and equalizers in the
category of locales, and hence all finite limits. It also constructs an initial locale:

\begin{lstlisting}
\instance LocaleCartesianPrecat.{u} : CartesianPrecat
  | Precat => LocaleCat.{u}
  | terminal => \new Product {
    | apex => discreteLocale (\Sigma) -- the locale of subsets of the unit type
    -- ...
  }
  | Bprod L M => \new Product { | apex => ProductLocale L M | -- ... }

\func localeEqualizer {L M : Locale} (f g : LocaleHom L M) : Equalizer {LocaleCat} f g
  -- the apex is the locale of fixed points of a nucleus on 'L'
  
\func InitialLocale : Locale (\Sigma)
\end{lstlisting}

The library also characterizes classes of maps: a map of locales is a regular
monomorphism (that is, exhibits
its domain as a sublocale) exactly when the corresponding frame homomorphism is
surjective, and there are two orthogonal factorization systems on $\mathsf{Locale}$,
(dense, closed) and (strongly dense, weakly closed).

Sites are related to locales by an adjunction. A \emph{preorder site} is a site whose
underlying category is a preorder; every locale gives one, and conversely every preorder
site \verb|P| generates a locale \verb|SiteLocale P| of its ideals, which is the
reflection of \verb|P| into locales:

\begin{lstlisting}
\class PreorderSite \extends Preorder, Site

\func LocaleSite (L : Locale) : PreorderSite L

\instance SiteLocale (P : PreorderSite) : Locale (Opens P)

\func FrameReflectiveSubcat.{u} : ReflectiveSubPrecat FrameCat.{u} PreorderSiteCat.{u}
\end{lstlisting}

This is what makes it practical to construct locales by generators and relations, which
is used, for instance, for the locale of reals and for the spectra of rings
(Section~\ref{sec:lib-alg}).

\textbf{Limits and derivatives.} Derivatives in $\arendlib$ are defined for maps
$f : X \to Y$ where $R$ is a certain kind of a topological ring and $X$, $Y$ are
topological $R$-modules. More specifically, $R$ is required to be a \emph{near skew field} which is a topological ring such that invertible elements $R_*$ are dense.
Moreover, for the uniqueness and other key properties of derivatives to hold, the 
codomain topological $R$-module $Y$ is required to be Hausdorff. 
\begin{lstlisting}
\class TopSpace \extends BaseSet {
  | isOpen : Set E -> \Prop -- here 'Set E' is the power set 'E -> \Prop'
  -- We omit properties of open sets
}  

\class TopSemigroup \extends TopSpace, Semigroup
  | *-cont : ContMap (ProductTopSpace \this \this) \this \lam s => s.1 * s.2
             -- '* : E x E -> E'  is continuous

\class TopMonoid \extends TopSemigroup, Monoid

\class TopAbGroup \extends TopSpace, AbGroup, UniformSpace {
  | +-cont : ContMap (ProductTopSpace \this \this) \this \lam s => s.1 + s.2
             -- '+ : E x E -> E'  is continuous
  | negative-cont : ContMap \this \this negative
             -- 'negative : E -> E' is continuous
  | neighborhood-uniform {C : Set (Set E)} : 
      isUniform C <-> ∃ (U : isOpen) (U 0) ∀ x ∃ (V : C) ∀ {y} (U (x - y) -> V y)
        -- ∃ and ∀ are metas simplifying the work with arguments
        -- 'isUniform C' is a predicate of 'UniformSpace' (see further sections for details)
        -- For abelian groups it can be defined in terms of neighborhoods of 0
        -- For technical (computational) reasons 'isUniform' is left unimplemented
  -- We omit implementations of fields from parent classes of 'TopAbGroup'
}  

\class TopRing \extends TopAbGroup, TopMonoid, Ring

\class NearSkewField \extends TopRing
  | inv-dense : IsDenseSet \lam (x : E) => Monoid.Inv x

\class NearField \extends NearSkewField, CRing
\end{lstlisting}

Examples of near skew fields of interest include topological fields $\mathbb{R}$, 
$\mathbb{C}$ and $\mathbb{Q}_p$. A non-field example: if $R$ is a near skew field
then the ring of dual numbers $R[x]/\langle x^2 \rangle$ is a near skew field as well.
Dual numbers arise, for example, as follows: if $M$ is a manifold over $R$ then the 
tangent manifold $TM$ is naturally a manifold over $R[x]/\langle x^2 \rangle$.

The directional derivative $f'_{a}(x)$ at a point $x \in X$ in a direction $a \in X$
is defined as $f'_{a}(x) := \lim\limits_{h\to 0}q_f(x, h, a)$ where 
$q_f(x, h, a) := h^{-1}\cdot (f (x + h a) - f (x))$, $h \in R_*$. The library
defines convergence of nets by the usual neighborhood condition:

\begin{lstlisting}
-- The following function is a dynamic member of '\class TopSpace'
-- 'f : I -> E' is a slight generalization of a sequence of points in '\this'
\func IsLimit {I : DirectedSet} (f : I -> E) (l : E) : \Prop
    => ∀ {U : isOpen} (U l) ∃ (N : I) ∀ {n} (N <= n -> U (f n))
\end{lstlisting}

In $\arendlib$ both $q_f$ and $f'_{a}(x)$ are defined relative to a subset $S \subseteq X$.

\begin{lstlisting}
-- Definition of qf
-- 'Set.Total S' for 'S : Set X' is the type of pairs (x : X, p : S x)
\func dquot {R : NearSkewField} {X Y : TopLModule R} {S : Set X} (f : Set.Total S -> Y) 
            {x : Set.Total S} (h : Inv {R}) (a : X) (\property s : S (x.1 + h.val *c a))
  => h.inv *c (f (_,s) - f x)
  
\func IsDirDerivAt {R : NearSkewField} {X Y : TopLModule R} {S : Set X} 
           (So : isOpen S) (f : Set.Total S -> Y) (x : Set.Total S) (a : X) (d : Y) : \Prop
  => Y.IsLimit {DirSet So} (\lam h => dquot f h.6.1 _ h.6.2) d
-- Denote G : R -> X, G := \lam r => x + ra. G^{-1}(G(R) ∧ S) ⊆ R is open and contains 0. 
-- 'DirSet' consists of open subsets U ⊆ R intersecting
--          G^{-1}(G(R) ∧ S) ∧ (\lam h => Monoid.Inv h) at some h_U and containing 0.
-- h.6.1 is precisely h_U.
\where {
    -- Proof that G^{-1}(G(R) ∧ S) is open (by continuity of G) 
    \private \lemma aux1 (So : isOpen S) 
                => func-cont {+-cont ∘ tuple (const x.1) (*c-cont ∘ tuple id (const a))} So

    -- Proof that G^{-1}(G(R) ∧ S) contains 0
    \private \lemma aux2 => transportInv S (pmap (x.1 +) (X.*c_zro-left {a}) *> zro-right) x.2

    \protected \func DirSet (So : isOpen S) => InvDirectedSet (aux1 So) aux2  
    -- InvDirectedSet {R : NearSkewField} {S : Set R} (So : isOpen S) (S0 : S 0) 
    -- Constructs a family of open sets U ⊆ R intersecting S ∧ (\lam h => Monoid.Inv h)
    -- and containing 0.
}
\end{lstlisting}

The predicate \verb|IsDerivAt| allows the direction to vary as well: for each
$a_0 \in X$, it requires $q_f(x,h,a) \to D_f(a_0)$ as $(h,a) \to (0,a_0)$,
with the base point $x$ fixed.

\begin{lstlisting}
\func IsDerivAt {R : NearSkewField} {X Y : TopLModule R} {S : Set X} (So : isOpen S) 
                (f : Set.Total S -> Y) (x : Set.Total S) (Df : X -> Y) : \Prop
  => \Pi (a0 : X) -> Y.IsLimit {DirSet So a0} (\lam h => dquot f h.6.1 _ h.6.2) (Df a0)
     -- In this case 'DirSet' consists of open subsets of R x X  
\end{lstlisting}

This condition alone does not imply linearity. The lemma \verb|deriv-linear|
proves linearity when the candidate $D_f$ is already a continuous additive map
and $Y$ is Hausdorff. The module \verb|Analysis.StrongDerivative| instead requires
a continuous extension of $q_f(x,h,a)$ to $h=0$, with $x$ also varying. For
Hausdorff $Y$, it constructs a linear derivative from this stronger condition.

\textbf{Topological hierarchy.} At the bottom of topological hierarchy in $\arendlib$ are normed spaces 
(classes \verb|NormedAbGroup|, \verb|NormedRing|, \verb|PseudoValuedRing| etc) and metric spaces 
(class \verb|MetricSpace|) together with a number of intermediate superclasses obtained
by dropping separatedness (\verb|PseudoMetricSpace|, \verb|PseudoNormedAbGroup|)
or admitting infinite distances (\verb|ExMetricSpace|, \verb|ExPseudoNormedAbGroup|),
and their complete variants (\verb|CompleteMetricSpace|, \verb|CompleteNormedAbGroup|,
\verb|BanachSpace|, \verb|RealBanachAlgebra|, \verb|StoneC*Algebra| etc). 

These are generalized to uniform spaces (class \verb|UniformSpace|), where one
can define uniformly continuous maps. A compatible uniform structure is specified
by a predicate \verb|isUniform C| on covers \verb|C| of the underlying set,
satisfying the uniformity axioms. The covers need not consist of open sets.
The axioms are distributed over three classes: the basic
closure properties of \verb|isUniform| live in \verb|PreuniformSpace|; the requirement that
the induced cover space be regular, so that uniform covers can be refined by star-smaller
ones, is added in \verb|RegularPreuniformSpace|; and the star refinement axiom itself,
which is what makes the induced cover space completely regular, lives in
\verb|UniformSpace|. The relation \verb|V <=* U| appearing below means that \verb|V|
is \emph{star-below} \verb|U|, i.e. that some uniform cover \verb|C| satisfies
$\bigcup\{W \in C \mid \exists x\,(x \in W \cap V)\} \subseteq U$.

\begin{lstlisting}
\class PreuniformSpace \extends PrecoverSpace {
  | isUniform : Set (Set E) -> \Prop
  | uniform-cover {C : Set (Set E)} : isUniform C -> \Pi (x : E) -> ∃ (U : C) (U x)
  | uniform-top : isUniform (single top)
  | uniform-refine {C D : Set (Set E)} : isUniform C -> Refines C D -> isUniform D
  | uniform-inter {C D : Set (Set E)} : isUniform C -> isUniform D -> isUniform (CoverInter C D)
  | uniform-cauchy {C : Set (Set E)} : isCauchy C <-> Closure isUniform C
     -- the Cauchy covers of the underlying cover space (see below) are exactly
     -- those generated by the uniform ones

  -- We omit the rest of the class, which deals with the parent class fields    
}  

\class RegularPreuniformSpace \extends PreuniformSpace, CoverSpace {
  | uniform-regular {C : Set (Set E)} : isUniform C -> isCauchy \lam V => ∃ (U : C) (V <=* U)
  -- ...
}

\class UniformSpace \extends RegularPreuniformSpace, CompletelyRegularCoverSpace {
  | uniform-star {C : Set (Set E)} 
      : isUniform C -> ∃ (D : isUniform) ∀ {V : D} ∃ (U : C) ∀ {W : D} (Given (V ∧ W) -> W ⊆ U)
  
  -- We omit the rest of the class, which deals with the parent class fields    
}  
\end{lstlisting}

With such a predicate one can define (locally) uniformly continuous maps as follows:

\begin{lstlisting}
\record LocallyUniformMap \extends CoverMap {
  \override Dom : RegularPreuniformSpace
  \override Cod : RegularPreuniformSpace

  | func-locally-uniform {E : Set (Set Cod)} : isUniform E
       -> ∃ (C : isUniform) ∀ {U : C} (isUniform \lam V => ∃ (W : E) (U ∧ V ⊆ func ^-1 W))
  -- We omit the rest of the class, which deals with the parent class fields    
}

\lemma uniform-subset {X : PreuniformSpace} {C D : Set (Set X)} (Cc : isUniform C)
            (e : \Pi {U : Set X} -> C U -> D U) : isUniform D 
 => uniform-refine Cc \lam {U} CU => inP (U, e CU, <=-refl)

\record UniformMap \extends LocallyUniformMap {
  | func-uniform {E : Set (Set Cod)} : isUniform E 
       -> isUniform \lam U => ∃ (V : E) (U = func ^-1 V)
  | func-locally-uniform Eu => inP (_, func-uniform Eu,
           \lam (inP (V,EV,p)) => uniform-subset (func-uniform Eu)
                                   \lam _ => inP $ later (V, EV, rewrite p meet-left))
}                                   
\end{lstlisting}

Constructively, completing $\mathbb{Q}$ with its usual metric does not by
itself establish the Heine--Borel property: one cannot in general prove
that every open cover of the resulting real interval $[0,1]$ has a finite
subcover. Localic completion provides another approach, yielding the
locale of real numbers, whose closed bounded intervals are compact as
locales~\cite{VickersLocalicCompletionII}. Maps between locales are
described by their action on opens. For constructions such as power
series, it can be more convenient to work directly with real numbers
and functions on them.

Isaev's \emph{cover spaces}~\cite{isaev24} retain this point-set approach.
A cover space equips a set with a distinguished collection of
\emph{Cauchy covers}, satisfying the axioms below. For a uniform space,
these covers are generated from its uniform covers. Morphisms are
functions whose inverse images take Cauchy covers to Cauchy covers.
Completion is constructed using Cauchy filters; for $\mathbb{Q}$ it
yields the Dedekind reals. A cover space is \emph{totally bounded} if
every Cauchy cover has a finite subcover, and \emph{compact} if it is also
complete. Closed bounded real intervals are compact in this sense
\cite[Section~9]{isaev24}. The distinction is that an arbitrary open
cover need not be a Cauchy cover constructively.

The connection with locales is concrete in this example: the locale
associated with the cover space of real numbers is the usual locale of
reals, and cover maps $\mathbb{R}\to\mathbb{R}$ correspond exactly to
morphisms of that locale~\cite[Proposition~8.16 and Corollary~8.17]{isaev24}.

The cover spaces are defined in $\arendlib$ as follows:

\begin{lstlisting}
\class PrecoverSpace \extends TopSpace {
  | isCauchy : Set (Set E) -> \Prop
  | cauchy-cover {C : Set (Set E)} : isCauchy C -> \Pi (x : E) -> ∃ (U : C) (U x)
  | cauchy-top : isCauchy (single top)
  | cauchy-refine {C D : Set (Set E)} : isCauchy C -> Refines C D -> isCauchy D
  | cauchy-glue {C : Set (Set E)} : isCauchy C ->
    \Pi {D : Set E -> Set (Set E)} -> (\Pi {U : Set E} -> C U -> isCauchy (D U))
    -> isCauchy (\lam U => ∃ (V W : Set E) (C V) (D V W) (U = V ∧ W))
  | cauchy-open {S : Set E} : isOpen S <-> ∀ {x : S} (isCauchy \lam U => U x -> U ⊆ S)
 
  \default isOpen S : \Prop => ∀ {x : S} (isCauchy \lam U => U x -> U ⊆ S)
  
  -- We omit the rest of the class
}  

\class CoverSpace \extends PrecoverSpace {
  | isRegular {C : Set (Set E)} : isCauchy C -> isCauchy \lam V => ∃ (U : C) (V <=< U)
}

\type \infix 4 <=< {X : PrecoverSpace} (V U : Set X) : \Prop
  => isCauchy \lam W => Given (V ∧ W) -> W ⊆ U
\end{lstlisting}

Completions, the relation to locales and other results from \cite{isaev24} have been
formalized in $\arendlib$, in the modules \verb|Complete| and \verb|Locale| under
\verb|Topology/CoverSpace|. Metric and normed spaces are integrated into this
hierarchy: pseudometric spaces are defined as an extension of \verb|UniformSpace| whose
uniform covers are those refined by the covers by balls of a fixed radius, so all of the
above applies to them. On this basis $\arendlib$ develops the constructive theory of real
numbers, series and power series, differentiation (both the topological notion of
Section~\ref{sec:lib-top} and a synthetic one), Banach spaces and algebras, commutative
$C^*$-algebras and the beginnings of measure theory.

\subsubsection{Algebra}\label{sec:lib-alg}

\textbf{Algebraic geometry.} The part of library $\arendlib$ devoted to algebraic geometry
contains basics of the theory of \emph{schemes}. The constructive theory of schemes
developed in $\arendlib$ is based on locales, which are better suited for constructive
algebraic geometry than topological spaces as explained in Section~\ref{sec:lib-top}.

Schemes in $\arendlib$ are defined in terms of locally ringed locales as follows:

\begin{enumerate}
\item A \emph{ringed locale} is a locale $L$ together with a ring-valued sheaf 
$\mathcal{O}_L$ on it:

\begin{lstlisting}
\record RingedLocale (\coerce L : Locale) (R : VSheaf CRingCat L)
-- Here 'VSheaf D C' is the type of D-valued sheaves on a site C (i.e. on a category
-- with a covering for objects). Note that 'Locale' extends 'Site'.
-- 'R a' abbreviates 'R.F a', the ring of sections over the open 'a : L', and
-- 'R.F.Func ba : RingHom (R a) (R b)' is the restriction along 'ba : b <= a'.
\end{lstlisting}

\item Standardly, locally ringed spaces are defined as ringed spaces 
$(X, \mathcal{O}_X)$ such that the stalk $\mathcal{O}_{X, x}$ at every point 
$x \in X$ is a local ring. This definition is not suitable for locales since
it refers to points and this does not work constructively as explained in 
Section~\ref{sec:lib-top}. Nevertheless, a \emph{locally ringed locale} can be 
defined by using an alternative formulation of the locality condition which is
classically equivalent to the standard one: for every section $x$ over $a$ the two
opens on which $x$, respectively $x + 1$, is invertible cover $a$.

\begin{lstlisting}
\record LocallyRingedLocale \extends RingedLocale {
  | isLRLNonTrivial (a : L) : 0 = {R a} 1 -> a <= bottom
  | isLocallyRinged (a : L) (x : R a) : a <= Join (\lam b => \Sigma (p : b <= a)
      (Inv (R.F.Func p x) || Inv (R.F.Func p (x + 1))))
  -- 'Join' here is the join of a subset of 'L' (see the previous section)
}
\end{lstlisting}

\item Earlier versions of $\arendlib$ defined a scheme as a locally ringed locale which
is locally \emph{equal} (by univalence, equivalently locally isomorphic) to the spectrum
of a ring. The current version avoids referring to the spectrum in the definition and
instead expresses affineness of an open $a \in L$ \emph{intrinsically}, in terms of
localizations of the ring of sections over $a$. The building block is the notion of
a \emph{principal} open: $b \leq a$ is the largest open on which a given section
$x$ over $a$ becomes invertible.

\begin{lstlisting}
-- Both definitions below are dynamic members of '\record RingedLocale'

\protected \func IsPrincipalOpenAt {a b : L} (ba : b <= a) (x : R a) : \Prop
  => Given (Inv (R.F.Func ba x)) ∀ {c} (ca : c <= a) (Inv (R.F.Func ca x)) (c <= b)

\type IsAffineOpen (a : L) => Given
  -- 1. the sections over a principal open 'b <= a' are the localization of 'R a'
  --    at the powers of the corresponding section
  (∀ {x : R a} {b} (ba : b <= a) (IsPrincipalOpenAt ba x)
     (Localization (powers x) (R b) (R.F.Func ba)))
  -- 2. if a principal open 'b <= a' is covered by 'U', then the corresponding
  --    section lies in the radical of the ideal generated by the sections
  --    cutting out the principal opens subordinate to 'U'
  (∀ {b} (ba : b <= a) (x : R a) (IsPrincipalOpenAt ba x) {U} (b<=U : b <= Join U)
     ((Ideal.sclosure \lam y => ∃ (d : U) (c : L) (c <= d) (cb : c <= b)
        (Localization (powers y) (R c) (R.F.Func (cb <=∘ ba)))).radical x))
  -- 3. principal opens form a basis below 'a'
  (∀ {b} (ba : b <= a) (b <= Join \lam c => Given (cb : c <= b)
     ∃ (x : R a) (IsPrincipalOpenAt (cb <=∘ ba) x)))

\func IsAffineScheme : \Prop => IsAffineOpen top
\end{lstlisting}

A \emph{scheme} is then simply a locally ringed locale covered by affine opens:

\begin{lstlisting}
\record Scheme \extends LocallyRingedLocale
  | isLocallyAffine : top <= Join IsAffineOpen
\end{lstlisting}

\end{enumerate}

Affine schemes themselves are constructed not by presenting the frame of opens of
$\mathsf{Spec}\, R$ directly, but through an intermediate notion of a \emph{scheme site}:
a preorder \verb|P| of ``formal principal opens'' together with a ring-valued presheaf
\verb|R| on it such that restriction along $b \leq a$ always exhibits \verb|R b| as a
localization of \verb|R a|, in a way stable under further restriction.

\begin{lstlisting}
\record SchemeSite (\coerce P : Preorder) (R : Functor P.op CRingBicat) {
  | loc-stable {a' a b : P} (r : a' <= a) (p : b <= a) : ∃ (b' : P) (p' : b' <= a')
      (b' <= b) (x : R a) (Localization (powers x) (R b) (R.Func p))
      (Localization (powers (R.Func r x)) (R b') (R.Func p'))

  \func IsLocalizationAt {a b : P} (ba : b <= a) (x : R a) : \Prop
    => Localization (powers x) (R b) (R.Func ba)

  -- 'P' is turned into a preorder site (see the previous section) whose basic covers
  -- are the families of principal opens whose defining sections generate the unit ideal
  \func toSite : PreorderSite \cowith
    | Preorder => P
    | isBasicCover a U => Ideal.sclosure (\lam x => ∃ (b c : P) (b <= c) (U c)
                                            (ba : b <= a) (IsLocalizationAt ba x)) 1
    | basic-cover-stable => -- ... (uses 'loc-stable')

  \func toLocale : Locale => SiteLocale toSite  -- the reflection of 'toSite' into locales
  \func toSheaf : VSheaf CRingCat toSite        -- 'R' is a sheaf for these covers
}

-- A scheme site is 'saturated' if every section already cuts out an object of 'P'
\record SatSchemeSite \extends SchemeSite {
  | loc-exists {a : P} (x : R a) : ∃ (b : P) (p : b <= a) (Localization (powers x) (R b) (R.Func p))

  \lemma basis-affineOpen {a : P} : toLRLSite.toLocale.IsAffineOpen (SiteLocale.embed a)

  \func toScheme : Scheme \cowith
    | LocallyRingedLocale => toLRLSite.toLocale
    | isLocallyAffine {x} _ => cover-refl $ inP (SiteLocale.embed x, basis-affineOpen, cover-refl idp)
}

-- Every scheme site can be saturated by adjoining formal localizations
\func SatSite (S : SchemeSite) : SatSchemeSite
\end{lstlisting}

The spectrum of a ring is obtained from the one-object scheme site. The projective
spectrum of a graded ring uses the preorder of positive-degree homogeneous elements
ordered by divisibility, with the degree-zero parts of their localizations as its rings
of sections:

\begin{lstlisting}
\func Spec (R : CRing) : Scheme
  => (satSite R).toScheme
  \where {
    \func site (R : CRing) : SchemeSite \cowith
      | P => TrivialPoset -- a single object
      | R { | F _ => R | Func _ => RingHom.id | Func-id => idp | Func-o => idp }
      | loc-stable _ _ => inP ((), (), (), 1, Localization.id1, Localization.id1)

    \func satSite (R : CRing) : SatSchemeSite => SatSite (site R)

    \lemma isAffineScheme : (Spec R).IsAffineScheme => -- ...
  }

\func Proj (R : GradedCRing) : Scheme => (site R).toScheme
\end{lstlisting}

Finally, $\mathsf{Spec}$ is shown to be right adjoint to the global sections functor,
with the affine schemes cut out exactly by the unit of this adjunction being an
isomorphism. This is the place where univalence for the category of locales
(the field \verb|univalence| of \verb|FrameCat| above) is used.

\begin{lstlisting}
-- The unit of the adjunction between affine schemes and locally ringed locales
\func specUnit {X : LocallyRingedLocale} : LocallyRingedLocaleHom X (Spec X.GlobalSections)

\lemma IsAffineScheme-char {X : LocallyRingedLocale} : X.IsAffineScheme <-> (specUnit {X}).IsIso

\func SpecFunctor : FullyFaithfulFunctor CRingCat.op SatSchemeSitePrecat
\func SchemeSiteFunctor : FullyFaithfulFunctor SatSchemeSitePrecat LocallyRingedLocaleCat
\func SpecLRLFunctor : ReflectiveSubPrecat CRingCat.op LocallyRingedLocaleCat
  -- affine schemes are a reflective subcategory of locally ringed locales,
  -- the reflector being 'GlobalSections' and the unit being 'specUnit'
\end{lstlisting}

\bigskip	

\textbf{Principal ideal domains.} As we mentioned earlier (Section~\ref{sec:classesLang}), domains are defined constructively
as rings with apartness relation \verb|x # y| satisfying \verb|x # 0 -> y # 0 -> (x * y) # 0|. 
According to the standard classical definition a domain $R$ is a \emph{Principal Ideal
Domain} (PID) if \emph{every} ideal $I \subseteq R$ is \emph{principal}, that is there exist
$a\in R$ such that $I$ is generated by $a$: $I = (a)$. This requirement is
too strong for the intended constructive examples. For an arbitrary
proposition $P$, consider the ideal
$I_P:=\{n\in\mathbb{Z}\mid n=0\lor P\}$. If $I_P=(d)$, decidable equality
in $\mathbb{Z}$ lets us distinguish $d=0$ from $d\ne0$. In the first case
$P$ is false, since $P$ would put $1$ in $(0)$; in the second case
$d\in I_P$ implies $P$. Conversely, a decision of $P$ makes $I_P$ either
$(1)$ or $(0)$. Thus the assertion that every ideal of $\mathbb{Z}$ is
principal implies excluded middle, and cannot be used to establish
constructively that $\mathbb{Z}$ is a PID.

A constructively more well-behaved property is the requirement that every \emph{finitely generated} ideal be principal. This is equivalent to the \emph{Bezout condition},
which is defined in $\arendlib$ as follows:

\begin{lstlisting}
\func IsBezout => \Pi (a b : E) -> ∃ (s t : E) (LDiv (s * a + t * b) a) (LDiv (s * a + t * b) b)

\class BezoutRing \extends CRing, GCDMonoid {
  | isBezout : IsBezout
  | isGCD a b => -- proof of TruncP (GCD a b)
  | gcd-ldistr c {a} {b} {z} (g : GCD a b z) => -- proof of TruncP (GCD (c * a) (c * b) (c * z))
}
\end{lstlisting}

where \verb|LDiv x y| is the type of left divisors and \verb|GCD| is the type of
greatest common divisors:
\begin{lstlisting}
\record DivBase {M : Monoid} (\coerce val : M) (elem inv : M)

\record LDiv \extends DivBase
    | inv-right : val * inv = elem 
    
\record GCD {M : CMonoid}
            (val1 val2 : M) (\coerce res : M)
            (res|val1 : LDiv res val1)
            (res|val2 : LDiv res val2)
            (res-univ : \Pi (g : M) -> LDiv g val1 -> LDiv g val2 -> LDiv g res)    
\end{lstlisting}

Now, if we assume LEM, then the classical definition of PID can be equivalently
reformulated as follows: PID is a domain 
satisfying Bezout condition and such that all ideals are finitely generated. 
The latter condition is one of equivalent definitions of \emph{Noetherian property}
assuming LEM. Constructively, however, this condition is too strong. Instead,
we say that a ring $R$ is \emph{weakly Noetherian} if for every ascending sequence 
$I_1 \subseteq I_2\subseteq \ldots$ of finitely generated ideals of $R$ there is 
$k\in \mathbb{N}$ such that $I_k = I_{k+1}$. 

\begin{lstlisting}
\class WeaklyNoetherianCRing \extends CRing
  | isWeaklyNoetherian (I : Nat -> Ideal \this) : (\Pi (n : Nat) -> (I n).IsFinitelyGenerated) -> Ideal.ChainCondition I   
 
-- function accompanying the class 'Ideal'  
\func ChainCondition {R : CRing} (I : Nat -> Ideal R) =>
    (\Pi (n : Nat) {a : R} -> I n a -> I (suc n) a) -> ∃ (n : Nat) ∀ {a} (I (suc n) a -> I n a)  
\end{lstlisting}

The condition above is the weakest of several constructive readings of the Noetherian
property, and it is the one PIDs are required to satisfy. The library also has a stronger
notion, phrased in terms of bar induction rather than of sequences, which avoids the
choice needed in passing from an indefinitely extendible finite chain to
an infinite sequence; this is
the class \verb|NoetherianCRing| extending \verb|WeaklyNoetherianCRing|, and it is what
Euclidean domains are shown to satisfy.

\begin{lstlisting}
\func IsNoetherian (R : CRing) : \Prop => Bar (PausesElem {R})  nil
-- 'PausesElem l' holds if some element of the array 'l' lies in the ideal generated
-- by the elements after it. The inductive predicate 'Bar P l' is generated by
-- P l, or by Bar P (a :: l) for every a. It implies that every infinite
-- extension meets P; the converse is not part of its definition.

\class NoetherianCRing \extends WeaklyNoetherianCRing
  | isNoetherian : IsNoetherian \this
  | isWeaklyNoetherian I Ifg Ic => -- derivation of the chain condition from the bar

\class EuclideanDomain \extends PID, NoetherianCRing
\end{lstlisting}

In $\arendlib$ PIDs are defined as domains $R$ satisfying Bezout condition, the weak
Noetherian property and such that equality $x =_R y$ is a decidable predicate. For Bezout rings
the weak Noetherian property admits an equivalent formulation which is more convenient to work
with. To state it, consider equivalence relation $ x \sim y := x\, |\, y\, \wedge\, y\,
|\, x$ on $R$ and consider the quotient monoid $R_{\sim}$. Then every sequence 
$\{a_i\}\subset R_{\sim}$ satisfying $a_{i+1}\, |\, a_i$ necessarily has 
$a_{k} = a_{k + 1}$ for some $k \in \mathbb{N}$ if and only if $R$ satisfies
the weak Noetherian condition.

The combination of the Bezout condition and the weak Noetherian property places PIDs in an
important class of domains known as \emph{Smith domains}. The distinguishing property
of Smith domains is existence of the Smith normal form for matrices over them. Namely,
for every $m \times n$, $n \geq m$ matrix $A$ over a Smith domain $R$ there exist
invertible matrices $S$, $T$ over $R$ such that 
$S\cdot A\cdot T = (diag(a_1, \ldots, a_k, 0^{m - k})\, |\, \textbf{0}^{m\times (n - m)})$
is $m\times m$ diagonal matrix followed by $m\times (n - m)$ zero matrix and 
$a_i\, |\, a_{i+1}$. In other words, if we say $A \sim B$ for matrices if there exist invertible $S$, $T$ such that $B = S\cdot A\cdot T$, then for every $A$ there exist
a matrix in the normal form $D$ such that $A \sim D$. Among discrete Bezout domains,
the Smith property is equivalent to the \emph{Kaplansky condition}.

\begin{lstlisting}
\class SmithRing \extends StrictBezoutRing {
  | isKaplansky : IsKaplansky \this
  
  -- Dynamic lemma (using \this parameter)
  \lemma toSmith {n m : Nat} (A : Matrix E n m) : ∃ (B : Matrix E n m) (IsSmith B) (A M~ B)
    => Smith-char 0 3 (isStrictBezout,isKaplansky) A
} \where {
  \type IsKaplansky (R : CRing) => \Pi (a b c : R) -> IsCoprimeArray (a,b,c) -> ∃ (t s : R) (IsCoprime (t * a) (t * b + s * c))

  -- The Smith normal form theorem is proven as part of the equivalence below
  \lemma Smith-char {R : CRing} : TFAE (
-- The four conditions below are equivalent
    \Sigma R.IsStrictBezout (IsKaplansky R),
    \Sigma R.IsStrictBezout (\Pi (A : Matrix R 2 2) -> A 1 0 = 0 -> IsCoprimeArray (A 0 0, A 0 1, A 1 1) -> ∃ (B : Matrix R 2 2) (IsDiagonal B) (A M~ B)), 
     -- here M~ is equivalence as in the Smith normal form
    \Pi {n m : Nat} -> n <= m -> m <= 2 -> \Pi (A : Matrix R n m) -> ∃ (B : Matrix R n m) (IsDiagonal B) (A M~ B),
    \Pi {n m : Nat} (A : Matrix R n m) -> ∃ (B : Matrix R n m) (IsSmith B) (A M~ B)
  ) => -- ...
}
  
\class SmithDomain \extends BezoutDomain.Dec, SmithRing   
\end{lstlisting}

Since PIDs are Smith domains, in $\arendlib$ they are defined as extension:

\begin{lstlisting}
\class PID \extends SmithDomain, WeaklyNoetherianCRing {
  | divChain : CMonoid.DivChain {DivQuotientMonoid nonZeroCMonoid}
  | isWeaklyNoetherian I Ifg => Ideal.fromMonoidChainCondition (IntegralDomain.Dec.fromNonZeroDivChain divChain) I \lam n => bezout_finitelyGenerated_principal.1 isBezout (Ifg n) -- f.g. ideals
--  are principal in Bezout rings 
  | isKaplansky => adequate_kaplansky \lam a/=0 => adequate'_adequate (domain_adequate' divChain a/=0)

  -- PIDs have Krull dimension at most one
  \lemma is1Dim : Dim<= 1
    => bezout_1Dim<->adequate'.2 \lam a/=0 => domain_adequate' divChain a/=0
}
\end{lstlisting}

The proof of Kaplansky condition goes via the proof of the \emph{adequate condition}
for PIDs first, following Helmer \cite{helmer43} and Kaplansky \cite{kaplansky49}. 
The dimension condition used above is defined as follows:

\begin{lstlisting}
-- Dynamic type (using \this parameter) in CRing class      
\type Dim<= (k : Nat) => \Pi (x : Array E (suc k)) -> 
                             ∃ (a : Array E (suc k)) (m : Array Nat (suc k)) (fold x a m = 0)
    \where
      \func fold {k : Nat} (xs as : Array E k) (ms : Array Nat k) : E \elim k, xs, as, ms
        | 0, nil, nil, nil => 1
        | suc k, x :: xs, a :: as, m :: ms => pow x m * (fold xs as ms - x * a)    
\end{lstlisting}

\bigskip	

\textbf{Polynomials of one or several variables.} Polynomials of one variable over any
pointed type (a type with $0$) have two equivalent representations in $\arendlib$:
as an inductive type with conditions \verb|Poly| and as a quotient set \verb|QPoly|.
Both are essentially arrays of coefficients quotiented by cutting trailing zeros, but 
sometimes it is more convenient to work with \verb|Poly| instead of \verb|QPoly|
and vice versa.

\begin{lstlisting}
\data Poly (R : AddPointed)
  | pzero
  | padd (Poly R) R
  | peq : padd pzero 0 = pzero

\func QPoly (R : AddPointed) => Quotient {Array R} (__ = __ ++ 0 :: nil)  
\end{lstlisting}

Multivariate polynomials are defined as a monoid algebra:

\begin{lstlisting}
\func MPoly (J : \Set) (R : CRing) => MonoidAlgebra (PermSetMonoid J) R

\instance PermSetMonoid (A : \Set) : CMonoid (PermSet A)

\type PermSet (A : \Set) => Quotient {Array A} EPerm
-- here the relation EPerm l l' holds iff l' can be obtained by permuting elements in l
\end{lstlisting}

\bigskip	

\textbf{Matrices over commutative rings.} The library $\arendlib$ defines matrices over 
any type $R$
as two-dimensional arrays and then formalizes operations on matrices and their 
properties for a variety of structures on $R$. In case of commutative rings $\arendlib$
has definitions of determinant, characteristic polynomial and proofs of a number of their
properties including the proof of Cayley-Hamilton theorem.

\begin{lstlisting}
\func determinant {R : CRing} {n : Nat} (M : Matrix R n n)
  => R.FinSum \lam (e : Sym n) => sign e * R.BigProd (\lam j => M (e j) j)
  
\func charPoly {R : CRing} {n : Nat} (M : Matrix R n n) : Poly R
  => determinant $ padd 1 0 *c matrix-map (padd 0) 1 - matrix-map (padd 0) M
-- det (x * I - M)  
-- 'matrix-map' maps in this case a matrix over 'R' to a matrix over 'Poly R'    

\lemma cayley-hamilton {R : CRing} {n : Nat} (A : Matrix R n n) 
                : polyMapEval (CAlgebra.coefHom {MatrixAlgebra R n}) (charPoly A) A = 0
    -- 'polyMapEval' converts a polynomial over R to polynomial over 'MatrixAlgebra R n'
    -- and then evaluates it at the matrix A                 
    => -- ...            
\end{lstlisting}

Among the properties of determinant formalized in $\arendlib$ is the cofactor expansion
and the existence of adjugate matrices.

\begin{lstlisting}
\func adjugate {R : CRing} {n : Nat} (M : Matrix R n n) : Matrix R n n \elim n
  | 0 => M
  | suc n => mkMatrix \lam i j => Monoid.pow -1 (j + i) * determinant (minor M j i)

\lemma adjugate-left {R : CRing} {n : Nat} {M : Matrix R n n} : adjugate M * M = determinant M *c 1
  => -- ...

\lemma adjugate-right {R : CRing} {n : Nat} {M : Matrix R n n} 
                                     : M * adjugate M = determinant M *c 1
  => -- ...  
\end{lstlisting}

As a corollary we have that $M$ is invertible if and only if $\mathsf{det}(M)$ is invertible. 

\begin{lstlisting}
\lemma determinant-inv {R : CRing} {n : Nat} {M : Matrix R n n} : Inv M <-> Inv (determinant M)
\end{lstlisting}

Another corollary is Nakayama's lemma, which says the following. Let $R$ be a 
commutative ring, $I \subseteq R$ an ideal and $M$ a finitely generated module
over $R$. If $I\cdot M = M$ then there exist $r\in R$ such that 
$r \equiv 1 (\mathsf{mod}\, I)$ such that $r\cdot M = 0$.
 
\begin{lstlisting}
\lemma nakayama {R : CRing} {M : LModule R} {I : Ideal R} (fg : M.IsFinitelyGenerated)
                (c : topSubLModule M <= I *m topSubLModule M) 
                : ∃ (e : R) (I (1 - e)) ∀ (m : M) (e *c m = 0)
\end{lstlisting}

\bigskip	

\textbf{Linear algebra over Smith domains.} Here we sketch some elements of linear 
algebra over Smith domains formalized in $\arendlib$.

\begin{enumerate}
\item Definition of the rank of a matrix over a Smith domain:

\begin{lstlisting}
\func rank {R : SmithDomain} {n m : Nat} (A : Matrix R n m) : Nat => -- ...
\end{lstlisting}

\item Let $R$ be a Smith domain. For any finite-dimensional free modules $U$, $V$ over 
$R$ and every linear $f : U \to V$ the image $\mathsf{Im}\, f$ and the kernel 
$\mathsf{Ker}\, f$ modules
are finite-dimensional free.

\begin{lstlisting}
\class FinModule \extends LModule
  | isFinModule : ∃ (l : Array E) (IsBasis l)
  
\instance image-fin {R : SmithDomain} {U V : FinModule R} (f : LinearMap U V) : FinModule
  | LModule => ImLModule f
  | isFinModule => \case U.isFinModule, V.isFinModule \with {
    | inP (lu,bu), inP (lv,bv) => TruncP.map (basis f bu bv) \lam s => (s.1,s.2)
  } 
\where {
    \protected \lemma basis {U V : LModule R} (f : LinearMap U V) {lu' : Array U} 
                      (bu' : U.IsBasis lu') {lv' : Array V} (bv' : V.IsBasis lv')
      : ∃ (l : Array (ImLModule f)) (IsBasis {ImLModule f} l) 
                        (l.len = rank (LinearMap.toMatrix lu' bv' f)) 
     => -- ...   
} 

\instance kernel-fin {R : SmithDomain} {U V : FinModule R} (f : LinearMap U V) : FinModule
  | LModule => IModule (KernelLModule f)
  | isFinModule => -- ...      
\end{lstlisting}

\item Let $R$ be a Smith domain and $U$ a finite-dimensional free module over $R$. 
Linear dependence for a set of vectors $u_1$, \ldots, $u_n$ in $U$ is decidable.

\begin{lstlisting}
\func dependency-dec {R : SmithDomain} {U : FinModule R} (l : Array U) 
                                      : Or (U.IsDependent l) (U.IsIndependent l)
\end{lstlisting}

\end{enumerate}

\bigskip	

\textbf{Algebraic closure of countable, discrete fields.} The library
$\arendlib$ defines integral extensions of rings, splitting fields for
polynomials over discrete fields and algebraic closure for countable, discrete fields.

\begin{lstlisting}
\func isIntegral (f : RingHom) (x : f.Cod) : \Prop
  => ∃ (p : Poly f.Dom) (isMonic p) (polyEval (polyMap f p) x = 0)
\end{lstlisting}

The library proves that an integral algebra over a discrete field is
$0$-dimensional. The condition \verb|Dim<= 0| can equivalently be expressed as
follows:

\begin{lstlisting}
-- Dynamic type (using \this parameter) in Ring class
\type IsZeroDimensional => \Pi (a : E) -> ∃ (b : E) (n : Nat) (pow a n = pow a (suc n) * b)
\end{lstlisting}

A discrete field is a commutative ring such that every element is either zero or
invertible (see Section~\ref{sec:classesLang} for a definition of a field).

\begin{lstlisting}
\class DiscreteField \extends Field, EuclideanDomain {
  | finv : E -> E
  | finv_zro : finv 0 = 0
  | finv-right {x : E} (x/=0 : x /= 0) : x * finv x = 1
  \field eitherZeroOrInv (x : E) : (x = 0) || Inv x
}

\func integralExt-zeroDim {K : DiscreteField} {R : CRing} {f : RingHom K R} (fi : isIntegralExt f) : R.IsZeroDimensional  
=> -- ...
\end{lstlisting}

The definition of a splitting field for a polynomial:

\begin{lstlisting}
\type IsSplittingField {k K : DiscreteField} (p : Poly k) (f : RingHom k K)
  => ∃ (l : Array K) (c : K) (RingHom.isAlgebraGenerated f l) 
  (polyMap f p = c PolyRing.*c Monoid.BigProd (map (\lam a => padd 1 (negative a)) l))
\end{lstlisting}

A ring $K$ is algebraically closed if every monic polynomial over $K$ of non-zero 
degree has a root.

\begin{lstlisting}
\type IsAlgebraicallyClosed (K : Ring)
  => \Pi {p : Poly K} -> ∃ (n : Nat) (n /= 0) (degree<= p n) (polyCoef p n = 1) -> ∃ (a : K) (polyEval p a = 0)
\end{lstlisting}

Every countable discrete field has an algebraic closure:

\begin{lstlisting}
\sfunc countableAlgebraicClosure {k : DiscreteField} (c : Countable k) 
: \Sigma (K : DiscreteField) (f : RingHom k K) (IsAlgebraicClosure f)
=> -- ...
\end{lstlisting}

\textbf{The Fundamental Theorem of Algebra.} The module
\verb|Arith/Complex/FTA| formalizes Richman's formulation~\cite{RichmanFTA}:
every monic polynomial of degree $n$ determines a point in the completion
of the space of unordered $n$-tuples of complex numbers. This represents
its roots with multiplicities, determines the polynomial, and depends
continuously on its coefficients. The construction uses Kneser's
approximate-root argument and Cauchy filters of approximate factorizations,
without choice. The module \verb|Arith/Algebraic/FTA| obtains actual roots
for nonconstant separable monic polynomials by Newton iteration, proves root existence for
nonconstant polynomials over discrete subfields of $\mathbb C$, and shows
that the algebraic numbers are algebraically closed. Finally,
\verb|Arith/Complex/FTA/Classical| derives root existence for arbitrary
monic nonconstant complex polynomials using countable choice.

\subsection{Other parts of the library}

\subsubsection{Synthetic homotopy theory}\label{sec:lib-hott}

All of the synthetic homotopy theory developed in the HoTT
Book~\cite{hottbook} has been formalized in $\arendlib$.
Parts of this formalization were carried out with AI assistance; the
resulting proofs are Arend terms subject to the same type checking as
manually written proofs. This development includes:
\begin{enumerate}
\item Pointed types, loop spaces, homotopy groups, and the Eckmann--Hilton argument.
\item Higher inductive constructions: the circle, spheres, the torus,
  suspensions, joins, and pushouts.
\item The hierarchy of $h$-levels, $n$-truncations, connected types and
  connected maps.
\item The equivalence $\Omega(\mathbb{S}^1)\simeq\Int$, the fundamental
  group of the circle, and covering spaces.
\item Fibrations and the long exact sequence of homotopy groups; the Hopf
  fibration with total space $\mathbb{S}^3$ and the calculation
  $\pi_3(\mathbb{S}^2)\cong\Int$.
\item The Freudenthal suspension theorem and the calculations
  $\pi_k(\mathbb{S}^n)=0$ for $0<k<n$ and
  $\pi_n(\mathbb{S}^n)\cong\Int$ for $n\geq1$.
\item The van Kampen theorem and Whitehead's theorem for $n$-types.
\item Eilenberg--MacLane spaces $K_1(G)$ and the equivalence
  $\Omega K_1(G)\simeq G$.
\item Localizations, modalities, and the generalized Blakers--Massey theorem.
\end{enumerate}

\subsubsection{Further topics}\label{sec:lib-rest}

The parts of $\arendlib$ selected above are far from exhaustive. We close with a very brief
list of the remaining major topics, to indicate the overall coverage of the library.

\begin{itemize}
\item \textbf{Group theory.} Subgroups, quotients, products, free groups, symmetric and
automorphism groups, Lagrange's theorem, group actions ($G$-sets) and their category, and
linear representations: intertwining maps, irreducibility, Maschke's lemma and characters.

\item \textbf{More commutative algebra.} Ideals (including graded ones) and factor rings,
localization, integral and integrally closed extensions, minimal polynomials, valuations,
Euclidean domains, GCD and Bezout domains, zero-dimensional and reduced rings, Noetherian
rings with the Hilbert basis theorem, graded rings, monoid rings, formal power series,
Boolean rings, and the Grothendieck ring.

\item \textbf{More linear algebra.} Modules and their category, submodules, bilinear maps,
linear maps and their matrices, tensor products, traces, and vector spaces.

\item \textbf{Numbers.} Natural numbers, integers and rationals; the constructive reals
(including lower and upper reals, roots, the intermediate value theorem, and $\pi$);
exponentials and logarithms; trigonometric functions; and the complex numbers, with the
argument, roots and Euler's formula. On the number-theoretic side: primality, Euler's
totient function, factorials, binomial coefficients and combinations.

\item \textbf{Logic and rewriting.} Classical principles and their interrelations,
\verb|TFAE| for stating chains of equivalent conditions, accessibility and well-founded
induction, bar induction and the fan theorem, first-order algebraic theories with their
models and free models, and a substantial development of term rewriting: abstract reduction
systems, confluence and termination, higher-order rewriting systems --- with the
$\lambda$-calculus as a worked example --- and modularity of confluence for unions of
rewriting systems.

\item \textbf{Sets, finiteness and data.} Subsets, filters, decidable and Kuratowski
finiteness, the pigeonhole principle, countability, Hedberg's theorem, and the category of
sets; lists, arrays with sorting and permutations, and sequential colimits.

\item \textbf{Order theory.} Partial, linear and strict orders, lattices and complete
lattices, Heyting and Boolean algebras, directed sets, and lexicographic orders.

\item \textbf{Solvers.} A family of metas discharging equational goals in a given algebraic
structure: semigroups, monoids, groups, semirings, rings and Boolean rings (in commutative
and non-commutative variants), linear arithmetic over ordered rings, linear algebra, and
composites of morphisms in a category.
\end{itemize}

\bibliographystyle{plain}
\bibliography{typetheory}
\end{document}

%% file: lstlean.tex
\lstdefinelanguage{lean} {

mathescape=false,
texcl=false,

morekeywords=[3]{
repeat, cases, rewrite, rw,
simp, simp_all, contradiction,
simplify, injection,
Given,
},

morecomment=[s][\color{commentcolor}]{/-}{-/},
morecomment=[l][\itshape \color{commentcolor}]{--},

showstringspaces=false,

keepspaces=true,

morestring=[b]",
morestring=[d]’,

tabsize=3,

extendedchars=true,

inputencoding=utf8,

literate=
  {∘}{{\ensuremath{\circ}}}1
  {∧}{{\ensuremath{\land}}}1
  {∨}{{\ensuremath{\lor}}}1
  {∃}{{\ensuremath{\exists}}}1
  {∀}{{\ensuremath{\forall}}}1
  {⊆}{{\ensuremath{\subseteq}}}1
  {→}{{\ensuremath{\rightarrow}}}1
  {↔}{{\ensuremath{\leftrightarrow}}}1
  {×}{{\ensuremath{\times}}}1
  {¬}{{\ensuremath{\neg}}}1
  {≤}{{\ensuremath{\leq}}}1
  {≥}{{\ensuremath{\geq}}}1
  {≠}{{\ensuremath{\neq}}}1
  {λ}{{\ensuremath{\lambda}}}1
  {α}{{\ensuremath{\alpha}}}1
  {β}{{\ensuremath{\beta}}}1
  {γ}{{\ensuremath{\gamma}}}1
  {δ}{{\ensuremath{\delta}}}1
  {ε}{{\ensuremath{\varepsilon}}}1
  {θ}{{\ensuremath{\theta}}}1
  {σ}{{\ensuremath{\sigma}}}1
  {τ}{{\ensuremath{\tau}}}1
  {φ}{{\ensuremath{\varphi}}}1
  {ψ}{{\ensuremath{\psi}}}1
  {ω}{{\ensuremath{\omega}}}1
  {Γ}{{\ensuremath{\Gamma}}}1
  {Δ}{{\ensuremath{\Delta}}}1
  {Σ}{{\ensuremath{\Sigma}}}1
  {Π}{{\ensuremath{\Pi}}}1
  {≡}{{\ensuremath{\equiv}}}1
  {≈}{{\ensuremath{\approx}}}1
  {∞}{{\ensuremath{\infty}}}1,

sensitive=true,

breaklines=true,
breakatwhitespace=true,

basicstyle=\ttfamily\small,

captionpos=b,

columns=[l]fullflexible,

identifierstyle={\ttfamily\color{black}},

keywordstyle=[1]{\ttfamily\color{keywordcolor}},

keywordstyle=[2]{\ttfamily\color{keywordcolor}},

keywordstyle=[3]{\ttfamily\color{tacticcolor}},

keywordstyle=[4]{\ttfamily\color{attributecolor}},

stringstyle=\ttfamily,

commentstyle={\ttfamily\footnotesize },

}

\begingroup
\catcode`\|=0
\catcode`\\=12
|lstdefinelanguage[arend]{lean}[]{lean}{%
  alsoletter={\},%
  morekeywords=[1]{\as,\axiom,\box,\case,\class,\classifying,%
   \coerce,\cons,\cowith,\data,\default,\elim,%
   \eval,\extends,\field,\fix,\fixl,\fixr,%
   \func,\have,\hiding,\hType,\import,\in,%
   \infix,\infixl,\infixr,\instance,\lam,\lemma,%
   \let,\level,\levels,\lh,\lp,\max,%
   \meta,\module,\new,\noclassifying,\open,\override,%
   \peval,\plevels,\postfix,\prefix,\private,\property,%
   \protected,\record,\return,\scase,\sfunc,\strict,%
   \suc,\this,\truncated,\type,\use,\using,%
   \where,\with,\Pi,\Sigma},%
  morekeywords=[2]{\Prop,\Set,\Type}%
}
|endgroup